\documentclass[journal]{IEEEtran}

\usepackage[T1]{fontenc}
\usepackage{textcomp}
\usepackage{blindtext}
\usepackage[utf8]{inputenc}
\usepackage[noadjust]{cite}
\usepackage[font=footnotesize]{caption}
\usepackage{dblfloatfix}
\usepackage{graphicx}
\usepackage{psfrag}
\usepackage{subfig}
\usepackage{amsmath,amssymb,amsthm,mathrsfs,amsfonts,dsfont}
\usepackage{color}
\usepackage[algoruled]{algorithm2e}
\usepackage{fixltx2e}
\usepackage{multirow}
\usepackage{multicol}
\usepackage[hyphens]{url}
\usepackage[hidelinks]{hyperref}
\usepackage[anythingbreaks]{breakurl}
\usepackage[normalem]{ulem}
\usepackage{acronym}
\usepackage[table,xcdraw]{xcolor}
\usepackage{accents}
\usepackage{mwe}
\usepackage{hhline}
\usepackage{trfsigns}
\usepackage{siunitx}
\usepackage{lscape}
\usepackage{wasysym}
\usepackage{bm}

\theoremstyle{definition}

\newacro{3GPP}{3rd Generation Partnership Project}
\newacro{5G}{fifth generation}
\newacro{5G NR}{5G New Radio}
\newacro{6G}{sixth generation}
\newacro{A/D}{analog-to-digital}
\newacro{ADC}{analog-to-digital converter}
\newacro{AFE}{analog front-end}
\newacro{AGV}{automatic guided vehicle}
\newacro{AWGN}{additive white Gaussian noise}
\newacro{B5G}{beyond \ac{5G}}
\newacro{BB}{baseband}
\newacro{BER}{bit error ratio}
\newacro{BPSK}{binary phase-shift keying}
\newacro{BP}{bandpass}
\newacro{BS}{base station}
\newacro{CDM}{code-division multiplexing}
\newacro{CFO}{carrier frequency offset}
\newacro{CFR}{channel frequency response}
\newacro{CIR}{channel impulse response}
\newacro{CoMP}{coordinated multipoint}
\newacro{CP}{cyclic prefix}
\newacro{CPE}{common phase error}
\newacro{CPO}{carrier phase offset}
\newacro{CRLB}{Cram\'er–Rao lower bound}
\newacro{CS}{chirp sequence}
\newacro{CW}{continuous wave}
\newacro{CZT}{chirp Z-transform}
\newacro{D/A}{digital-to-analog}
\newacro{DAC}{digital-to-analog converter}
\newacro{DDC}{digital down-conversion}
\newacro{DDS}{direct digital synthesis}
\newacro{DFRC}{dual-function radar-cunication or dual-functional radar-cunication}
\newacro{DFnT}{discrete Fresnel transform}
\newacro{DFT}{discrete Fourier transform}
\newacro{DUC}{digital up-conversion}
\newacro{EVM}{error vector magnitude}
\newacro{FDE}{frequency-domain equalization}
\newacro{FDM}{frequency-division multiplexing}
\newacro{FDZP}{frequency-domain zero padding}
\newacro{FMCW}{frequency-modulated continuous wave}
\newacro{FO}{frequency offset}
\newacro{FR2}{Frequency Range 2}
\newacro{gNB}{gNodeB}
\newacro{HP}{high-pass}
\newacro{IBFD}{in-band full duplex}
\newacro{ICI}{intercarrier interference}
\newacro{IDFT}{inverse discrete Fourier transform}
\newacro{IDFnT}{inverse discrete Fresnel transform}
\newacro{IF}{intermediate frequency}
\newacro{IHE}{Institute of Radio Frequency Engineering and Electronics}
\newacro{ISAC}{integrated sensing and communication}
\newacro{ISI}{intersymbol interference}
\newacro{ISLR}{integrated sidelobe level ratio}
\newacro{I/Q}{in-phase/quadrature}
\newacro{JCAS}{joint communication and sensing}
\newacro{KIT}{Karlsruhe Institute of Technology}
\newacro{LDPC}{low-density parity-check}
\newacro{LFSR}{linear-feedback shift register}
\newacro{LNA}{low-noise amplifier}
\newacro{LO}{local oscillator}
\newacro{LoS}{line-of-sight}
\newacro{LP}{low-pass}
\newacro{LS}{least squares}
\newacro{mmWave}{milimeter wave}
\newacro{MIMO}{multiple-input multiple-output}
\newacro{MLE}{maximum likelihood estimator}
\newacro{MLS}{maximum-length sequence}
\newacro{MUSIC}{multiple signal classification}
\newacro{NLoS}{non-line-of-sight}
\newacro{OCDM}{orthogonal chirp-division multiplexing}
\newacro{OFDM}{orthogonal frequency-division multiplexing}
\newacro{OOB}{out-of-band}
\newacro{OTA}{over-the-air}
\newacro{P/S}{paralell-to-serial}
\newacro{PA}{power amplifier}
\newacro{PACF}{periodic autocorrelation function}
\newacro{PCCF}{periodic cross-correlation function}
\newacro{PDF}{probability density function}
\newacro{PLC}{powerline cunication}
\newacro{PLL}{phase-locked loop}
\newacro{PMCW}{phase-modulated continuous wave}
\newacro{PMN}{perceptive mobile network}
\newacro{PN}{oscillator phase noise}
\newacro{PoC}{proof-of-concept}
\newacro{PPLR}{peak power loss ratio}
\newacro{PRBS}{pseudorandom binary sequence}
\newacro{PRS}{positioning reference signal}
\newacro{PSD}{power spectral density}
\newacro{PSLR}{peak sidelobe level ratio}
\newacro{PT-RS}{phase tracking reference signal}
\newacro{QAM}{quadrature amplitude modulation}
\newacro{QPSK}{quadrature phase-shift keying}
\newacro{RadCom}{radar-cunication}
\newacro{RCS}{radar cross section}
\newacro{RF}{radio-frequency}
\newacro{RIS}{reflective intelligent surface}
\newacro{RMS}{root mean square}
\newacro{RMSE}{root mean squared error}
\newacro{SC}[S\&C]{Schmidl \& Cox}
\newacro{SFO}{sampling frequency offset}
\newacro{SIC}{self-interference cancellation}
\newacro{SINR}{signal-to-interference-plus-noise ratio}
\newacro{SISO}{single-input single-output}
\newacro{SIR}{signal-to-interference ratio}
\newacro{SJ}{sampling jitter}
\newacro{SNR}{signal-to-noise ratio}
\newacro{SoC}{system-on-a-chip}
\newacro{STO}{symbol time offset}
\newacro{S/P}{serial-to-paralell}
\newacro{TDE}{time-domain equalization}
\newacro{TDM}{time-division multiplexing}
\newacro{TDR}{time-domain reflectometry}
\newacro{TITO}{tilt inference of time offset}
\newacro{TO}{time offset}
\newacro{UE}{user equipment}
\newacro{UWAC}{underwater acoustic cunication}
\newacro{V2V}{vehicle-to-vehicle}
\newacro{ZF}{zero forcing}
\newacro{ZP}{zero padding}
\usepackage{pgfplots}
\pgfplotsset{compat=newest}
\usetikzlibrary{plotmarks}
\usetikzlibrary{arrows.meta}
\usepgfplotslibrary{patchplots}
\usepackage{grffile}
\usepackage{amsmath}

\usepackage{calrsfs}

\usepackage{fancyhdr}
\fancypagestyle{empty}{fancy}

\begin{document}
	
	\title{\textcolor{black}{Analysis of Sensing in} OFDM-based\\ ISAC under the Influence of Sampling Jitter}
	
	\author{Lucas Giroto,~\IEEEmembership{Member,~IEEE}, \^{A}ndrei Camponogara, Yueheng Li~\IEEEmembership{Member,~IEEE},\\ Jiayi Chen,~\IEEEmembership{Graduate Student Member,~IEEE}, Lukas Sigg, Thomas Zwick,~\IEEEmembership{Fellow,~IEEE},\\ and Benjamin Nuss,~\IEEEmembership{Senior Member,~IEEE}
		\thanks{The authors acknowledge the financial support by the Federal Ministry of Education and Research of Germany in the projects ``KOMSENS-6G'' (grant number: 16KISK123) and ``Open6GHub'' (grant number: 16KISK010). \textit{(Corresponding author: Lucas Giroto.)}}
		\thanks{L. Giroto was with the Institute of Radio Frequency Engineering and Electronics (IHE), Karlsruhe Institute of Technology (KIT), 76131 Karlsruhe, Germany. He is now with Nokia Bell Labs, 70469 Stuttgart, Germany (e-mail: {lucas.giroto@nokia-bell-labs.com}).}
		\thanks{\^{A} Camponogara is with he Electrical Engineering Department,	Federal University of Paran\'a, Curitiba, PR 81530-000, Brazil (e-mail:  {andrei.camponogara@ufpr.br}).}
		\thanks{Y. Li \textcolor{black}{was} with the Institute of Radio Frequency Engineering and Electronics (IHE), Karlsruhe Institute of Technology (KIT), 76131 Karlsruhe, Germany. He is now with the Institute of Intelligent Communication Technology, Shandong University (SDU), 250100 Jinan, China (e-mail: {yueheng.li@sdu.edu.cn}).}
		\thanks{J. Chen and T. Zwick are with the Institute of Radio Frequency Engineering and Electronics (IHE), Karlsruhe Institute of Technology (KIT), 76131 Karlsruhe, Germany (e-mail: {jiayi.chen@kit.edu}, {thomas.zwick@kit.edu}).}
		\thanks{L. Sigg was with the Institute of Radio Frequency Engineering and Electronics (IHE), Karlsruhe Institute of Technology (KIT), 76131 Karlsruhe, Germany. He is now with MBDA Germany, 86529 Schrobenhausen, Germany (e-mail: {lukas.sigg@mbda-systems.de}).}
		\thanks{\textcolor{black}{{B. Nuss} was with the Institute of Radio Frequency Engineering and Electronics (IHE), Karlsruhe Institute of Technology (KIT), 76131 Karlsruhe, Germany. He is now with the Professorship of Microwave Sensors and Sensor Systems, Technical University of Munich, 80333 Munich, Germany. (e-mail: {benjamin.nuss@tum.de}).}}
	}
	
	
	\maketitle
	
	\begin{abstract}
		To enable integrated sensing and communication (ISAC) in cellular networks, a wide range of additional requirements and challenges are either imposed or become more critical. One such impairment is sampling jitter (SJ), which arises due to imperfections in the sampling instants of the clocks of digital-to-analog converters (DACs) and analog-to-digital converters (ADCs). While SJ is already well studied for communication systems based on orthogonal frequency-division multiplexing (OFDM), which is expected to be the waveform of choice for most sixth-generation (6G) scenarios where ISAC could be possible, the implications of SJ on the OFDM-based radar sensing must still be thoroughly analyzed. Considering that phase-locked loop (PLL)-based oscillators are used to derive sampling clocks, which leads to colored SJ, i.e., SJ with non-flat power spectral density, this article analyzes the resulting distortion of the  adopted digital constellation modulation and sensing performance in OFDM-based ISAC for both baseband (BB) and bandpass (BP) sampling strategies and different oversampling factors. For BB sampling, it is seen that SJ induces intercarrier interference (ICI), while for BP sampling, it causes carrier phase error and more severe ICI due to a phase noise-like effect at the digital intermediate frequency. \textcolor{black}{Obtained results for a single-input single-output OFDM-based ISAC system with various OFDM signal parameterizations} demonstrate that SJ-induced degradation becomes \textcolor{black}{non-negligible for both BB and BP sampling only for root mean square (RMS) SJ values above $\SI[parse-numbers = false]{10^{-11}}{\second}$ at both DAC and ADC, which corresponds to $0.5\times10^{-2}$ times the considered critical sampling period without oversampling.} \textcolor{black}{Based on the achieved results,  it can be concluded that state-of-the-art hardware enables sufficient communication and sensing robustness against SJ, as RMS SJ values in the femtosecond range can be achieved.}
	\end{abstract}
	
	
	\IEEEpeerreviewmaketitle
	

	\section{Introduction}\label{sec:introduction}
	
	\IEEEPARstart{W}{ith} the advent of \ac{6G} networks \cite{chafii2023}, \ac{ISAC} is expected to emerge as a pivotal technology \cite{lima2021,wild2021,liu2022,wymeersch2024}. The principle of \ac{ISAC} consists of simultaneously performing communication and radar sensing with shared hardware and spectral resources \cite{parssinen2021,giroto2021_tmtt}. With the added sensing functionality to the cellular infrastructure, a wide range of novel applications is expected to become possible. These include, e.g., environment sensing in urban scenarios to enable safe mobility and automation in intralogistics scenarios involving humans and \acp{AGV} \cite{kadelka2023,mandelli2023survey,shatov2024}. As a result, \ac{ISAC} is set to play a transformative role in next-generation cellular networks, with early developments anticipated in upcoming \ac{3GPP} \ac{6G} releases.
	
	One of the key challenges in \ac{ISAC} is enabling accurate radar sensing while meeting the main demand of \ac{6G} cellular networks, which is to support robust connectivity at high data rates \cite{viswanathan2020,rajatheva2020}. This is facilitated by the use of the multicarrier modulation scheme \ac{OFDM} \cite{hwang2009}, which is already employed in current \ac{5G} cellular systems and will most likely also be the choice for \ac{6G}. Besides a proper allocation of radio resources between communication and radar sensing, this requires jointly tackling impairments associated with one or both applications \cite{thomae2021,wild2023}. Among these are high isolation between closely located transmit and receive antenna arrays and \ac{SIC} requirements due to the required \ac{IBFD} operation for some communication applications and monostatic radar sensing \cite{barneto2019,nagulu2024,smida2024}, or hardware non-idealities which may lead to synchronization issues or signal impairments for communication and bistatic radar sensing \cite{thomae2019,bookSFO}. Among aforementioned non-idealities are time and frequency offset \cite{nguyen2002,speth1999}, \ac{SFO} \cite{tsai2005,wu2012}, \ac{PN} \cite{armada2001,syrjala2009}, and \ac{SJ} \cite{manoj2003,syrjala2010}, whose implications have already been widely studied for \ac{OFDM}-based communication systems. In the \ac{OFDM}-based radar and \ac{ISAC} context, recent studies have investigated the influence of most of these issues on radar sensing performance and proposed compensation strategies where feasible \cite{nuss2018,hakobyan2018,wang2023,pegoraro2024,brunner2024,aguilar2024,giroto2024,giroto2024PN}. However, the effects of \ac{SJ} on \ac{OFDM}-based sensing performance have been overlooked. 
	
	\ac{SJ} refers to the deviation in \ac{DAC} or \ac{ADC} sampling instants caused by imperfections in clock timing \cite{bookSFO}. According to \cite{loehning2007}, \ac{SJ} is composed of both aperture jitter, which represents the sampling time variations caused by noise in the sample-and-hold circuit of an \ac{ADC}, and clock jitter, which is raised by the clock generator used to feed the \ac{ADC} with a clock signal. The analysis in this study shows that, for short signal block durations as is typically the case in \ac{mmWave} \ac{OFDM}-based communication, clock jitter is the dominant effect. In addition, it is shown that increasing oversampling is ineffective against clock jitter-induced interference. Unlike aperture jitter that is spread over the whole digitization band, this happens since clock jitter interference is concentrated within the frequency band occupied by the input signal to an \ac{ADC}. Out of this reason, numerous studies consider only the effect of clock jitter. 
	
	When experienced in \ac{OFDM}-based systems with \ac{BB} sampling, \ac{SJ} leads to \ac{ICI} \cite{bookSFO}. For the case where \ac{BP} sampling is adopted, \cite{putra2009} and \cite{syrjala2010} presented an analysis of \ac{ADC} \ac{SJ} on an \ac{OFDM}-based communication system. These studies showed that, besides \ac{ICI}, \ac{SJ} results in an oscillator \ac{PN}-like effect due to the use of a digital \ac{IF}, leading to \ac{CPE}. For the \ac{BB} sampling case, \cite{manoj2003,yang2010} investigated the effect of oversampling on the \ac{ICI} induced by white \ac{ADC} \ac{SJ}, where the \ac{SJ} at different sampling time instants is uncorrelated as opposed to colored \ac{SJ} which has non-flat \ac{PSD}, without considering \ac{DAC} \ac{SJ}. Furthermore, \cite{onunkwo2006} derived closed-form expressions and upper bounds for the \ac{ICI} power due to white and colored \ac{ADC} \ac{SJ} only, also considering the effect of oversampling. It was shown that, for colored \ac{SJ}, oversampling is not as effective as for the white \ac{SJ} case, and that a residual \ac{ICI} is always present.
	
	Shifting the focus from the receiver to the transmitter side, \cite{alian2015} presented an analysis of white \ac{DAC} \ac{SJ}, showing that it leads to \ac{OOB} radiation and investigating spectral sidelobe supression techniques to compensate for it. Finally, \cite{gonem2021}, analyzed the effects of white and colored \ac{SJ} at both \ac{ADC} and \ac{DAC} for hybrid \ac{OFDM}-digital filter multiple access passive optical networks. Afterwards, the authors proposed a \ac{SJ} simulation model consisting of oversampling, a fractional delay filter at the increased sampling rate, and downsampling. Based on the obtained results, it was concluded that the effects of \ac{SJ} are virtually the same regardless of whether they take place at a \ac{DAC} or an \ac{ADC}. In addition, it was observed that oversampling at the \ac{ADC} is ineffective against \ac{DAC} jitter.
	
	In future \ac{6G} \ac{ISAC} systems, which are expected to operate at increasingly higher frequencies and require tighter synchronization margins, understanding the effects of \ac{SJ} and its implications on not only communication, but also on the radar sensing performance becomes crucial. In addition, to better understand how sampling jitter can affect the performance of \ac{OFDM}-based \ac{ISAC}, a larger variety of \ac{OFDM} signal parameters (i.e., number of subcarriers and modulation alphabet) than in the aforementioned studies needs to be considered. Furthermore, a \ac{SJ} model that better reflects the behavior of state-of-the-art hardware must be adopted. Such model should, e.g., consider the use of \ac{PLL}-based oscillators to derive \ac{DAC} and \ac{ADC} sampling clocks, which leads to colored \ac{SJ}.
	
	In this sense, this article investigates the impact of colored \ac{SJ} raised by the use of a \ac{PLL}-based oscillator to generate the sampling clock of \textcolor{black}{\ac{SISO}} \ac{OFDM}-based \ac{ISAC} systems. In this context, both communication and radar sensing \textcolor{black}{impairment} are assessed. Specifically, the degradation of the constellation shapes induced by \ac{SJ} at both \ac{DAC} and \ac{ADC} is analyzed using \ac{EVM} and \ac{SIR} as \textcolor{black}{parameters}. In addition, both \ac{BB} and \ac{BP} sampling strategies, as well as several distinct numbers of \ac{OFDM} subcarriers, modulation orders, and oversampling factors used in \ac{FDZP} are considered. Next, the resulting implications on the radar sensing capabilities of \ac{OFDM}-based \ac{ISAC} systems are analyzed. For that purpose, the degradation of main and sidelobes of radar target reflections, as well as the \ac{SIR} reduction due to \ac{SJ} is analyzed for the aforementioned sets of parameters.
	
	The contributions of this article can be summarized as follows:
	\begin{itemize}
		
		\item A mathematical formulation of the effects of \ac{SJ} is presented. A closed-form expression is provided that relates the received \ac{OFDM} symbols in the discrete-frequency domain to both the transmitted \ac{OFDM} symbols and the \ac{SJ} at the \ac{DAC} and \ac{ADC} stages.
		
		\item  A numerical analysis is conducted on the impact of colored \ac{SJ}, caused by \ac{PN} in \ac{PLL}-based oscillators, on \ac{OFDM}-based \ac{ISAC} systems with various waveform configurations. These include different constellation sizes, oversampling factors for \ac{FDZP} at the transmitter side, and subcarrier numbers. \ac{EVM} and \ac{SIR} are used to quantify constellation distortions, while \textcolor{black}{radar sensing} is assessed using \ac{PPLR}, \ac{PSLR} and \ac{ISLR} in both range and Doppler shift directions of radar images, and the \ac{SIR} of target reflections in radar images, henceforth referred to as image \ac{SIR}.
		
		\item A comparison of \ac{OFDM}-based \ac{ISAC} systems with \ac{BB} and \ac{BP} sampling strategies under the influence of \ac{SJ} is presented. While it is known in the literature that \ac{SJ} leads to \ac{ICI} when \ac{BB} sampling is adopted as well as in both \ac{ICI} and \ac{CPE} in the \ac{BP} sampling case as \ac{SJ} results in a \ac{PN}-like effect at the digital \ac{IF}, the consequences of these effects on the \textcolor{black}{sensing} is quantified and analyzed in this article.
		
	\end{itemize}
	
	The remainder of this article is organized as follows. Section~\ref{sec:sysModel} formulates the system model of an \ac{OFDM}-based \ac{ISAC} system impaired \ac{SJ}. Sections~\ref{sec:commPerf} and \ref{sec:sensPerf}  present analyses of the \ac{SJ}-induced \textcolor{black}{impairment} of the considered \ac{ISAC} system under \ac{SJ} influence for communication and radar sensing, respectively. To conclude the discussion, Section~\ref{sec:conclusion} provides final remarks and discusses the implications of the findings of this article.
	
	\begin{figure*}[!b]
		\setcounter{equation}{2}
		\hrulefill
		\vspace*{4pt}
		\begin{equation}\label{eq:x_t_SJ}
			x^\mathrm{SJ}_\mathrm{BB}(t)=\sum_{\nu=-\eta N_\mathrm{cp}}^{\eta N-1}x_\nu\frac{\sin\left(\pi\left[t-\left(\nu T_\mathrm{s}/\eta+\delta^\mathrm{SJ}_\mathrm{DAC}\left(\nu T_\mathrm{s}/\eta\right)\right)\right]/\left(T_\mathrm{s}/\eta\right)\right)}{\pi\left[t-\left(\nu T_\mathrm{s}/\eta+\delta^\mathrm{SJ}_\mathrm{DAC}\left(\nu T_\mathrm{s}/\eta\right)\right)\right]/\left(T_\mathrm{s}/\eta\right)}
		\end{equation}
	\end{figure*}
	\begin{figure*}[!b]
		\setcounter{equation}{5}
		\hrulefill
		\vspace*{4pt}
		\begin{equation}\label{eq:r_nu_2}
			y_{\nu'}=\sum_{\nu=-N_\mathrm{cp}}^{\eta N-1}s_\nu\frac{\sin\left(\pi\left[\left(\nu'T_\mathrm{s}/\eta+\delta^\mathrm{SJ}_\mathrm{ADC}\left(\nu' T_\mathrm{s}/\eta\right)\right)-\left(\nu T_\mathrm{s}/\eta+\delta^\mathrm{SJ}_\mathrm{DAC}\left(\nu T_\mathrm{s}/\eta\right)\right)\right]/\left(T_\mathrm{s}/\eta\right)\right)}{\pi\left[\left(\nu'T_\mathrm{s}/\eta+\delta^\mathrm{SJ}_\mathrm{ADC}\left(\nu' T_\mathrm{s}/\eta\right)\right)-\left(\nu T_\mathrm{s}/\eta+\delta^\mathrm{SJ}_\mathrm{DAC}\left(\nu T_\mathrm{s}/\eta\right)\right)\right]/\left(T_\mathrm{s}/\eta\right)}
		\end{equation}
	\end{figure*}
	\begin{figure*}[!b]
		\setcounter{equation}{7}
		\hrulefill
		\vspace*{4pt}
		\begin{equation}\label{eq:Yl_2}
			Y_l=\sum_{\nu'=0}^{\eta N-1}\left(\sum_{\nu=-\eta N_\mathrm{cp}}^{\eta N-1}x_\nu\frac{\sin\left(\pi\left[\left(\nu'-\nu\right)T_\mathrm{s}/\eta+\left(\delta^\mathrm{SJ}_\mathrm{ADC}\left(\nu' T_\mathrm{s}/\eta\right)-\delta^\mathrm{SJ}_\mathrm{DAC}\left(\nu T_\mathrm{s}/\eta\right)\right)\right]/\left(T_\mathrm{s}/\eta\right)\right)}{\pi\left[\left(\nu'-\nu\right)T_\mathrm{s}/\eta+\left(\delta^\mathrm{SJ}_\mathrm{ADC}\left(\nu' T_\mathrm{s}/\eta\right)-\delta^\mathrm{SJ}_\mathrm{DAC}\left(\nu T_\mathrm{s}/\eta\right)\right)\right]/\left(T_\mathrm{s}/\eta\right)}\right)\e^{\im2\pi l\nu'/\left(\eta N\right)}
		\end{equation}
	\end{figure*}
	
	\section{System Model}\label{sec:sysModel}
	
	At the transmitter side of the considered \textcolor{black}{\ac{SISO}} \ac{OFDM}-based \ac{ISAC} system, constellation symbols \mbox{$\mathbf{d}_\text{Tx}^\text{OFDM}\in\mathbb{C}^{N\times1}$} are mapped into a discrete-frequency domain transmit \ac{OFDM} symbol \mbox{$\mathbf{S}\in\mathbb{C}^{N\times 1}$} that contains \mbox{$N\in\mathbb{N}_{>0}$} subcarriers spaced by \mbox{$\Delta f=B/N$}, where $B$ is the \ac{BB} bandwidth later occupied by the \ac{OFDM} signal. It is worth emphasizing that, although the following formulation assumes a single \ac{OFDM} symbol for clarity and without loss of generality, a frame containing multiple \ac{OFDM} symbols is typically transmitted in practice. The transmitter processing chain continues with \ac{FDZP}\textcolor{black}{, which consists of appending empty subcarriers to the edges of the \ac{OFDM} symbol in the discrete-frequency domain,} to perform oversampling with factor \mbox{$\eta\in\mathbb{N}_{>0}$}, which results in the \ac{OFDM} symbol \mbox{$\mathbf{X}\in\mathbb{C}^{\eta N\times 1}$}. Next, by \ac{IDFT} is performed on $\mathbf{X}$ and \ac{CP} prepending to the resulting discrete-time domain \ac{OFDM} symbol. The resulting discrete-time domain transmit \ac{OFDM} symbol is denoted as \mbox{$\mathbf{x}_\mathrm{CP}\in\mathbb{C}^{\eta(N+N_\mathrm{CP})\times 1}$}, where $N_\mathrm{CP}$ is the \ac{CP} length without oversampling. $\mathbf{x}_\mathrm{CP}$ then undergoes \ac{P/S} conversion to form a stream of samples that finally undergoes \ac{D/A} conversion with sampling rate \mbox{$F^\eta_\mathrm{s}=\eta B$}. Consequently, a sampling period \mbox{$T^\eta_\mathrm{s}=1/F^\eta_\mathrm{s}=1/\left(\eta B\right)$}, which is $\eta$ times smaller than the critical sampling period \mbox{$T_\mathrm{s}=1/B$}, is obtained. The resulting \ac{BB} analog signal \mbox{$x_\mathrm{BB}(t)\in\mathbb{C}$} from \ac{D/A} conversion with an ideal, bandlimited \ac{DAC} to \mbox{$f\in[-F_\mathrm{s}/2,F_\mathrm{s}/2]$}, where $f$ denotes frequency, can be expressed as
	\setcounter{equation}{0}
	\begin{equation}\label{eq:x_t}
		x_\mathrm{BB}(t)=\sum_{\nu=-\eta N_\mathrm{cp}}^{\eta N-1}x_\nu\frac{\sin\left(\pi\left(t-\nu T_\mathrm{s}/\eta\right)/\left(T_\mathrm{s}/\eta\right)\right)}{\pi\left(t-\nu T_\mathrm{s}/\eta\right)/\left(T_\mathrm{s}/\eta\right)}
	\end{equation}
	for $-T_\mathrm{CP}\leq t<T$, where \mbox{$T_\mathrm{CP}=N_\mathrm{CP}T_\mathrm{s}$} and \mbox{$T=NT_\mathrm{s}$} are the durations of the \ac{CP} and the \ac{OFDM} symbol disregarding \ac{CP}, respectively. In \eqref{eq:x_t}, \mbox{$x_\nu\in\mathbb{C}$} is the $\nu\mathrm{th}$ sample, $\nu\in\{-\eta N_\mathrm{cp},-\eta N_\mathrm{cp}+1,\cdots,\eta N-1\}$, of $\mathbf{x}_\mathrm{CP}$.
	
	The expression in \eqref{eq:x_t} is only valid for ideal sampling instants. In practice, however, the \ac{DAC} sampling clock is usually derived from a \ac{PLL}-based oscillator, which has a \ac{PN} \mbox{$\theta^\mathrm{PN}_\mathrm{DAC}(t)\in\mathbb{R}$} with double-sided \ac{PSD} \mbox{$S_{\theta^\mathrm{PN}_\mathrm{DAC}}(f)\in\mathbb{R}_{\geq0}$} \cite{bookSFO,khanzadi2014,schweizer2018,walt2023}. The presence of the \ac{PN} $\theta^\mathrm{PN}_\mathrm{DAC}(t)$ in the oscillator used to derive the sampling clock ultimately leads to a \ac{SJ} \mbox{$\delta^\mathrm{DAC}_\mathrm{SJ}(t)\in\mathbb{R}$}, whose relationship to the aforementioned \ac{PN} can be expressed as \cite{bookSFO}
	\setcounter{equation}{1}
	\begin{equation}\label{eq:PN_jitter}
		\delta^\mathrm{DAC}_\mathrm{SJ}(t) = \theta^\mathrm{PN}_\mathrm{DAC}(t)/\left(2\pi F^\eta_\mathrm{s}\right).
	\end{equation}
	Considering the effect of \ac{SJ} \textcolor{black}{and neglecting further impairments such as \ac{SFO}}, \mbox{$x_\mathrm{BB}(t)\in\mathbb{C}$} becomes \mbox{$x^\mathrm{SJ}_\mathrm{BB}(t)\in\mathbb{C}$}, and \eqref{eq:x_t} can be rewritten as \eqref{eq:x_t_SJ}.
	
	Assuming a \textcolor{black}{noiseless,} ideal single-path channel \textcolor{black}{and still neglecting further impairments}, the receive \ac{BB} signal \mbox{$y_\mathrm{BB}(t)\in\mathbb{C}$} can simply be expressed as
	\setcounter{equation}{3}
	\begin{equation}\label{eq:r_t}
		y_\mathrm{BB}(t)=x^\mathrm{SJ}_\mathrm{BB}(t).
	\end{equation}
	After sampling at an \ac{ADC} with \ac{SJ} \mbox{$\delta^\mathrm{ADC}_\mathrm{SJ}(t)\in\mathbb{R}$} also resulting from the use of a \ac{PLL}-based oscillator to derive the sampling clock, the $\nu'\mathrm{th}$ sample \mbox{$y_{\nu'}\in\mathbb{C}$}, \mbox{$\nu'\in\{-\eta N_\mathrm{cp},-N_\mathrm{cp}+1,\cdots,\eta N-1\}$}, of the discrete-time domain receive \ac{OFDM} symbol \mbox{$\mathbf{y}_\mathrm{CP}\in\mathbb{C}^{\eta(N+N_\mathrm{CP})\times 1}$} is obtained as
	\setcounter{equation}{4}
	\begin{align}\label{eq:r_nu_1}
		y_{\nu'} &= y_\mathrm{BB}\left(\nu'T_\mathrm{s}/\eta+\delta^\mathrm{SJ}_\mathrm{ADC}\left(\nu' T_\mathrm{s}/\eta\right)\right)\nonumber\\
		&= x^\mathrm{SJ}_\mathrm{BB}\left(\nu'T_\mathrm{s}/\eta+\delta^\mathrm{SJ}_\mathrm{ADC}\left(\nu' T_\mathrm{s}/\eta\right)\right)
	\end{align}
	To expand \eqref{eq:r_nu_1}, \eqref{eq:r_t} and \eqref{eq:x_t_SJ} are used, which results in \eqref{eq:r_nu_2}. Next, the \ac{CP} is removed from $\mathbf{y}_\mathrm{CP}$, and a \ac{DFT} is performed to produce the discrete-frequency domain receive \ac{OFDM} symbol \mbox{$\mathbf{Y}\in\mathbb{C}^{N\times 1}$}, whose $l\mathrm{th}$ element, \mbox{$l\in\{-\eta N/2, -\eta N/2+1, \dots, \eta N/2-1\}$}, is expressed as
	\setcounter{equation}{6}
	\begin{equation}\label{eq:Yl_1}
		Y_l=\sum_{\nu'=0}^{\eta N-1}y_{\nu'}\exp^{\im2\pi\frac{l\nu'}{\eta N}}.
	\end{equation}
	The expression in \eqref{eq:r_nu_2} is then substituted into \eqref{eq:Yl_1} to yield \eqref{eq:Yl_2}. 
	\begin{figure*}[!t]
		\setcounter{equation}{8}
		\begin{equation}\label{eq:taylor1}
			\frac{\sin\left(\omega\left[\left(\nu'-\nu\right)T_\mathrm{s}/\eta+\left(\delta^\mathrm{SJ}_\mathrm{ADC}\left(\nu' T_\mathrm{s}/\eta\right)-\delta^\mathrm{SJ}_\mathrm{DAC}\left(\nu T_\mathrm{s}/\eta\right)\right)\right]\right)}{\omega\left[\left(\nu'-\nu\right)T_\mathrm{s}/\eta+\left(\delta^\mathrm{SJ}_\mathrm{ADC}\left(\nu' T_\mathrm{s}/\eta\right)-\delta^\mathrm{SJ}_\mathrm{DAC}\left(\nu T_\mathrm{s}/\eta\right)\right)\right]}.
		\end{equation}
		\vspace*{4pt}
		\hrulefill
	\end{figure*}
	\begin{figure*}[!t]
		\setcounter{equation}{9}
		\begin{equation}\label{eq:taylor2}
			\frac{\sin\left(\pi\left(\nu'-\nu\right)\right)}{\pi\left(\nu'-\nu\right)}-\left(\frac{\pi\left(\nu'-\nu\right)\cos\left(\pi\left(\nu'-\nu\right)\right)-\sin\left(\pi\left(\nu'-\nu\right)\right)}{\pi^2\left(\nu'-\nu\right)^2}\right)\omega\left(\delta^\mathrm{SJ}_\mathrm{ADC}\left(\nu' T_\mathrm{s}/\eta\right)-\delta^\mathrm{SJ}_\mathrm{DAC}\left(\nu T_\mathrm{s}/\eta\right)\right)
		\end{equation}
		\vspace*{4pt}
		\hrulefill
	\end{figure*}
	\begin{figure*}[!t]
		\setcounter{equation}{10}
		\begin{equation}\label{eq:taylor3}
			\varphi\left(\nu,\nu'\right) = \left\{\arraycolsep=3pt\def\arraystretch{1.3}
			\begin{array}{ll}
				1, & \nu'=\nu\\
				\frac{\cos\left(\pi\left(\nu'-\nu\right)\right)}{\pi\left(\nu'-\nu\right)}\omega\left(\delta^\mathrm{SJ}_\mathrm{ADC}\left(\nu' T_\mathrm{s}/\eta\right)-\delta^\mathrm{SJ}_\mathrm{DAC}\left(\nu T_\mathrm{s}/\eta\right)\right), & \nu'\neq\nu
			\end{array}\right.
		\end{equation}
		\vspace*{4pt}
		\hrulefill
	\end{figure*}
	\begin{figure*}[!t]
		\setcounter{equation}{11}
		\begin{equation}\label{eq:Yl_3}
			Y_l=X_l+\sum_{\nu'=0}^{\eta N-1}\left(\sum_{\nu=-N_\mathrm{cp},\nu\neq\nu'}^{\eta N-1}x_\nu\frac{\cos\left(\pi\left(\nu'-\nu\right)\right)}{\pi\left(\nu'-\nu\right)}\omega\left(\delta^\mathrm{SJ}_\mathrm{ADC}\left(\nu' T_\mathrm{s}/\eta\right)-\delta^\mathrm{SJ}_\mathrm{DAC}\left(\nu T_\mathrm{s}/\eta\right)\right)\right)\e^{\im2\pi l\nu'/\left(\eta N\right)}
		\end{equation}
		\vspace*{4pt}
		\hrulefill
	\end{figure*}
	
	Assuming $\omega=\pi/\left(T_\mathrm{s}/\eta\right)$, the fraction in \eqref{eq:Yl_2} can be rewritten as the function \mbox{$\varphi\left(\nu,\nu'\right)$} of $\nu$ and $\nu'$ in \eqref{eq:taylor1}. Expanding \eqref{eq:taylor1} via Taylor series and keeping only the first two terms yields \eqref{eq:taylor2}.	The latter equation can be further rewritten as \eqref{eq:taylor3}, which allows rewriting \eqref{eq:Yl_2} as \eqref{eq:Yl_3}. After discarding the subcarriers added at the transmitter side for \ac{FDZP}, the discrete-frequency domain receive \ac{OFDM} symbol \mbox{$\mathbf{R}\in\mathbb{C}^{N\times 1}$} is obtained. Since $(\eta-1)N$ inactive subcarriers are added at the transmitter side for \ac{FDZP}, $(\eta-1)N/2$ to the left and $(\eta-1)N/2$ to the right of the discrete-frequency domain spectrum, the $k\mathrm{th}$ element \mbox{$R_k\in\mathbb{C}$}, \mbox{$k\{-N/2, -N/2+1, \dots, N/2-1\}$}, of $\mathbf{R}$ is expressed as
	\begin{equation}\label{eq:Rk_1}
		R_k = Y_{k+\eta N/2+1}.
	\end{equation}
		
	An analysis of the obtained expressions in \eqref{eq:Yl_3} and \eqref{eq:Rk_1} reveals that the non-ideal sampling results in an \ac{ICI} term that depends on all discrete-time domain samples, including the ones in the \ac{CP}, as well as their mutual leakage due to \ac{SJ} at both \ac{DAC} and \ac{ADC}.	If only \ac{SJ} at the \ac{DAC} is considered, it is known that unwanted \ac{OOB} radiation will occur \cite{alian2015}. Combined with \ac{ADC} \ac{SJ}, this will not only result in \ac{ICI} among the subcarriers of interest, but in more severe \ac{ICI} due to the leakage into the subcarriers originally used for \ac{FDZP} at the transmitter side that were supposed to be inactive. The aforementioned issues are the only impairments induced by \ac{SJ} when \ac{BB} sampling is performed. When \ac{BP} sampling is adopted instead, the transmit samples undergo \ac{DUC} to an \ac{IF} before \ac{D/A} conversion at the transmitter side, and the receive samples undergo \ac{DDC} from the \ac{IF} into the baseband after \ac{A/D} conversion at the receiver side. Consequently, not only \ac{OOB} radiation due to \ac{DAC} \ac{SJ} and \ac{ICI} due to both \ac{DAC} and \ac{ADC} \ac{SJ} occur, but also a \ac{PN}-like effect is induced to the \ac{IF} as discussed in \cite{putra2009}, which in turn results in more \ac{ICI} and \ac{CPE} \cite{giroto2024PN}. The implications of the discussed degradations induced by \ac{SJ} and \textcolor{black}{their effects on the communication and radar sensing} in \ac{OFDM}-based \ac{ISAC} systems with both \ac{BB} and \ac{BP} sampling are analyzed in Sections~\ref{sec:commPerf} and \ref{sec:sensPerf}, respectively.	
	
	\section{Numerical Results for Communication}\label{sec:commPerf}
	
	To analyze the \ac{SJ}-induced \textcolor{black}{impairments} in an \ac{mmWave} \ac{OFDM}-based \ac{ISAC} system, the set of \ac{OFDM} signal parameters listed in Table~\ref{tab:ofdmParameters} are adopted, being the specifically considered parameter values are mentioned in the following subsections. The digital \acp{IF} \mbox{$f_\mathrm{IF}=\SI{0}{\hertz}$} and \mbox{$f_\mathrm{IF}=\SI{1}{\giga\hertz}$} are for \ac{BB} and \ac{BP} sampling, respectively. As for the adopted \ac{RF} carrier and frequency bandwidth in Table~\ref{tab:ofdmParameters}, they are close to what is typically adopted at initial part of the \ac{5G NR} \ac{FR2}. The somewhat deviating values from the \ac{5G NR} were chosen to yield reasonable performance due to the constraints in the measurement setup for \ac{ISAC} measurements adopted in previous works \cite{li2024,giroto2024,giroto2024PN}.
	
	\begin{table}[!t]
		\renewcommand{\arraystretch}{1.5}
		\arrayrulecolor[HTML]{708090}
		\setlength{\arrayrulewidth}{.1mm}
		\setlength{\tabcolsep}{4pt}
		
		\centering
		\captionsetup{width=43pc,justification=centering,labelsep=newline}
		\caption{\textsc{Adopted OFDM signal parameters}}
		\label{tab:ofdmParameters}
		\resizebox{\columnwidth}{!}{
			\begin{tabular}{|cc|}
				\hhline{|==|}
				\multicolumn{1}{|c|}{\textbf{Carrier frequency} ($f_\text{c}$)}      & $\SI{26.2}{\giga\hertz}$ \\ \hline
				\multicolumn{1}{|c|}{\textbf{Bandwidth} ($B$)}      & $\SI{500}{\mega\hertz}$ \\ \hline
				\multicolumn{1}{|c|}{\textbf{Critical sampling period} ($T_\mathrm{s}$)}      & $\SI{2}{\nano\second}$ \\ \hline				
				\multicolumn{1}{|c|}{\textbf{Digital IF} ($f_\mathrm{IF}$)}      & $\{\SI{0}{\hertz},\SI{1}{\giga\hertz}\}$ \\ \hline
				\multicolumn{1}{|c|}{\textbf{Oversampling factor} ($\eta$)}      & $\{1, 2, 4, 8\}$ \\ \hline
				\multicolumn{1}{|c|}{\textbf{No. of subcarriers} ($N$)}      & $\{256, 512, 1024, 2048, 4096, 8192, 16384\}$ \\ \hline
				\multicolumn{1}{|c|}{\textbf{CP length} ($N_\mathrm{CP}$)}      & $\{0, N/4, N\}$ \\ \hline
				\multicolumn{1}{|c|}{\textbf{No. of OFDM symbols} ($M$)}      & $\{128\}$ \\ \hline
				\multicolumn{1}{|c|}{\textbf{Modulation alphabet}}      & \{QPSK, 16-QAM, 64-QAM, 256-QAM\} \\  \hhline{|==|}	
				
			\end{tabular}
		}
	\end{table}
	
\begin{figure}[!t]
		\centering
		\subfloat[ ]{

			\psfrag{22}[c][c]{\scriptsize $10^2$}
			\psfrag{44}[c][c]{\scriptsize $10^4$}
			\psfrag{66}[c][c]{\scriptsize $10^6$}
			\psfrag{88}[c][c]{\scriptsize $10^8$}
			
			\psfrag{-180}[c][c]{\scriptsize -$180$}
			\psfrag{-150}[c][c]{\scriptsize -$150$}
			\psfrag{-120}[c][c]{\scriptsize -$120$}
			\psfrag{-90}[c][c]{\scriptsize -$90$}
			
			\psfrag{Frequency offset (Hz)}[c][c]{\footnotesize Frequency offset (Hz)}
			\psfrag{Phase noise PSD (dBc/Hz)}[c][c]{\vspace{-0.1cm}\footnotesize PN PSD (dBc/Hz)}
			
			\includegraphics[width=3.75cm]{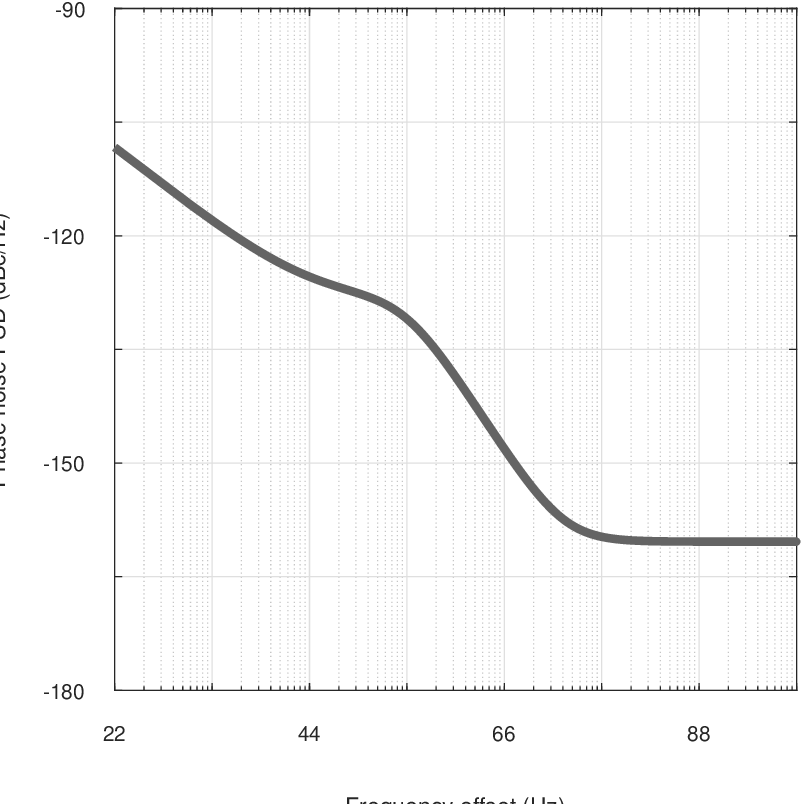}\label{fig:samplingJitter_PNpsd}			
		}\hspace{0.1cm}
		\subfloat[ ]{

			\psfrag{44}[c][c]{\scriptsize $10^4$}
			\psfrag{55}[c][c]{\scriptsize $10^5$}
			\psfrag{66}[c][c]{\scriptsize $10^6$}
			\psfrag{77}[c][c]{\scriptsize $10^7$}
			\psfrag{88}[c][c]{\scriptsize $10^8$}
			\psfrag{99}[c][c]{\scriptsize $10^9$}
			
			\psfrag{-85}[c][c]{\scriptsize -$85$}
			\psfrag{-80}[c][c]{\scriptsize -$80$}
			\psfrag{-75}[c][c]{\scriptsize -$75$}
			\psfrag{-70}[c][c]{\scriptsize -$70$}
			\psfrag{-65}[c][c]{\scriptsize -$65$}
			
			\psfrag{Frequency offset (Hz)}[c][c]{\footnotesize Frequency offset (Hz)}
			\psfrag{Int. phase noise level (dBc)}[c][c]{\footnotesize Int. PN level (dBc)}
			
			\includegraphics[width=3.75cm]{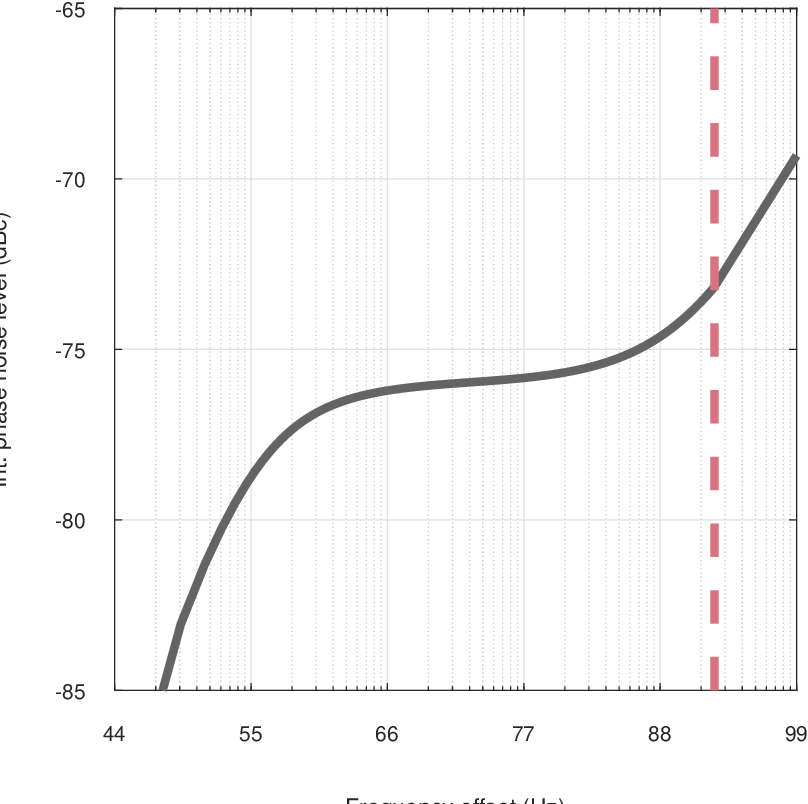}\label{fig:samplingJitter_intPN}			
		}\\
		\subfloat[ ]{
			
			\psfrag{-1}[c][c]{\scriptsize -$1$}
			\psfrag{-0.5}[c][c]{\scriptsize -$0.5$}
			\psfrag{0}[c][c]{\scriptsize $0$}
			\psfrag{0.5}[c][c]{\scriptsize $0.5$}
			\psfrag{1}[c][c]{\scriptsize $1$}
			
			\psfrag{0}[c][c]{\scriptsize $0$}
			\psfrag{1}[c][c]{\scriptsize $1$}
			\psfrag{2}[c][c]{\scriptsize $2$}
			\psfrag{3}[c][c]{\scriptsize $3$}
			
			\psfrag{Phase noise (rad)}[c][c]{\footnotesize Phase offset ($\times 10^{-3}$ rad)}
			\psfrag{Probability density}[c][c]{\footnotesize Probability density($\times 10^{3}$)}
			
			\includegraphics[width=3.75cm]{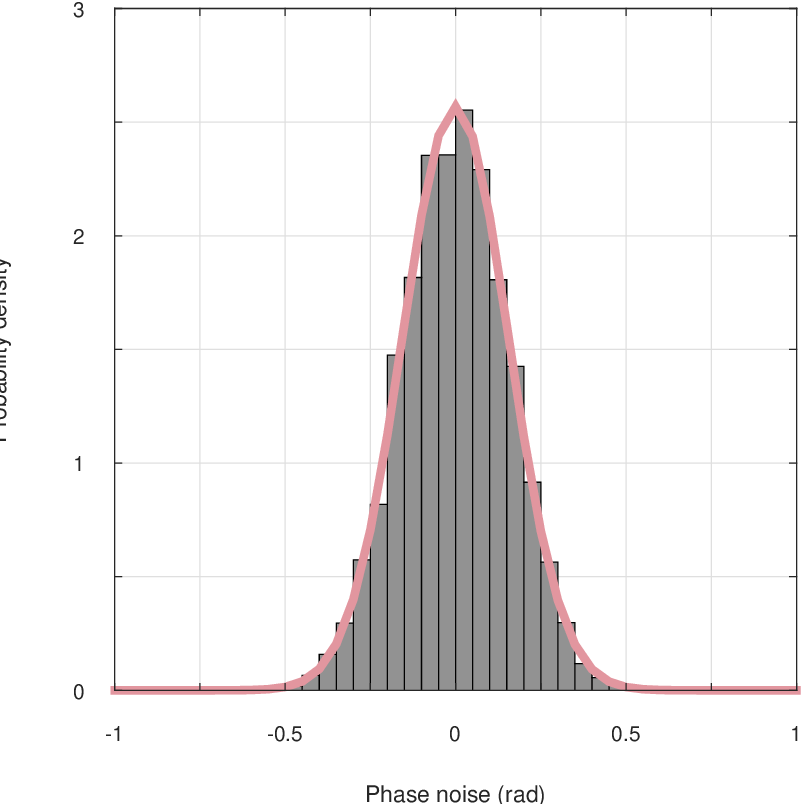}\label{fig:samplingJitter_PNdist}			
		}\hspace{0.1cm}
		\subfloat[ ]{
			
			\psfrag{-2}[c][c]{\scriptsize -$1$}
			\psfrag{-1}[c][c]{\scriptsize -$0.5$}
			\psfrag{0}[c][c]{\scriptsize $0$}
			\psfrag{1}[c][c]{\scriptsize $0.5$}
			\psfrag{2}[c][c]{\scriptsize $1$}
			
			\psfrag{0}[c][c]{\scriptsize $0$}
			\psfrag{4}[c][c]{\scriptsize $4$}
			\psfrag{8}[c][c]{\scriptsize $8$}
			\psfrag{12}[c][c]{\scriptsize $12$}
			\psfrag{16}[c][c]{\scriptsize $16$}
			\psfrag{20}[c][c]{\scriptsize $20$}
			
			\psfrag{RMSJ}[c][c]{\footnotesize Norm. SJ ($\times 10^{-4}T_\mathrm{s}$)}
			\psfrag{Probability density}[c][c]{\footnotesize Probability density ($\times 10^{3}$)}
			
			\includegraphics[width=3.75cm]{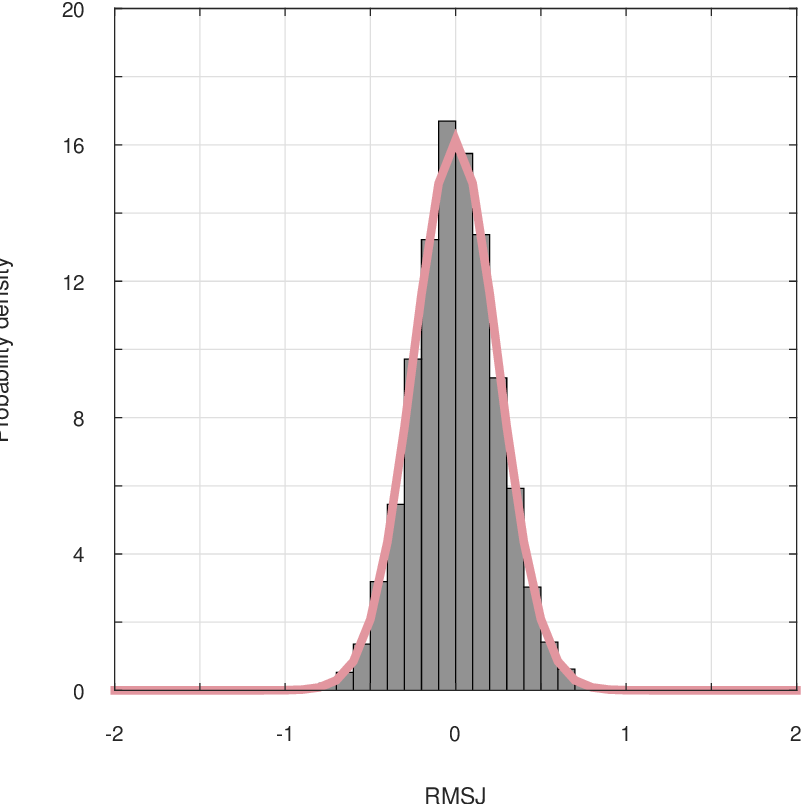}\label{fig:samplingJitter_SJtseries}			
		}

		\captionsetup{justification=raggedright,labelsep=period,singlelinecheck=false}
		\caption{\ Adopted PN model based on the Texas Instruments LMX2594 \ac{RF} synthesizer \cite{ti2019}. The double-sided PN PSD is shown in (a), while the integrated PN level as a function of frequency offset (double-sided) for the frequency range of interest assuming BB sampling and \mbox{$\eta=1$} is presented in (b). In (b), \mbox{$B/2=\SI{250}{\mega\hertz}$} is highlighted ({\color[rgb]{0.8471,0.4510,0.4980}\textbf{\textendash~\textendash}}), as this offset corresponds to the bandwidth \mbox{$B=\SI{500}{\mega\hertz}$} listed in Table~\ref{tab:ofdmParameters}. Finally, the PDF of the PN time series is shown in (c) and the corresponding SJ PDF in (d). Since a critical sampling period of \mbox{$T_\mathrm{s}=\SI{2}{\nano\second}$} was considered, the PDF in (d) is associated with an RMS SJ of \SI{49.44}{\femto\second}.}\label{fig:samplingJitter_PNpsd_RMS}
		
	\end{figure}

	Besides the aforementioned \ac{OFDM} signal parameters, the \ac{PLL}-based oscillator \ac{PN} model from \cite{bookSFO} is adopted. Its parameters were chosen to match the \ac{PN} \ac{PSD} reported for the \ac{PLL} contained in the Texas Instruments LMX2594 \ac{RF} synthesizer \cite{ti2019} for the closest frequency to the \SI{122.88}{\mega\hertz} reference oscillator. Both the aforementioned \ac{PLL} and sampling clock frequency are used for the reference oscillator from which the sampling clocks of \ac{DAC} and \ac{ADC} tiles are generated in the Zynq UltraScale+ RFSoC ZCU111 \ac{SoC} platform \cite{rfsoc2021}. In reality, however, the oscillator signal passes through a Texas Instruments LMK04208 low-noise clock jitter cleaner with dual loop \acp{PLL} \cite{ti2016}, which shapes the \ac{PN} \ac{PSD} and eventually reduces the resulting \ac{SJ}. For simplicity, the jitter cleaner LMK04208 is disregarded from the following analyses, the \ac{RMS} \ac{SJ} associated with the LMX2594 is varied instead to allow analyzing the effect of a wider range of \ac{SJ} RMS values on the communication and radar sensing performance in the considered \ac{OFDM}-based \ac{ISAC} system. The adopted \ac{PN} model, which was based on data from the LMX2594 data sheet \cite{ti2019}, has the double-sided \ac{PN} \ac{PSD} shown in Fig.~\ref{fig:samplingJitter_PNpsd} and the integrated \ac{PN} level shown in Fig.~\ref{fig:samplingJitter_intPN}. A \ac{PN} time series was then generated based on the aforementioned \ac{PSD}, and the resulting \ac{SJ} from the aforementioned \ac{PLL}-based oscillator \ac{PN} calculated according to \eqref{eq:PN_jitter}. The resulting \ac{SJ} \ac{PDF} is shown in Fig.~\ref{fig:samplingJitter_SJtseries}. Due to the fact that the considered \ac{PN} \ac{PSD} from Fig.~\ref{fig:samplingJitter_PNpsd} is not flat, the \ac{SJ} will be colored. Moreover, since individual \ac{RF} synthesizers and therefore \acp{PLL} are used for the \ac{DAC} and \ac{ADC} tiles in the Zynq UltraScale+ RFSoC ZCU111, the \ac{DAC} and \ac{ADC} \acp{SJ} $\delta^\mathrm{DAC}_\mathrm{SJ}(t)$ and $\delta^\mathrm{ADC}_\mathrm{SJ}(t)$ are henceforth assumed to be uncorrelated regardless of whether a monostatic or bistatic \ac{ISAC} architecture is assumed and communication or sensing is performed.
	
	\begin{figure*}[!t]
		\centering
		\subfloat[ ]{
			\psfrag{8}[c][c]{\scriptsize $8$}
			\psfrag{9}[c][c]{\scriptsize $9$}
			\psfrag{10}[c][c]{\scriptsize $10$}
			\psfrag{11}[c][c]{\scriptsize $11$}
			\psfrag{12}[c][c]{\scriptsize $12$}
			\psfrag{13}[c][c]{\scriptsize $13$}
			\psfrag{14}[c][c]{\scriptsize $14$}
			
			\psfrag{-70}[c][c]{\scriptsize -$70$}
			\psfrag{-60}[c][c]{\scriptsize -$60$}
			\psfrag{-50}[c][c]{\scriptsize -$50$}
			\psfrag{-40}[c][c]{\scriptsize -$40$}
			\psfrag{-30}[c][c]{\scriptsize -$30$}
			\psfrag{-20}[c][c]{\scriptsize -$20$}
			
			\psfrag{log2N}[c][c]{\footnotesize $\log_2(N)$}
			\psfrag{EVM (dB)}[c][c]{\footnotesize $\mathrm{EVM (dB)}$}
			
			\includegraphics[width=3.75cm]{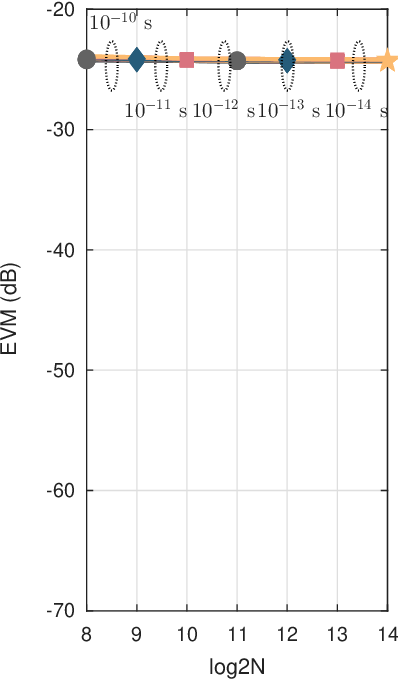}
		}
		\subfloat[ ]{
			\psfrag{8}[c][c]{\scriptsize $8$}
			\psfrag{9}[c][c]{\scriptsize $9$}
			\psfrag{AA}[c][c]{\scriptsize $10$}
			\psfrag{11}[c][c]{\scriptsize $11$}
			\psfrag{12}[c][c]{\scriptsize $12$}
			\psfrag{13}[c][c]{\scriptsize $13$}
			\psfrag{14}[c][c]{\scriptsize $14$}
			
			\psfrag{-70}[c][c]{}
			\psfrag{-60}[c][c]{}
			\psfrag{-50}[c][c]{}
			\psfrag{-40}[c][c]{}
			\psfrag{-30}[c][c]{}
			\psfrag{-20}[c][c]{}
			\psfrag{-10}[c][c]{}
			
			\psfrag{log2N}[c][c]{\footnotesize $\log_2(N)$}
			\psfrag{EVM (dB)}[c][c]{}
			
			\includegraphics[width=3.75cm]{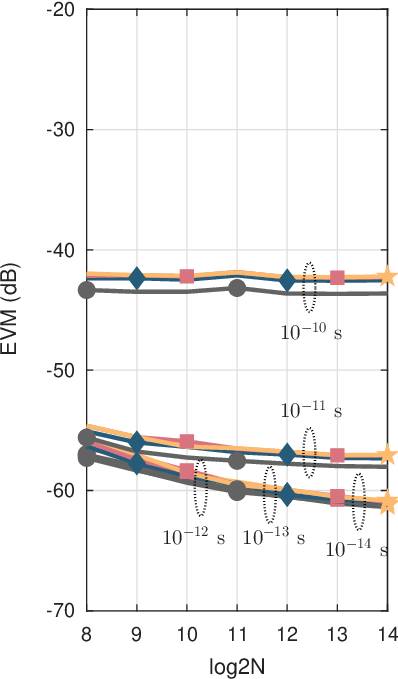}
		}
		\subfloat[ ]{
			\psfrag{8}[c][c]{\scriptsize $8$}
			\psfrag{9}[c][c]{\scriptsize $9$}
			\psfrag{AA}[c][c]{\scriptsize $10$}
			\psfrag{11}[c][c]{\scriptsize $11$}
			\psfrag{12}[c][c]{\scriptsize $12$}
			\psfrag{13}[c][c]{\scriptsize $13$}
			\psfrag{14}[c][c]{\scriptsize $14$}
			
			\psfrag{-70}[c][c]{}
			\psfrag{-60}[c][c]{}
			\psfrag{-50}[c][c]{}
			\psfrag{-40}[c][c]{}
			\psfrag{-30}[c][c]{}
			\psfrag{-20}[c][c]{}
			\psfrag{-10}[c][c]{}
			
			\psfrag{log2N}[c][c]{\footnotesize $\log_2(N)$}
			\psfrag{EVM (dB)}[c][c]{}
			
			\includegraphics[width=3.75cm]{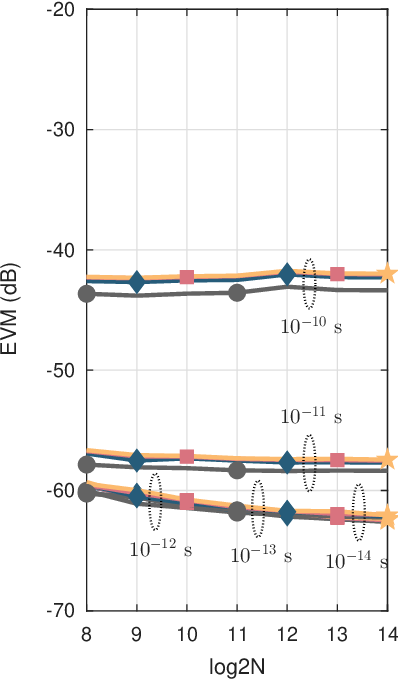}
		}
		\subfloat[ ]{
			\psfrag{8}[c][c]{\scriptsize $8$}
			\psfrag{9}[c][c]{\scriptsize $9$}
			\psfrag{AA}[c][c]{\scriptsize $10$}
			\psfrag{11}[c][c]{\scriptsize $11$}
			\psfrag{12}[c][c]{\scriptsize $12$}
			\psfrag{13}[c][c]{\scriptsize $13$}
			\psfrag{14}[c][c]{\scriptsize $14$}
			
			\psfrag{-70}[c][c]{}
			\psfrag{-60}[c][c]{}
			\psfrag{-50}[c][c]{}
			\psfrag{-40}[c][c]{}
			\psfrag{-30}[c][c]{}
			\psfrag{-20}[c][c]{}
			\psfrag{-10}[c][c]{}
			
			\psfrag{log2N}[c][c]{\footnotesize $\log_2(N)$}
			\psfrag{EVM (dB)}[c][c]{}
			
			\includegraphics[width=3.75cm]{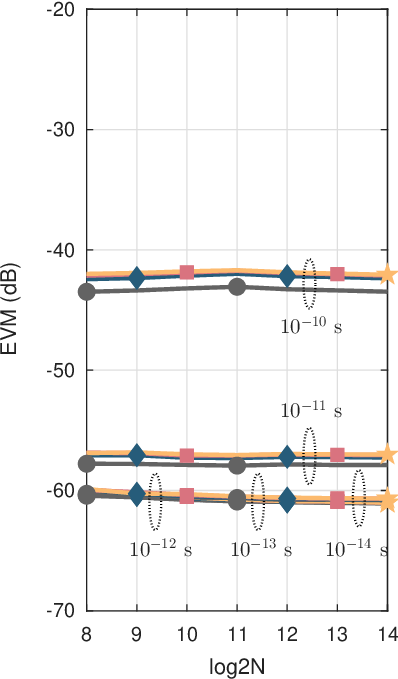}
		}
		\captionsetup{justification=raggedright,labelsep=period,singlelinecheck=false}
		\caption{\ Mean EVM obtained with \ac{BB} sampling as a function of the number of subcarriers $N$. The performed simulations considered $M=128$ OFDM symbols, and SJ RMS values of  \textcolor{black}{$\SI[parse-numbers = false]{10^{-14}}{\second}$, $\SI[parse-numbers = false]{10^{-13}}{\second}$, $\SI[parse-numbers = false]{10^{-12}}{\second}$, $\SI[parse-numbers = false]{10^{-11}}{\second}$, and $\SI[parse-numbers = false]{10^{-10}}{\second}$} at both transmitter and receiver. These values correspond to RMS SJs of  \textcolor{black}{$0.5\times10^{-5}T_\mathrm{s}$, $0.5\times10^{-4}T_\mathrm{s}$, $0.5\times10^{-3}T_\mathrm{s}$, $0.5\times10^{-2}T_\mathrm{s}$, and $0.5\times10^{-1}T_\mathrm{s}$}, respectively. In addition, \ac{QPSK} ({\color[rgb]{0.3922,0.3922,0.3922}$\CIRCLE$}), \mbox{16-\ac{QAM}} ({\color[rgb]{0.1490,0.3569,0.4824}$\blacklozenge$}), \mbox{64-\ac{QAM}} ({\color[rgb]{0.8471,0.4510,0.4980}$\blacksquare$}), and \mbox{256-\ac{QAM}} ({\color[rgb]{0.9882,0.7333,0.4275}$\bigstar$}) modulations were considered. For the results in (a), (b), (c), and (d), $\eta=1$, $\eta=2$, $\eta=4$, and $\eta=8$ were used, respectively.}\label{fig:samplingJitter_BB_EVM}		
		
	\end{figure*}
	
	Next, the \ac{SJ}-induced communication performance degradation in an \ac{OFDM}-based \ac{ISAC} system over an ideal, noiseless channel as assumed in Section~\ref{sec:sysModel} is analyzed. For that, \ac{EVM} and \ac{SIR} are adopted as performance parameters. Results and discussions are presented for both \ac{BB} and \ac{BP} sampling strategies in Sections~\ref{subsec:commBB} and \ref{subsec:commBP}, respectively, and remarks on the presented simulation results are given in Section~\ref{subsec:commRemarks}.
	
	\subsection{Baseband Sampling}\label{subsec:commBB}
	In Fig.~\ref{fig:samplingJitter_BB_EVM}, the achieved mean \ac{EVM} is shown as a function of the number of subcarriers $N$ for all digital modulation schemes from Table~\ref{tab:ofdmParameters}, namely \ac{QPSK}, \mbox{16-\ac{QAM}}, \mbox{64-\ac{QAM}} and \mbox{256-\ac{QAM}}. In addition, the case where no \ac{FDZP} was performed, i.e., $\eta=1$, as well as the cases with \ac{FDZP} oversampling factors of $\eta\in\{2,4,8\}$ were analyzed. The achieved results assume the use of the \ac{SJ} model with \ac{PDF} depicted in Fig.~\ref{fig:samplingJitter_SJtseries}, which was derived from the \ac{PN} \ac{PSD} in Fig.~\ref{fig:samplingJitter_PNpsd}, but with varying \ac{RMS} values to better analyze the effects of different \ac{SJ} levels. The adopted \ac{SJ} \ac{RMS} values, which were assumed to be the same for both \ac{DAC} and \ac{ADC}, were \textcolor{black}{$\SI[parse-numbers = false]{10^{-14}}{\second}$, $\SI[parse-numbers = false]{10^{-13}}{\second}$, $\SI[parse-numbers = false]{10^{-12}}{\second}$, $\SI[parse-numbers = false]{10^{-11}}{\second}$, and $\SI[parse-numbers = false]{10^{-10}}{\second}$}. Since \mbox{$T_\mathrm{s}=1/B=\SI{2}{\nano\second}$}, these numbers correspond to \textcolor{black}{$0.5\times10^{-5}T_\mathrm{s}$, $0.5\times10^{-4}T_\mathrm{s}$, $0.5\times10^{-3}T_\mathrm{s}$, $0.5\times10^{-2}T_\mathrm{s}$, and $0.5\times10^{-1}T_\mathrm{s}$}, respectively. It can be seen in Fig.~\ref{fig:samplingJitter_BB_EVM}(a) that \textcolor{black}{all considered parameterizations result in the same \ac{EVM} of \mbox{$\SI{24.13}{dB}$}.} \begin{figure*}[!t]
		\centering
		\subfloat[ ]{
			
			\psfrag{-1.5}[c][c]{\scriptsize -$1.5$}
			\psfrag{-0.5}[c][c]{\scriptsize -$0.5$}
			\psfrag{0.5}[c][c]{\scriptsize $0.5$}
			\psfrag{1.5}[c][c]{\scriptsize $1.5$}
			
			\psfrag{AA}[c][c]{\scriptsize -$1.5$}
			\psfrag{BB}[c][c]{\scriptsize -$0.5$}
			\psfrag{CC}[c][c]{\scriptsize $0.5$}
			\psfrag{DD}[c][c]{\scriptsize $1.5$}
			
			\psfrag{I}[c][c]{\scriptsize $I$}
			\psfrag{Q}[c][c]{\scriptsize $Q$}
			
			\includegraphics[width=2.5cm]{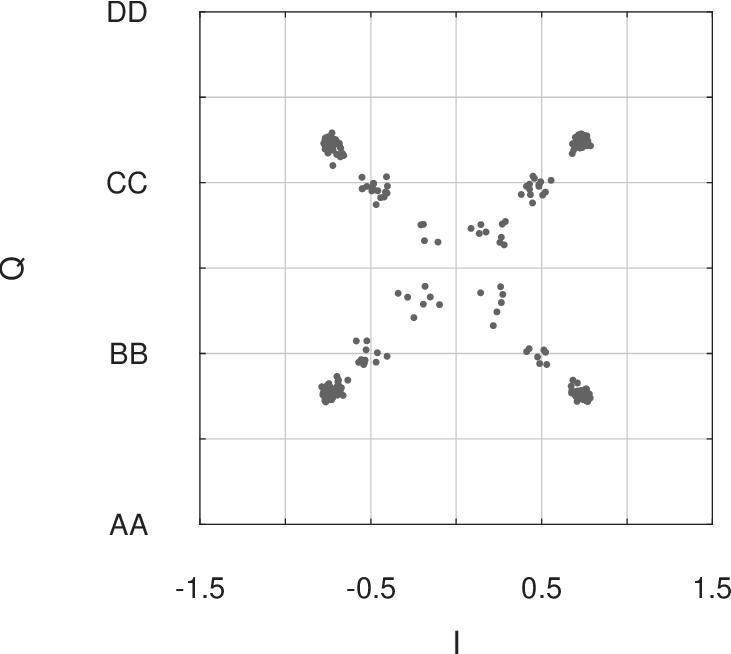}
		}\hspace{-0.2cm}
		\subfloat[ ]{
			
			\psfrag{-1.5}[c][c]{\scriptsize -$1.5$}
			\psfrag{-0.5}[c][c]{\scriptsize -$0.5$}
			\psfrag{0.5}[c][c]{\scriptsize $0.5$}
			\psfrag{1.5}[c][c]{\scriptsize $1.5$}
			
			\psfrag{AA}[c][c]{ }
			\psfrag{BB}[c][c]{ }
			\psfrag{CC}[c][c]{ }
			\psfrag{DD}[c][c]{ }
			
			\psfrag{I}[c][c]{\scriptsize $I$}
			\psfrag{Q}[c][c]{ }
			
			\includegraphics[width=2.5cm]{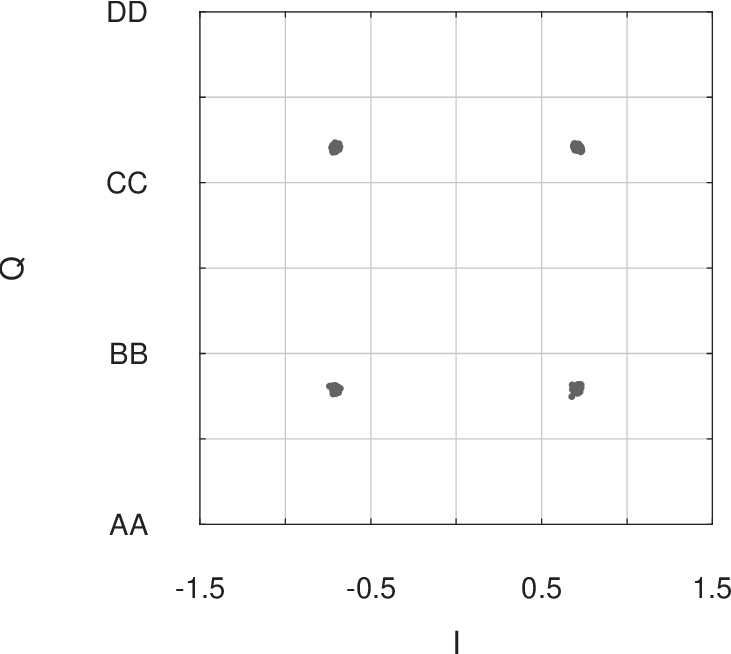}
		}\hspace{-0.2cm}
		\subfloat[ ]{
			
			\psfrag{-1.5}[c][c]{\scriptsize -$1.5$}
			\psfrag{-0.5}[c][c]{\scriptsize -$0.5$}
			\psfrag{0.5}[c][c]{\scriptsize $0.5$}
			\psfrag{1.5}[c][c]{\scriptsize $1.5$}
			
			\psfrag{AA}[c][c]{ }
			\psfrag{BB}[c][c]{ }
			\psfrag{CC}[c][c]{ }
			\psfrag{DD}[c][c]{ }
			
			\psfrag{I}[c][c]{\scriptsize $I$}
			\psfrag{Q}[c][c]{ }
			
			\includegraphics[width=2.5cm]{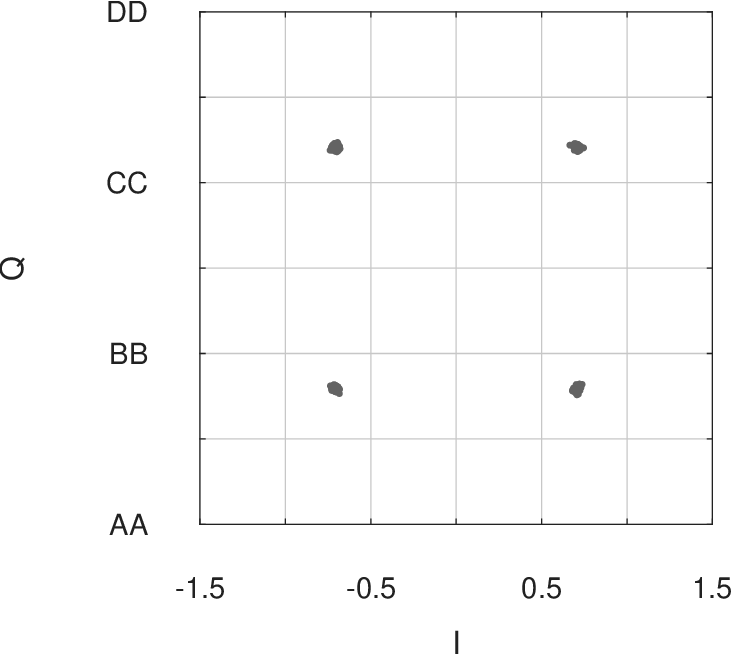}
		}\hspace{-0.2cm}
		\subfloat[ ]{
			
			\psfrag{-1.5}[c][c]{\scriptsize -$1.5$}
			\psfrag{-0.5}[c][c]{\scriptsize -$0.5$}
			\psfrag{0.5}[c][c]{\scriptsize $0.5$}
			\psfrag{1.5}[c][c]{\scriptsize $1.5$}
			
			\psfrag{AA}[c][c]{ }
			\psfrag{BB}[c][c]{ }
			\psfrag{CC}[c][c]{ }
			\psfrag{DD}[c][c]{ }
			
			\psfrag{I}[c][c]{\scriptsize $I$}
			\psfrag{Q}[c][c]{ }
			
			\includegraphics[width=2.5cm]{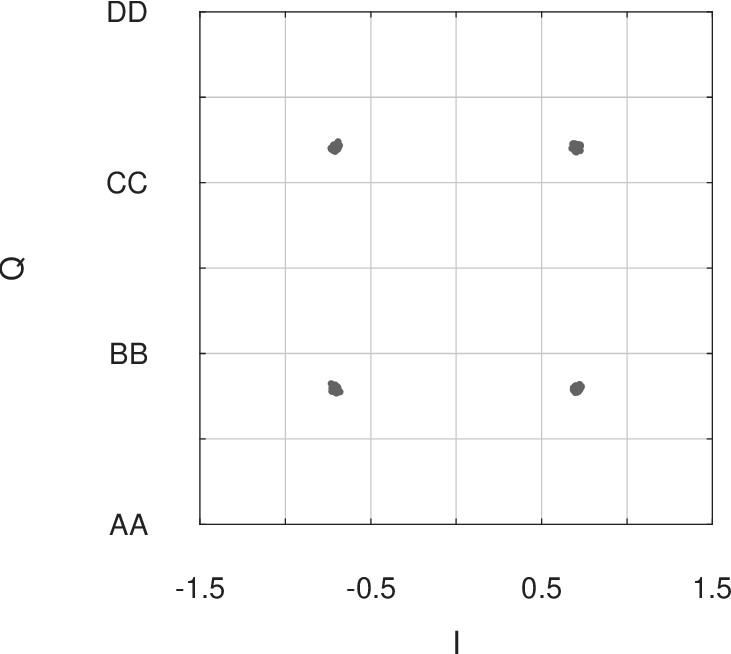}
		}\\
		\subfloat[ ]{
			
			\psfrag{-1.5}[c][c]{\scriptsize -$1.5$}
			\psfrag{-0.5}[c][c]{\scriptsize -$0.5$}
			\psfrag{0.5}[c][c]{\scriptsize $0.5$}
			\psfrag{1.5}[c][c]{\scriptsize $1.5$}
			
			\psfrag{AA}[c][c]{\scriptsize -$1.5$}
			\psfrag{BB}[c][c]{\scriptsize -$0.5$}
			\psfrag{CC}[c][c]{\scriptsize $0.5$}
			\psfrag{DD}[c][c]{\scriptsize $1.5$}
			
			\psfrag{I}[c][c]{\scriptsize $I$}
			\psfrag{Q}[c][c]{\scriptsize $Q$}
			
			\includegraphics[width=2.5cm]{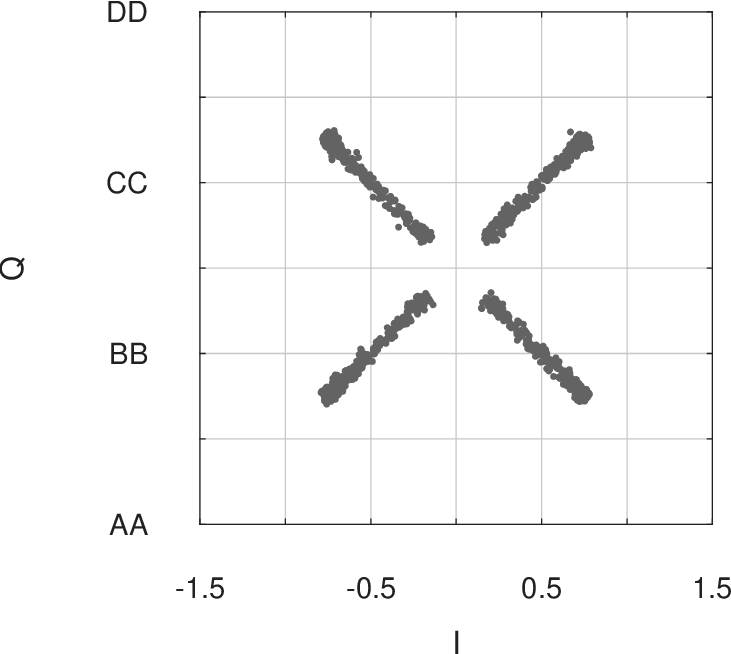}
		}\hspace{-0.2cm}
		\subfloat[ ]{
			
			\psfrag{-1.5}[c][c]{\scriptsize -$1.5$}
			\psfrag{-0.5}[c][c]{\scriptsize -$0.5$}
			\psfrag{0.5}[c][c]{\scriptsize $0.5$}
			\psfrag{1.5}[c][c]{\scriptsize $1.5$}
			
			\psfrag{AA}[c][c]{ }
			\psfrag{BB}[c][c]{ }
			\psfrag{CC}[c][c]{ }
			\psfrag{DD}[c][c]{ }
			
			\psfrag{I}[c][c]{\scriptsize $I$}
			\psfrag{Q}[c][c]{ }
			
			\includegraphics[width=2.5cm]{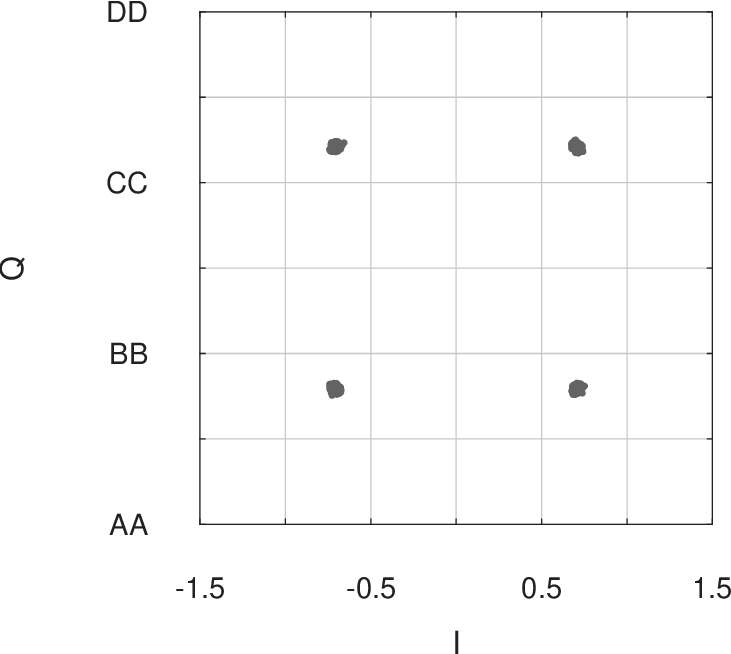}
		}\hspace{-0.2cm}
		\subfloat[ ]{
			
			\psfrag{-1.5}[c][c]{\scriptsize -$1.5$}
			\psfrag{-0.5}[c][c]{\scriptsize -$0.5$}
			\psfrag{0.5}[c][c]{\scriptsize $0.5$}
			\psfrag{1.5}[c][c]{\scriptsize $1.5$}
			
			\psfrag{AA}[c][c]{ }
			\psfrag{BB}[c][c]{ }
			\psfrag{CC}[c][c]{ }
			\psfrag{DD}[c][c]{ }
			
			\psfrag{I}[c][c]{\scriptsize $I$}
			\psfrag{Q}[c][c]{ }
			
			\includegraphics[width=2.5cm]{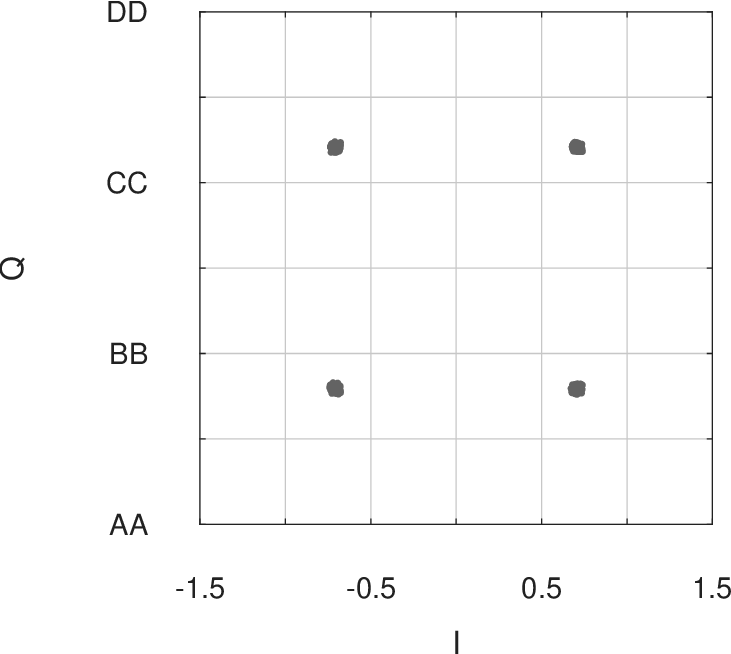}
		}\hspace{-0.2cm}
		\subfloat[ ]{
			
			\psfrag{-1.5}[c][c]{\scriptsize -$1.5$}
			\psfrag{-0.5}[c][c]{\scriptsize -$0.5$}
			\psfrag{0.5}[c][c]{\scriptsize $0.5$}
			\psfrag{1.5}[c][c]{\scriptsize $1.5$}
			
			\psfrag{AA}[c][c]{ }
			\psfrag{BB}[c][c]{ }
			\psfrag{CC}[c][c]{ }
			\psfrag{DD}[c][c]{ }
			
			\psfrag{I}[c][c]{\scriptsize $I$}
			\psfrag{Q}[c][c]{ }
			
			\includegraphics[width=2.5cm]{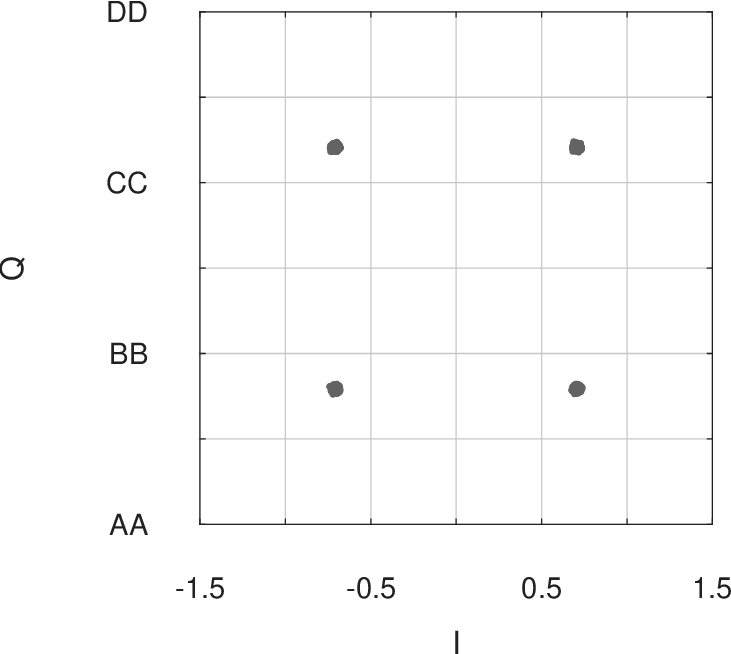}
		}\\
		\subfloat[ ]{
			
			\psfrag{-1.5}[c][c]{\scriptsize -$1.5$}
			\psfrag{-0.5}[c][c]{\scriptsize -$0.5$}
			\psfrag{0.5}[c][c]{\scriptsize $0.5$}
			\psfrag{1.5}[c][c]{\scriptsize $1.5$}
			
			\psfrag{AA}[c][c]{\scriptsize -$1.5$}
			\psfrag{BB}[c][c]{\scriptsize -$0.5$}
			\psfrag{CC}[c][c]{\scriptsize $0.5$}
			\psfrag{DD}[c][c]{\scriptsize $1.5$}
			
			\psfrag{I}[c][c]{\scriptsize $I$}
			\psfrag{Q}[c][c]{\scriptsize $Q$}
			
			\includegraphics[width=2.5cm]{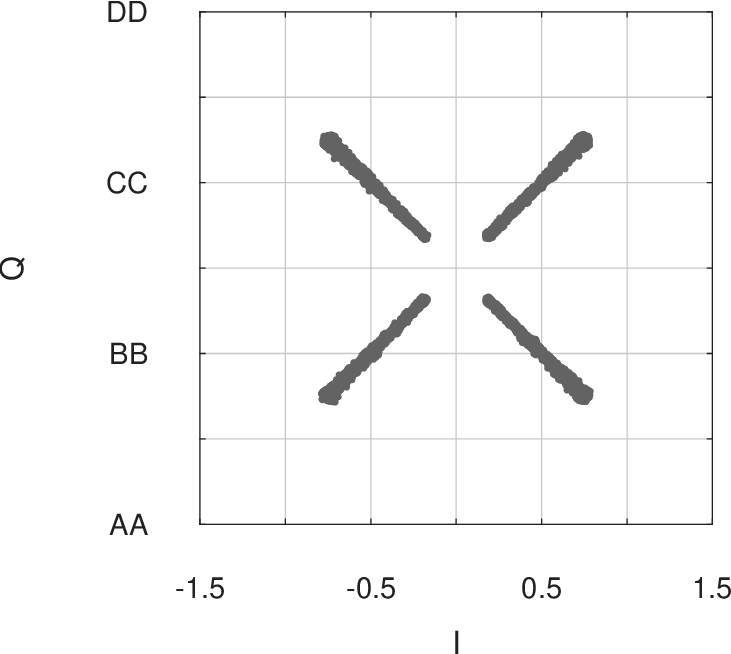}
		}\hspace{-0.2cm}
		\subfloat[ ]{
			
			\psfrag{-1.5}[c][c]{\scriptsize -$1.5$}
			\psfrag{-0.5}[c][c]{\scriptsize -$0.5$}
			\psfrag{0.5}[c][c]{\scriptsize $0.5$}
			\psfrag{1.5}[c][c]{\scriptsize $1.5$}
			
			\psfrag{AA}[c][c]{ }
			\psfrag{BB}[c][c]{ }
			\psfrag{CC}[c][c]{ }
			\psfrag{DD}[c][c]{ }
			
			\psfrag{I}[c][c]{\scriptsize $I$}
			\psfrag{Q}[c][c]{ }
			
			\includegraphics[width=2.5cm]{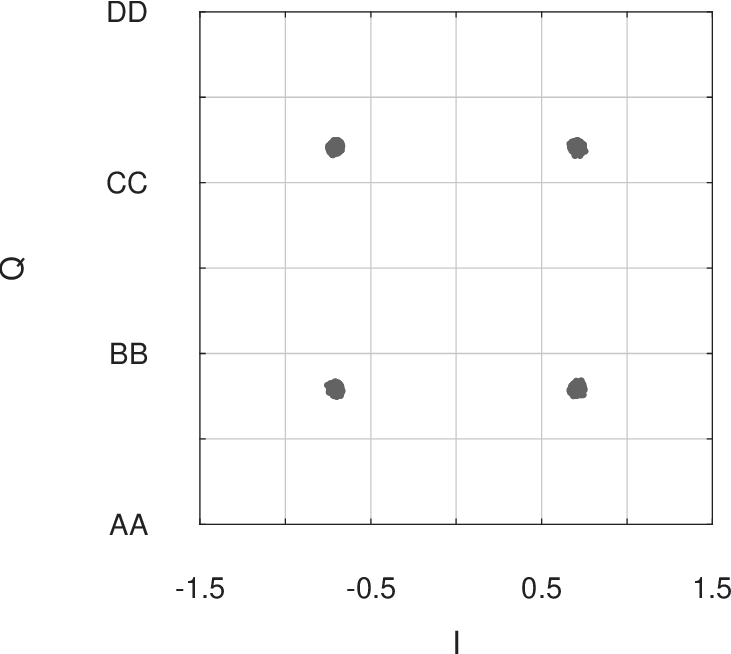}
		}\hspace{-0.2cm}
		\subfloat[ ]{
			
			\psfrag{-1.5}[c][c]{\scriptsize -$1.5$}
			\psfrag{-0.5}[c][c]{\scriptsize -$0.5$}
			\psfrag{0.5}[c][c]{\scriptsize $0.5$}
			\psfrag{1.5}[c][c]{\scriptsize $1.5$}
			
			\psfrag{AA}[c][c]{ }
			\psfrag{BB}[c][c]{ }
			\psfrag{CC}[c][c]{ }
			\psfrag{DD}[c][c]{ }
			
			\psfrag{I}[c][c]{\scriptsize $I$}
			\psfrag{Q}[c][c]{ }
			
			\includegraphics[width=2.5cm]{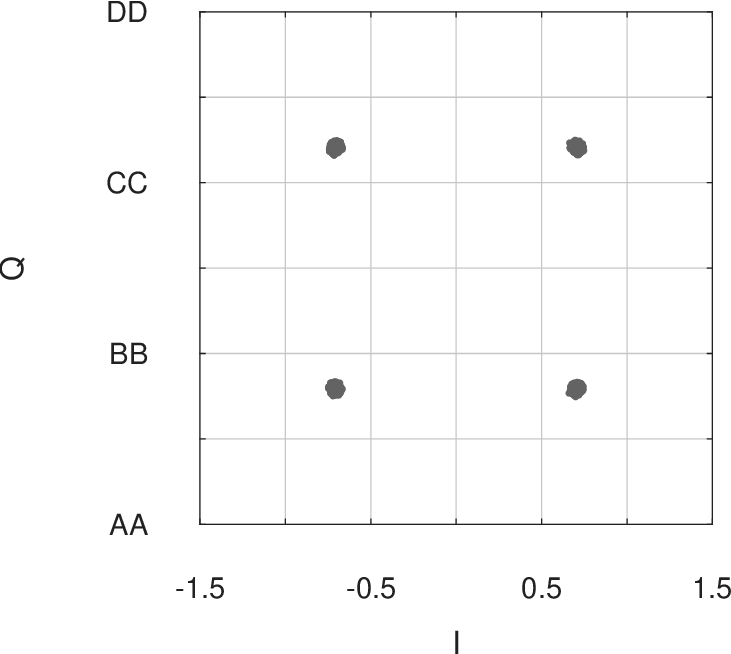}
		}\hspace{-0.2cm}
		\subfloat[ ]{
			
			\psfrag{-1.5}[c][c]{\scriptsize -$1.5$}
			\psfrag{-0.5}[c][c]{\scriptsize -$0.5$}
			\psfrag{0.5}[c][c]{\scriptsize $0.5$}
			\psfrag{1.5}[c][c]{\scriptsize $1.5$}
			
			\psfrag{AA}[c][c]{ }
			\psfrag{BB}[c][c]{ }
			\psfrag{CC}[c][c]{ }
			\psfrag{DD}[c][c]{ }
			
			\psfrag{I}[c][c]{\scriptsize $I$}
			\psfrag{Q}[c][c]{ }
			
			\includegraphics[width=2.5cm]{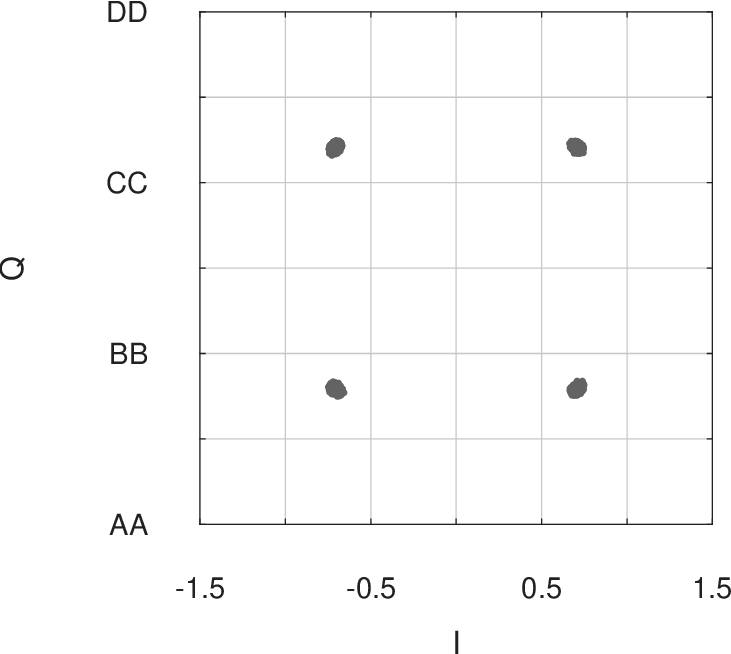}
		}

		\captionsetup{justification=raggedright,labelsep=period,singlelinecheck=false}
		\caption{\ Normalized receive QPSK constellations obtained with \ac{BB} sampling, $M=128$ OFDM symbols, and an RMS SJ of \textcolor{black}{$\SI[parse-numbers = false]{10^{-10}}{\second}$ (i.e., $0.5\times10^{-1}T_\mathrm{s}$)}. Results are shown for different numbers of subcariers and \ac{ZP} factors, namely $N=256$ and (a) $\eta=1$, (b) $\eta=2$, (c) $\eta=4$, (d) $\eta=8$, $N=2048$ and (e) $\eta=1$, (f) $\eta=2$, (g) $\eta=4$, (h) $\eta=8$, and $N=16384$ and (i) $\eta=1$, (j) $\eta=2$, (k) $\eta=4$, (l) $\eta=8$.}\label{fig:QPSKconst_BB_N8_11_14_jm8}
		
	\end{figure*}When $\eta=2$ is adopted in Fig.~\ref{fig:samplingJitter_BB_EVM}(b), a significantly improved \ac{EVM} is obtained in comparison to the case for $\eta=1$ in Fig.~\ref{fig:samplingJitter_BB_EVM}, especially for \ac{RMS} \ac{SJ} values equal to or smaller than \textcolor{black}{$\SI[parse-numbers = false]{10^{-11}}{\second}$ (i.e., $0.5\times10^{-2}T_\mathrm{s}$)}. When the oversampling factor is further increased to $\eta=4$\textcolor{black}{, an improvement of around \mbox{$\SI{2}{dB}$} or less for \mbox{$N\leq1024$} is observed for an \ac{RMS} \ac{SJ} of $\SI[parse-numbers = false]{10^{-11}}{\second}$. For \ac{RMS} \ac{SJ}  values equal to or lower than $\SI[parse-numbers = false]{10^{-12}}{\second}$, an improvement of up to \mbox{$\SI{3}{dB}$} is observed for the whole considered $N$ range.} \textcolor{black}{As for the oversampling factor $\eta=8$, similar performance to the case with $\eta=4$ is experienced, with exception for \ac{RMS} \ac{SJ} values equal to or lower than $\SI[parse-numbers = false]{10^{-12}}{\second}$ (i.e., $0.5\times10^{-3}T_\mathrm{s}$). At the aforementioned \ac{SJ} levels, up to around \mbox{$\SI{1.4}{dB}$} higher \ac{EVM} is experienced with $\eta=8$. This is due to the fact that the experienced deviations from the correct sampling points are more relevant in this case than for $\eta=4$, where the two times higher oversampled sampling period is experienced, i.e., \mbox{$T^\eta_\mathrm{s}=T_\mathrm{s}/4$} instead of \mbox{$T^\eta_\mathrm{s}=T_\mathrm{s}/8$}. This result aligns the fact that colored makes oversampling less effective against the \ac{SJ} than it would be in a case with white \ac{DAC} and \ac{ADC} \ac{SJ}.}
	
	\textcolor{black}{For all oversampling factors other than \mbox{$\eta=1$}, i.e., \mbox{$\eta\in\{2,4,8\}$}, increasing the number of subcarriers $N$ tends to result in lower \ac{EVM} for \ac{RMS} \ac{SJ} equal to or lower than $\SI[parse-numbers = false]{10^{-11}}{\second}$ (i.e., $0.5\times10^{-2}T_\mathrm{s}$). This is due to the smaller subcarrier spacing \mbox{$\Delta f=B/N$}, which leads to a more pronounced \ac{ICI} effect and less relevant \ac{CPE} Since, however, \ac{ICI} is rather low for these \ac{RMS} \ac{SJ} values, the greater robustness against \ac{CPE} with increasing $N$ stands out. In addition, an \ac{EVM} degradation along with $N$ of between approximately \mbox{$\SI{13}{dB}$} and \mbox{$\SI{15}{dB}$} is experienced between the \ac{RMS} \ac{SJ} values of $\SI[parse-numbers = false]{10^{-11}}{\second}$ and $\SI[parse-numbers = false]{10^{-10}}{\second}$ (i.e., $0.5\times10^{-2}T_\mathrm{s}$ and $0.5\times10^{-1}T_\mathrm{s}$), showing that the \ac{RMS} \ac{SJ} point where performance degradation starts becoming relevant lies between these two values. At this point, one relevant factor for performance degradation is the fact that increasing \ac{DAC} \ac{SJ} leads to increased \ac{OOB} radiation \cite{alian2015}, which will ultimately result in additional \ac{ICI} after sampling at an \ac{ADC} also impaired by \ac{SJ}.}
	
	\begin{figure*}[!t]
		\centering
		\subfloat[ ]{
			
			\psfrag{-60}[c][c]{\scriptsize -$60$}
			\psfrag{-45}[c][c]{\scriptsize -$45$}
			\psfrag{-30}[c][c]{\scriptsize -$30$}
			\psfrag{-15}[c][c]{\scriptsize -$15$}
			\psfrag{AA}[c][c]{\scriptsize $0$}
			
			\psfrag{-N/2}[c][c]{\scriptsize -$N/2$}
			\psfrag{-N/4}[c][c]{\scriptsize -$N/4$}
			\psfrag{0}[c][c]{\scriptsize $0$}
			\psfrag{N/4}[c][c]{\scriptsize $N/4$}
			\psfrag{N/2-1}[c][c]{\scriptsize $N/2\tiny{-}1$}
			
			\psfrag{n}[c][c]{\footnotesize $n$}
			\psfrag{EVM (dB)}[c][c]{\footnotesize $\mathrm{EVM~(dB)}$}
			
			\includegraphics[width=4.35cm]{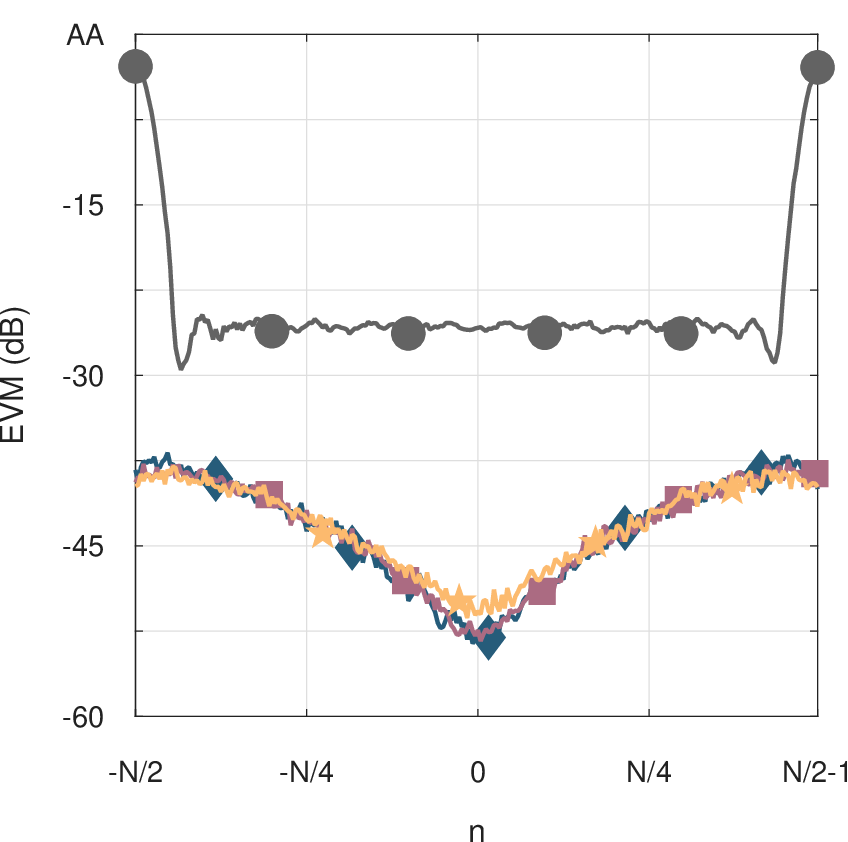}\label{fig:BBsubcEVM_N8}			
		}
		\subfloat[ ]{
			
			\psfrag{-60}[c][c]{}
			\psfrag{-45}[c][c]{}
			\psfrag{-30}[c][c]{}
			\psfrag{-15}[c][c]{}
			\psfrag{AA}[c][c]{}
			
			\psfrag{-N/2}[c][c]{\scriptsize -$N/2$}
			\psfrag{-N/4}[c][c]{\scriptsize -$N/4$}
			\psfrag{0}[c][c]{\scriptsize $0$}
			\psfrag{N/4}[c][c]{\scriptsize $N/4$}
			\psfrag{N/2-1}[c][c]{\scriptsize $N/2\tiny{-}1$}
			
			\psfrag{n}[c][c]{\footnotesize $n$}
			\psfrag{EVM (dB)}[c][c]{}
			
			\includegraphics[width=4.35cm]{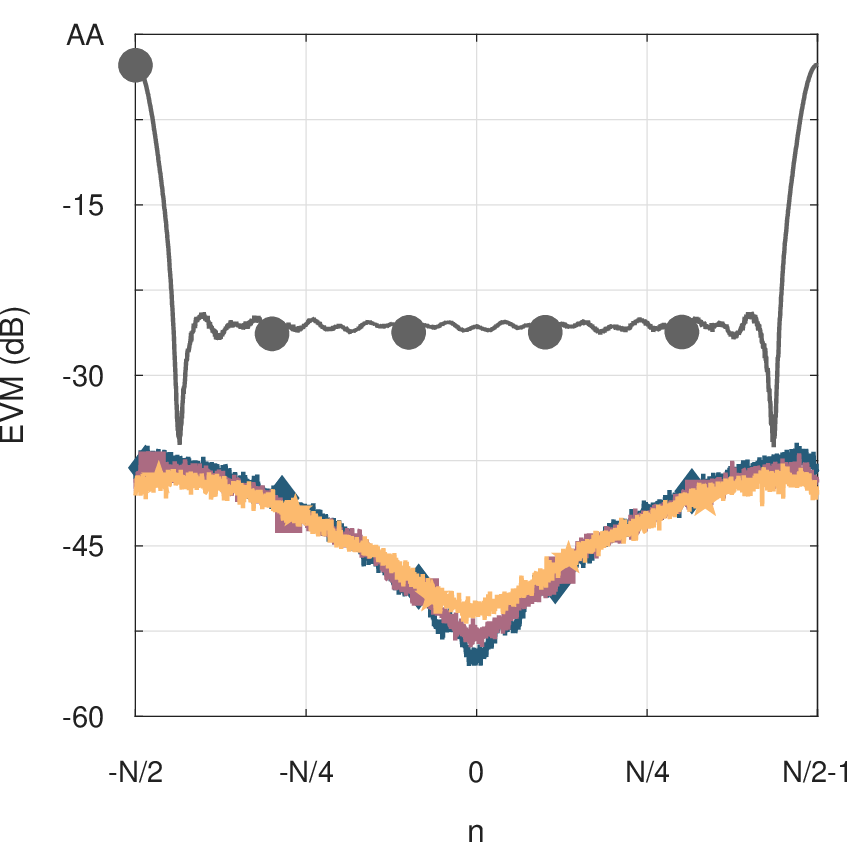}\label{fig:BBsubcEVM_N11}			
		}
		\subfloat[ ]{
			
			\psfrag{-60}[c][c]{}
			\psfrag{-45}[c][c]{}
			\psfrag{-30}[c][c]{}
			\psfrag{-15}[c][c]{}
			\psfrag{AA}[c][c]{}
			
			\psfrag{-N/2}[c][c]{\scriptsize -$N/2$}
			\psfrag{-N/4}[c][c]{\scriptsize -$N/4$}
			\psfrag{0}[c][c]{\scriptsize $0$}
			\psfrag{N/4}[c][c]{\scriptsize $N/4$}
			\psfrag{N/2-1}[c][c]{\scriptsize $N/2\tiny{-}1$}
			
			\psfrag{n}[c][c]{\footnotesize $n$}
			\psfrag{EVM (dB)}[c][c]{}
			
			\includegraphics[width=4.35cm]{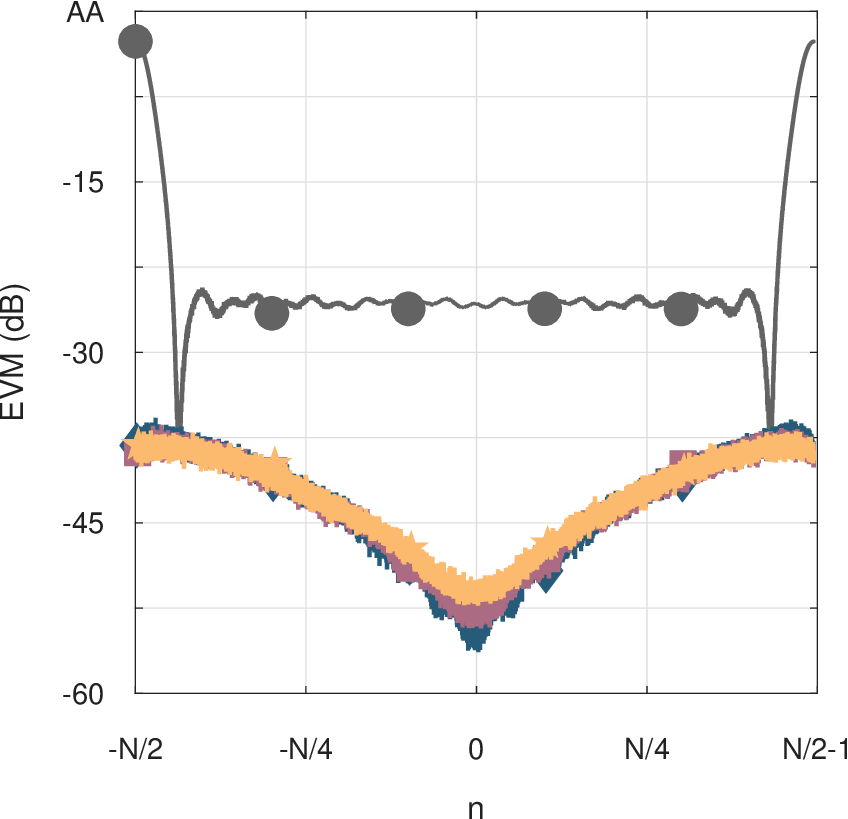}\label{fig:BBsubcEVM_N14}			
		}
		
		\captionsetup{justification=raggedright,labelsep=period,singlelinecheck=false}
		\caption{\ EVM at the $n\mathrm{th}$ subcarrier obtained with \ac{BB} sampling and an RMS SJ of \textcolor{black}{$\SI[parse-numbers = false]{10^{-10}}{\second}$ (i.e., $0.5\times10^{-1}T_\mathrm{s}$)}, QPSK modulation, and: (a) \mbox{$N=256$}, (b) \mbox{$N=2048$}, and (c) \mbox{$N=16384$}. In all cases, results are shown for $\eta=1$ ({\color[rgb]{0.3922,0.3922,0.3922}$\CIRCLE$}), $\eta=2$ ({\color[rgb]{0.1490,0.3569,0.4824}$\blacklozenge$}), $\eta=4$ ({\color[rgb]{0.8471,0.4510,0.4980}$\blacksquare$}), and $\eta=8$ ({\color[rgb]{0.9882,0.7333,0.4275}$\bigstar$}).}\label{fig:BBsubcEVM}
		
	\end{figure*}
	\begin{figure*}[!t]
		\centering
		\subfloat[ ]{
			
			\psfrag{10}[c][c]{\scriptsize $10$}
			\psfrag{30}[c][c]{\scriptsize $30$}
			\psfrag{50}[c][c]{\scriptsize $50$}
			\psfrag{70}[c][c]{\scriptsize $70$}
			
			\psfrag{-17}[c][c]{\scriptsize -$17$}
			\psfrag{-15.25}[c][c]{\scriptsize -$15.25$}
			\psfrag{-13.5}[c][c]{\scriptsize -$13.5$}
			\psfrag{-11.75}[c][c]{\scriptsize -$11.75$}	
			\psfrag{-10}[c][c]{\scriptsize -$10$}
			
			\psfrag{RMSJ}[c][c]{\footnotesize $\log_{10}\left(\delta_\mathrm{SJ,RMS}\right)$}
			\psfrag{SIR (dB)}[c][c]{\footnotesize $\mathrm{SIR~(dB)}$}
			
			\includegraphics[width=3.6cm]{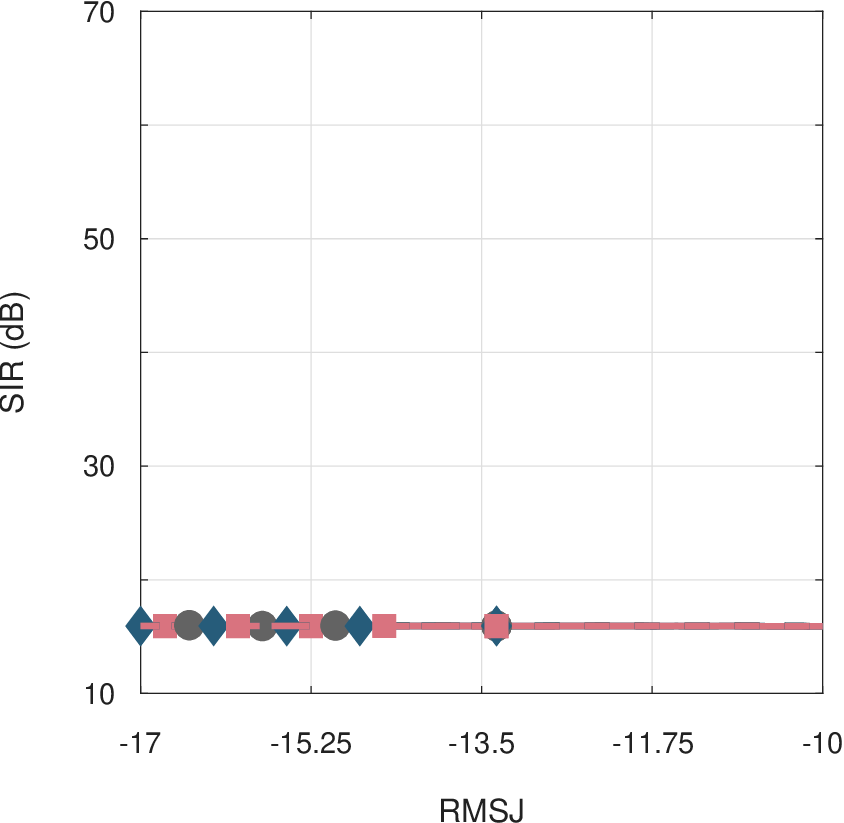}\label{fig:BB_SIR_OS1}			
		}
		\subfloat[ ]{
			
			\psfrag{10}[c][c]{}
			\psfrag{30}[c][c]{}
			\psfrag{50}[c][c]{}
			\psfrag{70}[c][c]{}
			
			\psfrag{-17}[c][c]{\scriptsize -$17$}
			\psfrag{-15.25}[c][c]{\scriptsize -$15.25$}
			\psfrag{-13.5}[c][c]{\scriptsize -$13.5$}
			\psfrag{-11.75}[c][c]{\scriptsize -$11.75$}	
			\psfrag{-10}[c][c]{\scriptsize -$10$}
			
			\psfrag{RMSJ}[c][c]{\footnotesize $\log_{10}\left(\delta_\mathrm{SJ,RMS}\right)$}
			\psfrag{SIR (dB)}[c][c]{}
			
			\includegraphics[width=3.6cm]{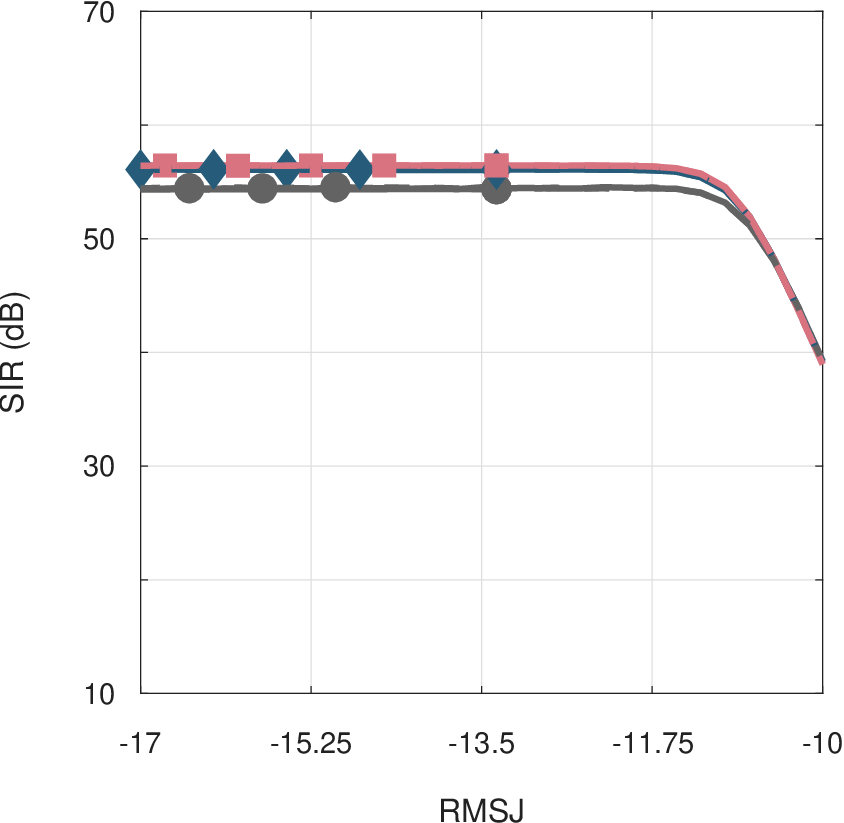}\label{fig:BB_SIR_OS2}			
		}
		\subfloat[ ]{
			
			\psfrag{10}[c][c]{}
			\psfrag{30}[c][c]{}
			\psfrag{50}[c][c]{}
			\psfrag{70}[c][c]{}
			
			\psfrag{-17}[c][c]{\scriptsize -$17$}
			\psfrag{-15.25}[c][c]{\scriptsize -$15.25$}
			\psfrag{-13.5}[c][c]{\scriptsize -$13.5$}
			\psfrag{-11.75}[c][c]{\scriptsize -$11.75$}	
			\psfrag{-10}[c][c]{\scriptsize -$10$}
			
			\psfrag{RMSJ}[c][c]{\footnotesize $\log_{10}\left(\delta_\mathrm{SJ,RMS}\right)$}
			\psfrag{SIR (dB)}[c][c]{}
			
			\includegraphics[width=3.6cm]{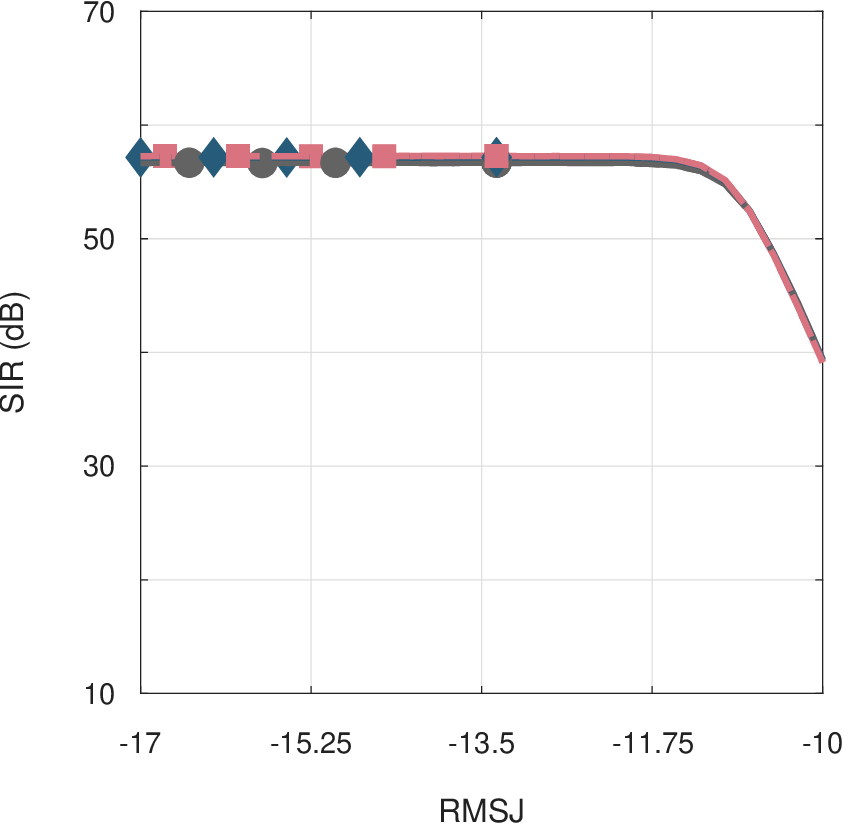}\label{fig:BB_SIR_OS4}			
		}
		\subfloat[ ]{
			
			\psfrag{10}[c][c]{}
			\psfrag{30}[c][c]{}
			\psfrag{50}[c][c]{}
			\psfrag{70}[c][c]{}
			
			\psfrag{-17}[c][c]{\scriptsize -$17$}
			\psfrag{-15.25}[c][c]{\scriptsize -$15.25$}
			\psfrag{-13.5}[c][c]{\scriptsize -$13.5$}
			\psfrag{-11.75}[c][c]{\scriptsize -$11.75$}	
			\psfrag{-10}[c][c]{\scriptsize -$10$}
			
			\psfrag{RMSJ}[c][c]{\footnotesize $\log_{10}\left(\delta_\mathrm{SJ,RMS}\right)$}
			\psfrag{SIR (dB)}[c][c]{}
			
			\includegraphics[width=3.6cm]{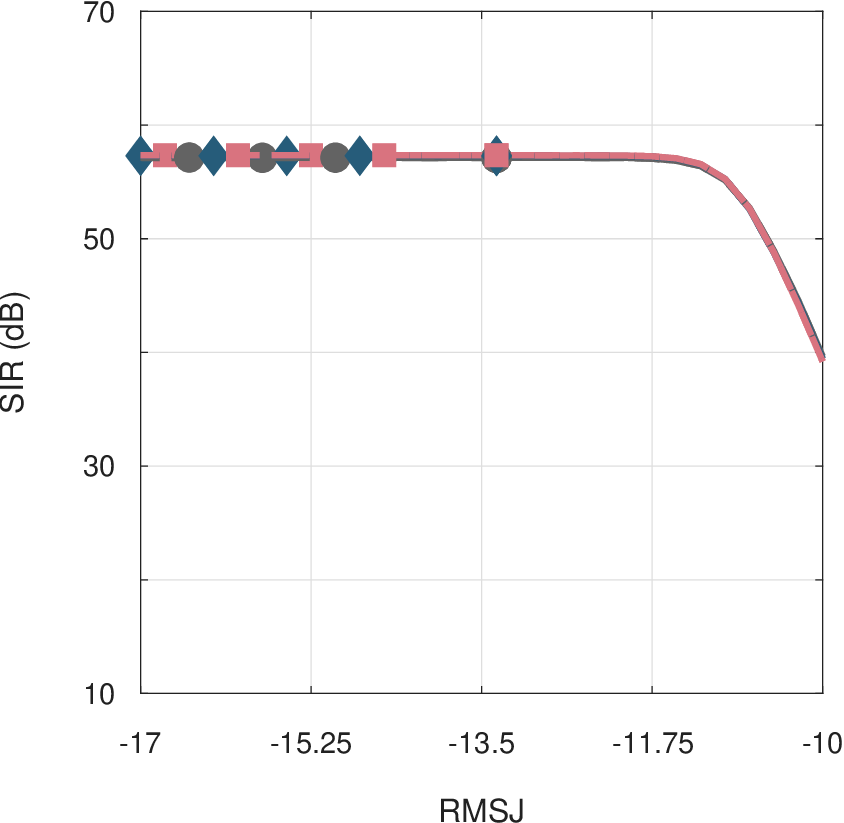}\label{fig:BB_SIR_OS8}			
		}
		
		\captionsetup{justification=raggedright,labelsep=period,singlelinecheck=false}
		\caption{\ SIR as a function of the RMS SJ obtained with \ac{BB} sampling, $M=128$ OFDM symbols, QPSK modulation, and (a) \mbox{$\eta=1$}, (b) \mbox{$\eta=2$}, (c) \mbox{$\eta=4$}, and (d) \mbox{$\eta=8$}. In all cases, results are shown for \mbox{$N=256$} ({\color[rgb]{0.3922,0.3922,0.392}$\CIRCLE$}), \mbox{$N=2048$} ({\color[rgb]{0.1490,0.3569,0.4824}$\blacklozenge$}), and \mbox{$N=16384$} ({\color[rgb]{0.8471,0.4510,0.4980}$\blacksquare$}).}\label{fig:BB_SIR}
		
	\end{figure*}
	\begin{figure*}[!t]
		\captionsetup[subfigure]{labelformat=empty}
		
		\centering
		
		\psfrag{55}{\footnotesize (a)}
		\psfrag{22}{\footnotesize (b)}
		\psfrag{33}{\footnotesize (c)}
		\psfrag{44}{\footnotesize (d)}
		
		\psfrag{-17}[c][c]{\scriptsize -$17$}
		\psfrag{-13.5}[c][c]{\scriptsize -$13.5$}
		\psfrag{-10}[c][c]{\scriptsize -$10$}
		
		\psfrag{10}[c][c]{\scriptsize $10$}
		\psfrag{30}[c][c]{\scriptsize $30$}
		\psfrag{50}[c][c]{\scriptsize $50$}
		\psfrag{70}[c][c]{\scriptsize $70$}
		
		\psfrag{RMSJT}[c][c]{\footnotesize $\log_{10}(\delta^\mathrm{Tx}_\mathrm{SJ,RMS})$}
		\psfrag{RMSJR}[c][c]{\footnotesize $\log_{10}(\delta^\mathrm{Rx}_\mathrm{SJ,RMS})$}
		\psfrag{SIR (dB)}[c][c]{\footnotesize $\mathrm{SIR~(dB)}$}
		
		\includegraphics[height=4cm]{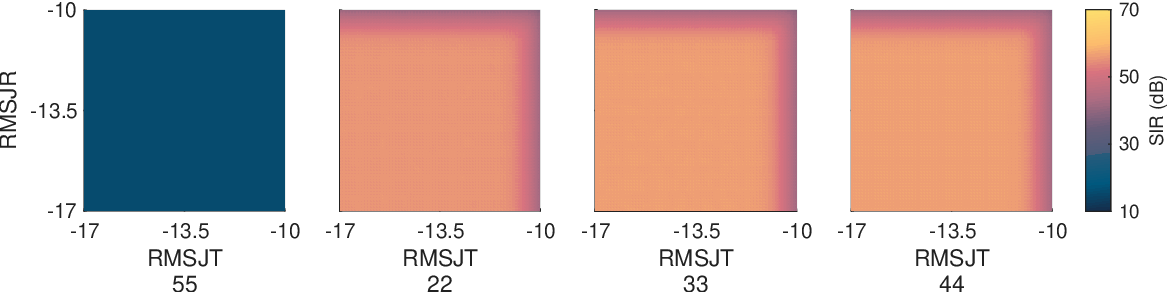}
		
		\captionsetup{justification=raggedright,labelsep=period,singlelinecheck=false}
		\caption{\ SIR obtained with \ac{BB} sampling, $M=128$ OFDM symbols, QPSK modulation, and different RMS SJ values at the DAC and ADC. Results are shown for $N=2048$ and (a) \mbox{$\eta=1$}, (b) \mbox{$\eta=2$}, (c) \mbox{$\eta=4$}, and (d) \mbox{$\eta=8$}.}\label{fig:samplingJitter_BB_SIR_diffTxRx_4QAM_N2048}
		
	\end{figure*}
	
	 To illustrate the influence of \ac{DAC} and \ac{ADC} \ac{SJ} on the receive constellations, Fig.~\ref{fig:QPSKconst_BB_N8_11_14_jm8} shows normalized \ac{QPSK} constellations obtained with the transmission of \ac{OFDM} symbols with \mbox{$N\in\{256,2048,16384\}$}, \mbox{$N_\mathrm{CP}=0$} since an ideal channel is assumed, and \mbox{$\eta\in\{1,2,4,8\}$}. For all cases, $M=128$ \ac{OFDM} symbols \textcolor{black}{and an \ac{RMS} \ac{SJ} of $\SI[parse-numbers = false]{10^{-10}}{\second}$ (i.e., $0.5\times10^{-1}T_\mathrm{s}$)} were considered. The obtained constellations confirm that the highest \ac{EVM} values are observed for $\eta=1$, \textcolor{black}{besides an increasing \ac{EVM} trend from $\eta=2$ to $\eta=8$, while $N$ plays no significant role. While this would be different for lower \ac{RMS} \ac{SJ} values as previously discussed, rather low \ac{EVM} values would be experienced and their effect would not be clearly visible in the \ac{QPSK} constellations.} To enable a better understanding of the mean \ac{EVM} behavior and the constellation shapes, Fig.~\ref{fig:BBsubcEVM} shows the \ac{EVM} per subcarrier for the same parameters assumed for the results presented in Fig.~\ref{fig:QPSKconst_BB_N8_11_14_jm8}. The obtained results reveal that $\eta=1$ results in pronounced distortion at the edge subcarriers due to the use of the critical sampling frequency. It can also be seen that the \ac{EVM} pattern is \textcolor{black}{clearly improved with oversampling factors \mbox{$\eta\in\{2,4,8\}$}. The obtained values are still frequency-selective, which, however, is due to the considered colored \ac{SJ}. As expected from the results in Fig.~\ref{fig:samplingJitter_BB_EVM} for the \ac{RMS} \ac{SJ} of $\SI[parse-numbers = false]{10^{-10}}{\second}$, no significant difference is observed for different $N$ values for the considered \ac{RMS} \ac{SJ} value.}
	 	 
	 To further quantify the \ac{SJ}-induced communication performance degradation in the considered \ac{OFDM}-based \ac{ISAC} system, Fig.~\ref{fig:BB_SIR} shows the obtained \ac{SIR} as a function of the \ac{RMS} \ac{SJ}, which is assumed to be equal at both \ac{DAC} and \ac{ADC}. For the performed simulations, the same numbers of subcarriers $N$, \ac{CP} length, number of \ac{OFDM} symbols, digital modulation schemes, and oversampling factors was adopted as for the results in Figs.~\ref{fig:QPSKconst_BB_N8_11_14_jm8} and \ref{fig:BBsubcEVM}, i.e., \mbox{$N\in\{256,2048,16384\}$}, \mbox{$N_\mathrm{CP}=0$}, \mbox{$M=128$}, \ac{QPSK} and \mbox{$\eta\in\{1,2,4,8\}$}. The achieved results show a \textcolor{black}{constant \ac{SIR} of \SI{15.92}{dB}} for \mbox{$\eta=1$} regardless of the number of subcarriers $N$, which is mainly due to the severe distortion of the edge subcarriers. For ${\eta=2}$, in turn, a maximum \ac{SIR} of around \textcolor{black}{\SI{56.44}{dB}} is observerd, whereas {$\eta=4$} and {$\eta=8$} yield maximum \ac{SIR} values of approximately \textcolor{black}{\SI{57.27}{dB} and \SI{57.33}{dB}}, respectively. \textcolor{black}{The obtained results agree with the fact that oversampling has limited effectiveness against \ac{ADC} \ac{SJ} and the \ac{OOB} radiation due to \ac{DAC} \ac{SJ}, as previously discussed in Section~\ref{sec:sysModel}.} With increasing \ac{RMS} \ac{SJ} \textcolor{black}{after $\SI[parse-numbers = false]{10^{-11}}{\second}$ (i.e., $0.5\times10^{-12}T_\mathrm{s}$)}, however, the performance of the cases with higher $\eta$ among \mbox{$\eta\in\{2,4,8\}$} tend to \textcolor{black}{start degrading. In addition, slightly higher} robustness against increasing \ac{RMS} \ac{SJ} is observed for a higher number of subcarriers $N$ \textcolor{black}{for these oversampling factors}, as was the case in the previous \ac{EVM} analyses. Next, Fig.~\ref{fig:samplingJitter_BB_SIR_diffTxRx_4QAM_N2048} shows the obtained \ac{SIR} considering different \ac{RMS} \ac{SJ} values for the \ac{DAC} and the \ac{ADC}, \mbox{$N=2048$}, and \ac{QPSK} modulation. It can be seen that, apart from $\eta=1$, which achieves the worst \ac{SIR} performance at low \ac{RMS} \acp{SJ}, \textcolor{black}{a slight \ac{SIR} improvement is observed when increasing $\eta$ from $2$ to $4$, but no further noticeable enhancement is observable for \mbox{$\eta=8$},} which \textcolor{black}{supports} the previous results. In addition, the obtained results reveal that the adopted colored \ac{SJ} model leads to similar effects when applied at either \ac{DAC} or \ac{ADC}, which agrees with the result from \eqref{eq:Yl_3}.
	 
	 \begin{figure}[!t]
	 	\centering
	 	\psfrag{8}[c][c]{\scriptsize $8$}
	 	\psfrag{9}[c][c]{\scriptsize $9$}
	 	\psfrag{10}[c][c]{\scriptsize $10$}
	 	\psfrag{11}[c][c]{\scriptsize $11$}
	 	\psfrag{12}[c][c]{\scriptsize $12$}
	 	\psfrag{13}[c][c]{\scriptsize $13$}
	 	\psfrag{14}[c][c]{\scriptsize $14$}
	 	
	 	\psfrag{-70}[c][c]{\scriptsize -$70$}
	 	\psfrag{-60}[c][c]{\scriptsize -$60$}
	 	\psfrag{-50}[c][c]{\scriptsize -$50$}
	 	\psfrag{-40}[c][c]{\scriptsize -$40$}
	 	\psfrag{-30}[c][c]{\scriptsize -$30$}
	 	\psfrag{-20}[c][c]{\scriptsize -$20$}
	 	
	 	\psfrag{log2N}[c][c]{\footnotesize $\log_2(N)$}
	 	\psfrag{EVM (dB)}[c][c]{\footnotesize $\mathrm{EVM (dB)}$}
	 	
	 	\includegraphics[width=3.75cm]{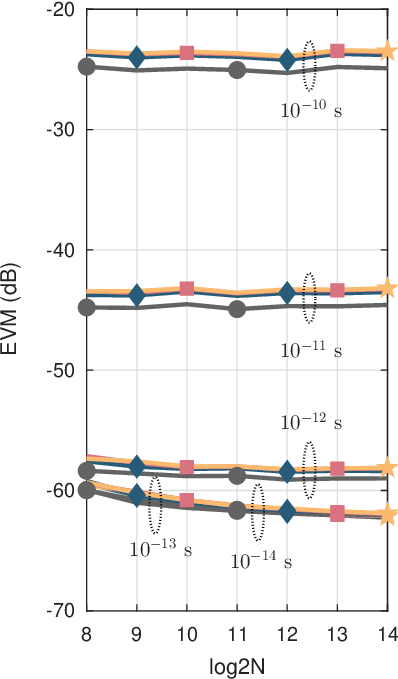}
	 	
	 	\captionsetup{justification=raggedright,labelsep=period,singlelinecheck=false}
	 	\caption{\ Mean EVM obtained with \ac{BP} sampling at a digital IF $f_\mathrm{IF}=\SI{1}{\giga\hertz}$ as a function of the number of subcarriers $N$. For the obtained results, $\eta=8$, $M=128$ OFDM symbols, and SJ RMS values of \textcolor{black}{$\SI[parse-numbers = false]{10^{-14}}{\second}$, $\SI[parse-numbers = false]{10^{-13}}{\second}$, $\SI[parse-numbers = false]{10^{-12}}{\second}$, $\SI[parse-numbers = false]{10^{-11}}{\second}$, and $\SI[parse-numbers = false]{10^{-10}}{\second}$ (i.e., $0.5\times10^{-5}T_\mathrm{s}$, $0.5\times10^{-4}T_\mathrm{s}$, $0.5\times10^{-3}T_\mathrm{s}$, $0.5\times10^{-2}T_\mathrm{s}$, and $0.5\times10^{-1}T_\mathrm{s}$, respectively)} at both the DAC and the ADC were considered. As in Fig.~\ref{fig:samplingJitter_BB_EVM}, \ac{QPSK} ({\color[rgb]{0.3922,0.3922,0.3922}$\CIRCLE$}), \mbox{16-\ac{QAM}} ({\color[rgb]{0.1490,0.3569,0.4824}$\blacklozenge$}), \mbox{64-\ac{QAM}} ({\color[rgb]{0.8471,0.4510,0.4980}$\blacksquare$}), and \mbox{256-\ac{QAM}} ({\color[rgb]{0.9882,0.7333,0.4275}$\bigstar$}) modulations were considered.}\label{fig:samplingJitter_BP_EVM}
	 	
	 \end{figure}
	 \begin{figure}[!t]
	 	\centering
	 	\subfloat[ ]{
	 		
	 		\psfrag{-1.5}[c][c]{\scriptsize -$1.5$}
	 		\psfrag{-0.5}[c][c]{\scriptsize -$0.5$}
	 		\psfrag{0.5}[c][c]{\scriptsize $0.5$}
	 		\psfrag{1.5}[c][c]{\scriptsize $1.5$}
	 		
	 		\psfrag{AA}[c][c]{\scriptsize -$1.5$}
	 		\psfrag{BB}[c][c]{\scriptsize -$0.5$}
	 		\psfrag{CC}[c][c]{\scriptsize $0.5$}
	 		\psfrag{DD}[c][c]{\scriptsize $1.5$}
	 		
	 		\psfrag{I}[c][c]{\scriptsize $I$}
	 		\psfrag{Q}[c][c]{\scriptsize $Q$}
	 		
	 		\includegraphics[width=2.5cm]{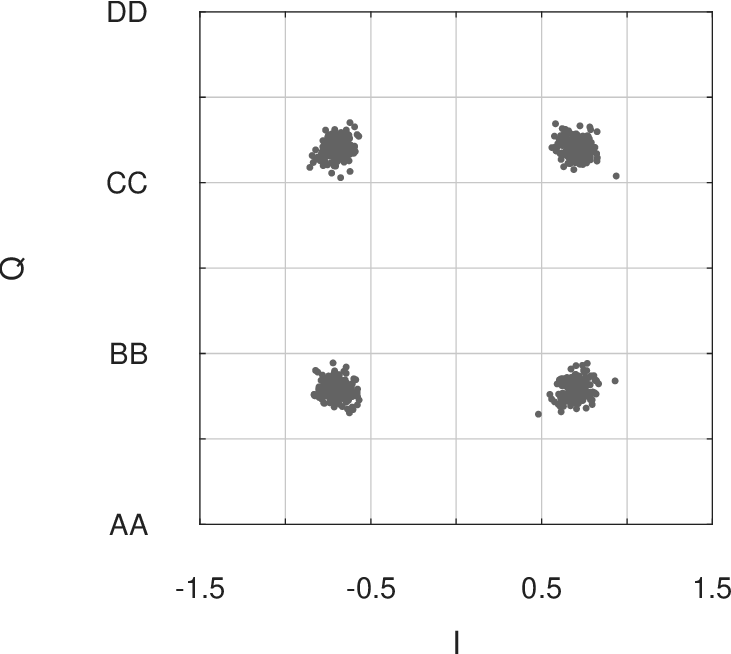}
	 	}\hspace{-0.2cm}
	 	\subfloat[ ]{
	 		
	 		\psfrag{-1.5}[c][c]{\scriptsize -$1.5$}
	 		\psfrag{-0.5}[c][c]{\scriptsize -$0.5$}
	 		\psfrag{0.5}[c][c]{\scriptsize $0.5$}
	 		\psfrag{1.5}[c][c]{\scriptsize $1.5$}
	 		
	 		\psfrag{AA}[c][c]{ }
	 		\psfrag{BB}[c][c]{ }
	 		\psfrag{CC}[c][c]{ }
	 		\psfrag{DD}[c][c]{ }
	 		
	 		\psfrag{I}[c][c]{\scriptsize $I$}
	 		\psfrag{Q}[c][c]{ }
	 		
	 		\includegraphics[width=2.5cm]{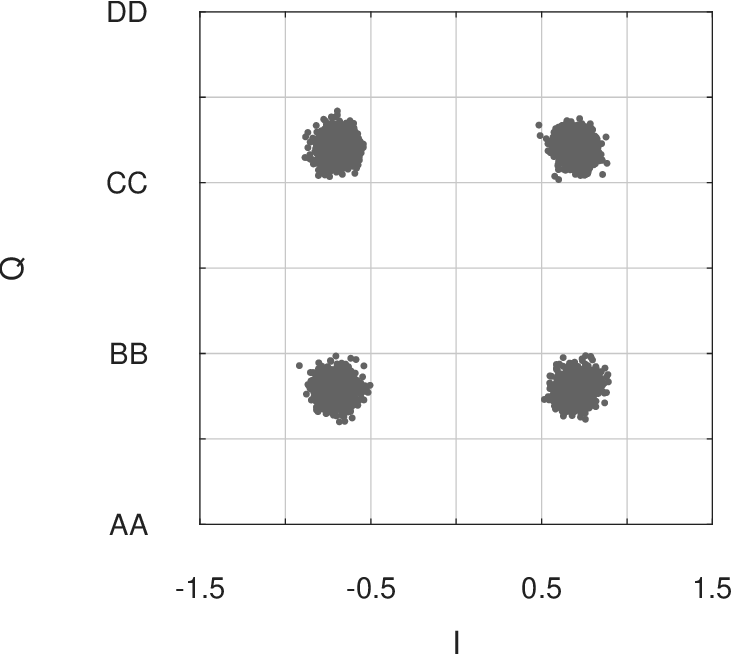}
	 	}\hspace{-0.2cm}
	 	\subfloat[ ]{
	 		
	 		\psfrag{-1.5}[c][c]{\scriptsize -$1.5$}
	 		\psfrag{-0.5}[c][c]{\scriptsize -$0.5$}
	 		\psfrag{0.5}[c][c]{\scriptsize $0.5$}
	 		\psfrag{1.5}[c][c]{\scriptsize $1.5$}
	 		
	 		\psfrag{AA}[c][c]{ }
	 		\psfrag{BB}[c][c]{ }
	 		\psfrag{CC}[c][c]{ }
	 		\psfrag{DD}[c][c]{ }
	 		
	 		\psfrag{I}[c][c]{\scriptsize $I$}
	 		\psfrag{Q}[c][c]{ }
	 		
	 		\includegraphics[width=2.5cm]{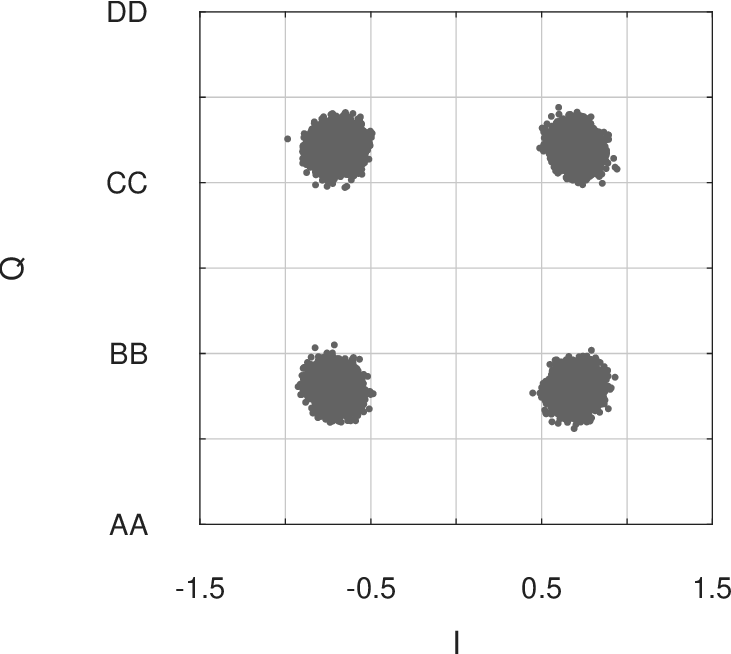}
	 	}		
	 	
	 	\captionsetup{justification=raggedright,labelsep=period,singlelinecheck=false}
	 	\caption{\ Normalized receive QPSK constellations obtained with \ac{BP} sampling at a digital IF $f_\mathrm{IF}=\SI{1}{\giga\hertz}$, $\eta=8$, $M=128$ OFDM symbols, and an RMS SJ of \textcolor{black}{$\SI[parse-numbers = false]{10^{-10}}{\second}$ (i.e., $0.5\times10^{-1}T_\mathrm{s}$)} at both the DAC and the ADC. Results are shown for (a) $N=256$, (b) $N=2048$, and (c) $N=16384$ total number of subcarriers.}\label{fig:QPSKconst_BP_N8_11_14_jm8}
	 	
	 \end{figure}
	 \begin{figure}[!t]
	 	\centering
	 	
	 	\psfrag{-60}[c][c]{\scriptsize -$60$}
	 	\psfrag{-45}[c][c]{\scriptsize -$45$}
	 	\psfrag{-30}[c][c]{\scriptsize -$30$}
	 	\psfrag{-15}[c][c]{\scriptsize -$15$}
	 	\psfrag{0}[c][c]{\scriptsize $0$}
	 	
	 	\psfrag{-N/2}[c][c]{\scriptsize $\tiny{-}N/2$}
	 	\psfrag{-N/4}[c][c]{\scriptsize $\tiny{-}N/4$}
	 	\psfrag{0}[c][c]{\scriptsize $0$}
	 	\psfrag{N/4}[c][c]{\scriptsize $N/4$}
	 	\psfrag{N/2-1}[c][c]{\scriptsize $N/2\tiny{-}1$}
	 	
	 	\psfrag{n}[c][c]{\footnotesize $n$}
	 	\psfrag{EVM (dB)}[c][c]{\footnotesize $\mathrm{EVM~(dB)}$}
	 	
	 	\includegraphics[width=4cm]{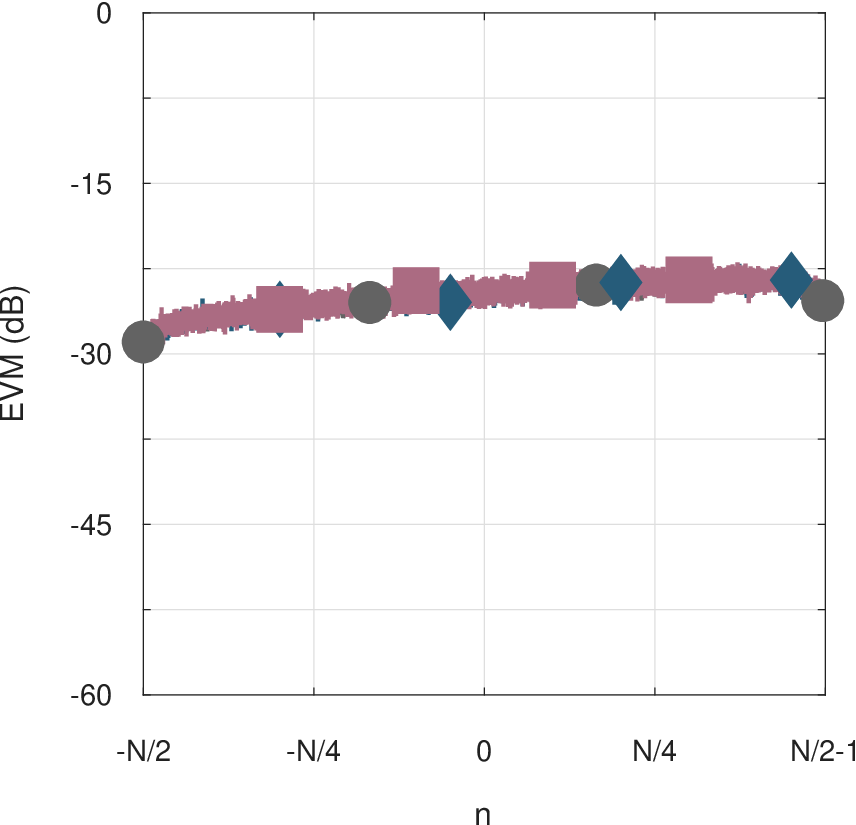}
	 	
	 	\captionsetup{justification=raggedright,labelsep=period,singlelinecheck=false}
	 	\caption{\ EVM at the $n\mathrm{th}$ subcarrier obtained with \ac{BP} sampling at a digital IF $f_\mathrm{IF}=\SI{1}{\giga\hertz}$, $\eta=8$, $M=128$ OFDM symbols, QPSK modulation and an RMS SJ of \textcolor{black}{$\SI[parse-numbers = false]{10^{-10}}{\second}$ (i.e., $0.5\times10^{-1}T_\mathrm{s}$)} at both the DAC and the ADC. Results are shown for \mbox{$N=256$} ({\color[rgb]{0.3922,0.3922,0.392}$\CIRCLE$}), \mbox{$N=2048$} ({\color[rgb]{0.1490,0.3569,0.4824}$\blacklozenge$}), and \mbox{$N=16384$} ({\color[rgb]{0.8471,0.4510,0.4980}$\blacksquare$}) total number of subcarriers.}\label{fig:BPsubcEVM}
	 	
	 \end{figure}
	 
	\subsection{Bandpass Sampling}\label{subsec:commBP}
	
	In \ac{BP} sampling, the transmit signal must first undergo oversampling before \ac{DUC} to the digital \ac{IF}. The inverse is done at the receiver side, i.e., \ac{DDC} followed by downsampling. In this subsection, a digital \ac{IF} of \mbox{$f_\mathrm{IF}=\SI{1}{\giga\hertz}$} and an oversampling factor \mbox{$\eta=8$} used during \ac{FDZP} are adopted to analyze the effect of \ac{SJ} on the communication performance of an \ac{OFDM}-based \ac{ISAC} system with \ac{BP} sampling. Given these parameters, Fig.~\ref{fig:samplingJitter_BP_EVM} shows the achieved mean \ac{EVM} as a function of the number of subcarriers for the same digital modulations and \ac{RMS} \ac{SJ} values considered for the results for \ac{BB} sampling in Fig.~\ref{fig:samplingJitter_BB_EVM}. Compared to the results for \mbox{$\eta=8$} in the \ac{BB} sampling case, considerably higher \ac{EVM} values are achieved with \ac{BP} sampling.
		
	\textcolor{black}{The aforementioned results are illustrated by the normalized receive \ac{QPSK} constellations in Fig.~\ref{fig:QPSKconst_BP_N8_11_14_jm8} for the \ac{RMS} \ac{SJ} of $\SI[parse-numbers = false]{10^{-10}}{\second}$ (i.e., $0.5\times10^{-1}T_\mathrm{s}$). The observed results show no difference in the constellations densities for different $N$ as expected from Fig.~\ref{fig:samplingJitter_BP_EVM}. Only less points are seeing for \mbox{$N=14$} as the same number of \ac{OFDM} symbols, i.e., \mbox{$M=128$}, was assumed in all cases. Compared to the constellations in the \ac{BB} case with \mbox{$\eta=8$} in Fig.~\ref{fig:QPSKconst_BB_N8_11_14_jm8}, a clear degradation of the constellation is observed both in the form of spreading due to \ac{ICI} and slight rotation of some points due to \ac{CPE}, which is mainly due to two factors. The first is the more severe \ac{ICI} effect, as the \ac{OFDM} signal is transmitted at higher frequencies where the \ac{PN} \ac{PSD} from which the \ac{SJ} is derived has accumulated a higher level as seen in Fig.~\ref{fig:samplingJitter_intPN}. The second factor, which is the source of the constellation rotation, is the \ac{CPE} that is experienced due to the \ac{PN}-like effect induced by the \ac{SJ} on the digital \ac{IF} used to transmit the \ac{OFDM} signal as discussed in \cite{putra2009}. These results are linked to the ones in Fig.~\ref{fig:BPsubcEVM}, where the obtained \ac{EVM} per subcarrier for the same signal parameters are shown. Unlike the results for \ac{BB} sampling in Fig.~\ref{fig:BBsubcEVM}, a smaller \ac{EVM} variation with an overall higher level is observed in all cases with \ac{BP} sampling, which is due to the aforementioned accumulated \ac{PN} level at that region and \ac{SJ}-induced \ac{CPE} to the digital \ac{IF}.}
	
\begin{figure}[!t]
		\centering
		\subfloat[ ]{
			
			\psfrag{10}[c][c]{\scriptsize $10$}
			\psfrag{30}[c][c]{\scriptsize $30$}
			\psfrag{50}[c][c]{\scriptsize $50$}
			\psfrag{70}[c][c]{\scriptsize $70$}
			
			\psfrag{-17}[c][c]{\scriptsize -$17$}
			\psfrag{-15.25}[c][c]{\scriptsize -$15.25$}
			\psfrag{-13.5}[c][c]{\scriptsize -$13.5$}
			\psfrag{-11.75}[c][c]{\scriptsize -$11.75$}	
			\psfrag{-10}[c][c]{\scriptsize -$10$}
			
			\psfrag{RMSJ}[c][c]{\footnotesize $\log_{10}\left(\delta_\mathrm{SJ,RMS}\right)$}
			\psfrag{SIR (dB)}[c][c]{\footnotesize $\mathrm{SIR~(dB)}$}
			
			\includegraphics[width=3.75cm]{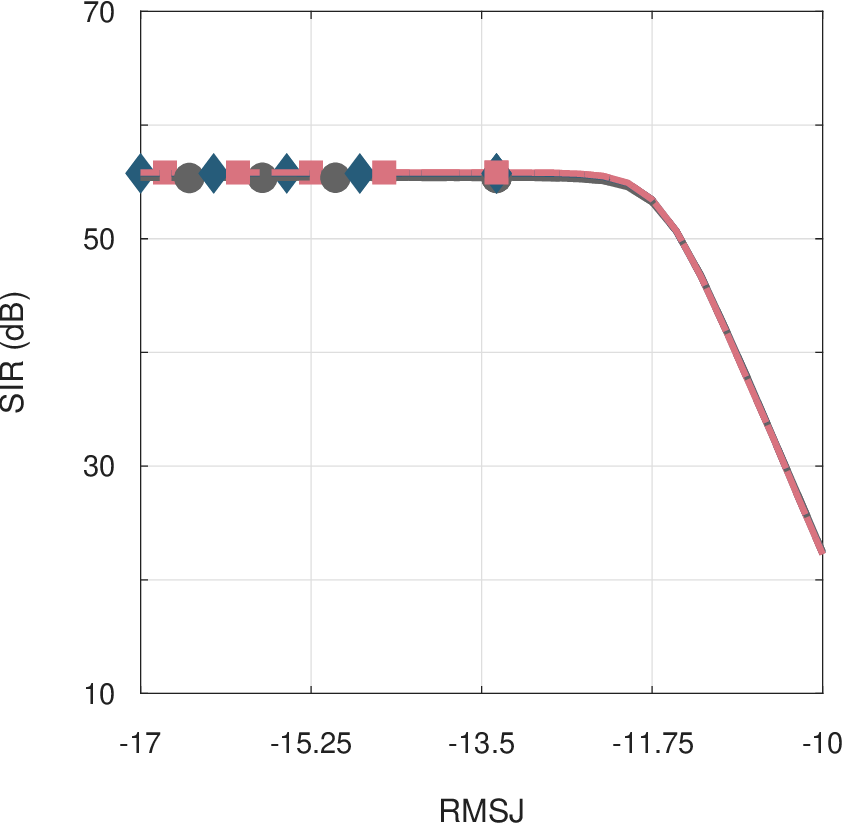}\label{fig:BP_SIR_OS8}			
		}\hspace{0.1cm}
		\subfloat[ ]{
			
			\psfrag{10}[c][c]{\scriptsize $10$}
			\psfrag{30}[c][c]{\scriptsize $30$}
			\psfrag{50}[c][c]{\scriptsize $50$}
			\psfrag{70}[c][c]{\scriptsize $70$}
			
			\psfrag{-17}[c][c]{\scriptsize -$17$}
			\psfrag{-15.25}[c][c]{\scriptsize -$15.25$}
			\psfrag{-13.5}[c][c]{\scriptsize -$13.5$}
			\psfrag{-11.75}[c][c]{\scriptsize -$11.75$}	
			\psfrag{-10}[c][c]{\scriptsize -$10$}
			
			\psfrag{RMSJ}[c][c]{\footnotesize $\log_{10}\left(\delta_\mathrm{SJ,RMS}\right)$}
			\psfrag{SIR (dB)}[c][c]{}
			
			\includegraphics[width=3.75cm]{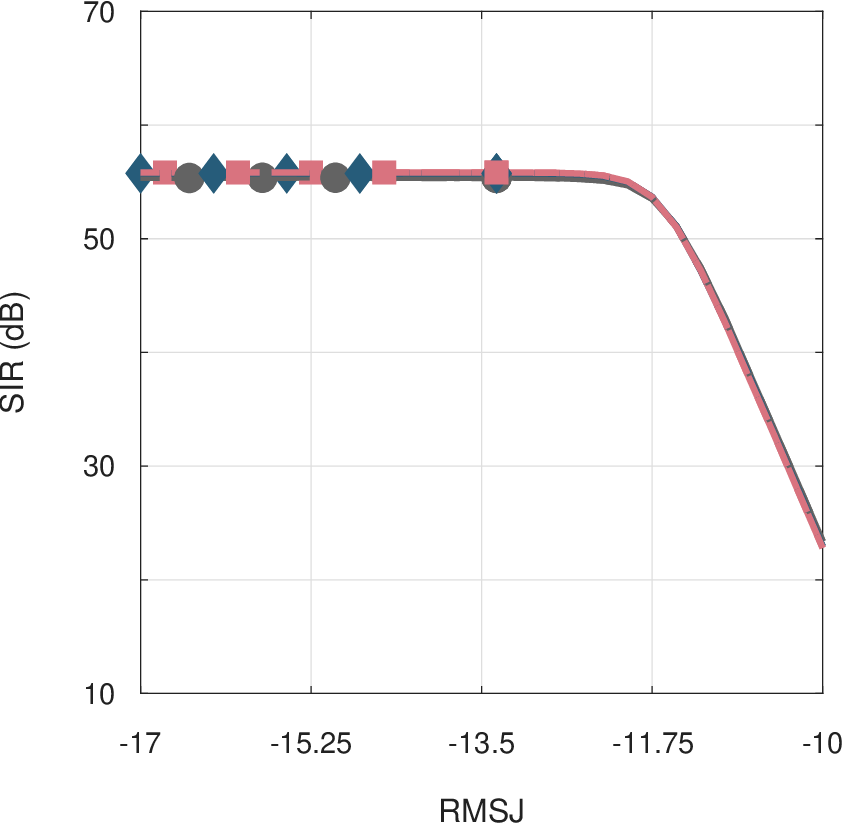}\label{fig:BP_SIR_OS8_CPEcorr}			
		}
		
		\captionsetup{justification=raggedright,labelsep=period,singlelinecheck=false}
		\caption{\ SIR as a function of the RMS SJ (same at both the DAC and the ADC) obtained with \ac{BP} sampling at a digital IF $f_\mathrm{IF}=\SI{1}{\giga\hertz}$, \mbox{$\eta=8$}, and $M=128$ OFDM symbols (a) without and (b) with CPE correction. In all cases, results are shown for \mbox{$N=256$} ({\color[rgb]{0.3922,0.3922,0.392}$\CIRCLE$}), \mbox{$N=2048$} ({\color[rgb]{0.1490,0.3569,0.4824}$\blacklozenge$}), and \mbox{$N=16384$} ({\color[rgb]{0.8471,0.4510,0.4980}$\blacksquare$}), as well as QPSK ({\color[rgb]{0,0,0}\textbf{\textemdash}}) and 256-QAM ({\color[rgb]{0,0,0}\textbf{\textendash~\textendash}}) modulations. \textcolor{black}{In the obtained results, an SIR improvement of only around \SI{0.57}{dB} is seen when CPE compensation is performed.} In addition, no significant difference is observed between the experienced SIR with QPSK and 256-QAM.}\label{fig:BP_SIR}
		
	\end{figure}

\begin{figure}[!t]
		\captionsetup[subfigure]{labelformat=empty}
		
		\centering
		
		\psfrag{55}{\footnotesize (a)}
		\psfrag{22}{\footnotesize (b)}
		
		\psfrag{-17}[c][c]{\scriptsize -$17$}
		\psfrag{-13.5}[c][c]{\scriptsize -$13.5$}
		\psfrag{-10}[c][c]{\scriptsize -$10$}
		
		\psfrag{10}[c][c]{\scriptsize $10$}
		\psfrag{30}[c][c]{\scriptsize $30$}
		\psfrag{50}[c][c]{\scriptsize $50$}
		\psfrag{70}[c][c]{\scriptsize $70$}
		
		\psfrag{RMSJT}[c][c]{\footnotesize $\log_{10}(\delta^\mathrm{Tx}_\mathrm{SJ,RMS})$}
		\psfrag{RMSJR}[c][c]{\footnotesize $\log_{10}(\delta^\mathrm{Rx}_\mathrm{SJ,RMS})$}
		\psfrag{SIR (dB)}[c][c]{\footnotesize $\mathrm{SIR~(dB)}$}

		\includegraphics[height=4.08cm]{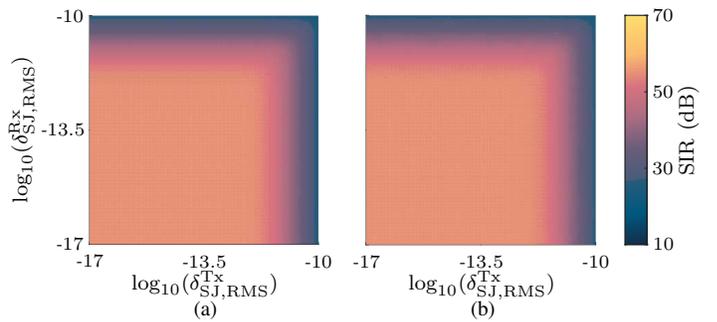}
		
		\captionsetup{justification=raggedright,labelsep=period,singlelinecheck=false}
		\caption{\ SIR as a function of the RMS SJ (same at both the DAC and the ADC) obtained with \ac{BP} sampling at a digital \ac{IF} of $f_\mathrm{IF}=\SI{1}{\giga\hertz}$, \mbox{$\eta=8$}, $N=2048$ subcarriers, and $M=128$ OFDM symbols, and QPSK modulation (a) without and (b) with CPE correction. \textcolor{black}{As in Fig.~\ref{fig:BP_SIR}, only around only around \SI{0.57}{dB} SIR improvement due to CPE compensation was observed.}}\label{fig:BP_SIR_diffTxRx}
		
	\end{figure}

	Next, Fig.~\ref{fig:BP_SIR} shows the experienced \ac{SIR} as a function of the \ac{RMS} \ac{SJ} for \ac{QPSK} modulation and the same remaining signal parameters assumed for the results in Figs.~\ref{fig:QPSKconst_BP_N8_11_14_jm8} and \ref{fig:BPsubcEVM}. The obtained results show that an \textcolor{black}{maximum \ac{SIR} of around \SI{55.82}{dB} is achieved for \textcolor{black}{all considered $N$ values}, which is around \SI{1.51}{dB}} lower than in the \ac{BB} sampling case with \mbox{$\eta=8$} in Fig.~\ref{fig:BB_SIR}(d). If \ac{CPE} estimation and correction is performed, e.g., as described in \cite{giroto2024PN}, the \ac{SIR} is improved by only around \textcolor{black}{\SI{0.57}{dB} for \ac{RMS} \acp{SJ} above $\SI[parse-numbers = false]{10^{-11}}{\second}$ (i.e., $0.5\times10^{-2}T_\mathrm{s}$)}, which is where the \ac{SIR} \textcolor{black}{starts degrading}. The reason for this is that, when \ac{CPE} becomes relevant, the \ac{ICI} induced by \ac{SJ} is already severe enough to impair the \ac{CPE} estimation and make its correction less effective. Results assuming different \ac{RMS} \ac{SJ} values at the \ac{DAC} and \ac{ADC} are finally shown for a reduced \ac{RMS} \ac{SJ} range in Fig.~\ref{fig:BP_SIR_diffTxRx} for the cases with and without \ac{CPE} compensation. As in the \ac{BB} sampling case for \mbox{$\eta=8$} in Fig.~\ref{fig:samplingJitter_BB_SIR_diffTxRx_4QAM_N2048}(d), similar effects are observed for both \ac{DAC} and \ac{ADC} \acp{SJ}. With \ac{BP} sampling, however, a significant \ac{SIR} degradation is experienced as previously discussed, and the \ac{CPE} correction presents the same slight \ac{SIR} improvement of only around \textcolor{black}{\SI{0.57}{dB} for \ac{RMS} \acp{SJ} above $\SI[parse-numbers = false]{10^{-11}}{\second}$}. This allows concluding that, for the adopted colored \ac{SJ} model, \ac{ICI} is the dominant impairment induced by \ac{SJ}.
	
	\subsection{Remarks on simulation results}\label{subsec:commRemarks}
	
	As evidenced by the obtained expressions \eqref{eq:Yl_3} and \eqref{eq:Rk_1}, besides related studies in the literature, it was shown that \ac{SJ} exclusively causes \ac{ICI} if \ac{BB} sampling is adopted. If \ac{BP} sampling is adopted instead, \ac{SJ} also leads to \ac{CPE} and additional \ac{ICI} due to a \ac{PN}-like effect on the digital \ac{IF}. Consequently, \ac{BP} tends to result in more severely degraded communication performance, yielding higher \ac{EVM} and \ac{SIR} than in the \ac{BB} case as the receive constellations are not only spread due to \ac{ICI}, but also rotated due to \ac{CPE}. It was also shown that the absence of oversampling in the \ac{BB} sampling case, i.e., \mbox{$\eta=1$}, leads to the distortion of the edge subcarriers and therefore the highest \ac{EVM} and lowest \ac{SIR}. While performing oversampling prevents that issue, the results have shown that adopting increasing oversampling factors among the considered values $\eta\in\{2,4,8\}$ only leads to improvement at rather low \ac{RMS} \ac{SJ} at both \ac{DAC} and \ac{ADC}. For higher \ac{RMS} \ac{SJ}, however, more significant \ac{OOB} radiation due to \ac{DAC} \ac{SJ} happens, which combined with the \ac{ICI} induced by both \ac{DAC} and \ac{ADC} \ac{SJ}, leads to more degraded communication performance, both in terms of \ac{EVM} and \ac{SIR}, than with a lower oversampling factor among the considered ones. Finally, it can be concluded that \ac{RMS} \ac{SJ} values of up to around \textcolor{black}{$\SI[parse-numbers = false]{10^{-10}}{\second}$ (i.e., $0.5\times10^{-1}T_\mathrm{s}$)} at both \ac{DAC} and \ac{ADC} result in tolerable communication performance degradation for \ac{BB} sampling with all considered oversampling factors, ensuring \ac{EVM} \textcolor{black}{below \SI{-40}{dB} and maximum \ac{SIR} of around \SI{56}{dB} to \SI{57}{dB}. The reason why better values are not obtained with even lower \ac{RMS} \ac{SJ} values is due to limited accuracy in the \ac{SJ} simulation, which is done with via resampling with a Farrow filter \cite{farrow1988} preceded by cubic interpolation \cite{erup1993}.} For the \ac{BP} sampling scenario at an \ac{IF} of \mbox{$f_\mathrm{IF}=\SI{1}{\giga\hertz}$} and with oversampling \mbox{$\eta=8$}, \textcolor{black}{the same \ac{RMS} \ac{SJ} upper bound applies.}


	\section{Numerical Results for Radar Sensing}\label{sec:sensPerf}
	
	In this section, the effects of \ac{SJ} on \textcolor{black}{radar sensing in} \ac{OFDM}-based \ac{ISAC} systems is analyzed. For that purpose, the same \ac{SJ} model from Section~\ref{sec:commPerf} is adopted. To analyze how the \ac{EVM} and \ac{SIR} degradation ultimately translate into degradation of target reflections in a radar image, several performance parameters are used. These include \ac{PPLR}, \ac{PSLR}, and \ac{ISLR} \cite{lellouch2016,giroto2021_tmtt,liao2024}, as well as \ac{SIR} \cite{giroto2024PN}. 
	To better understand the effects of \ac{SJ}, the \ac{PSLR} and \ac{ISLR} parameters are evaluated for both range and Doppler shift directions of the simulated radar images. Finally, \ac{SIR} is used as a measure of the ratio between a target peak and the interference floor raised by \ac{SJ} in the radar image. 
	
	The obtained results assuming \ac{BB} and \ac{BP} sampling are discussed in Sections~\ref{subsec:sensBB} and \ref{subsec:sensBP}, respectively, with remarks on the presented simulation results given in Section~\ref{subsec:sensRemarks}. For the performed simulations, the parameters from Table~\ref{tab:ofdmParameters} are considered. For conciseness, however, only $N\in\{256,2048,16384\}$, besides \ac{QPSK} and \mbox{256-\ac{QAM}} are henceforth considered. In addition, \textcolor{black}{a scenario with a single static target was considered and} all other typical impairments in \ac{OFDM}-based \ac{ISAC} systems such as \ac{AWGN}, external interference and self-interference are not considered to focus on the \ac{SJ} effects. Since the assumed \ac{SJ} model is uncorrelated in both monostatic and bistatic \ac{ISAC} architectures, the same degradation due to \ac{SJ} is experienced in both cases. Therefore, the following radar sensing performance analyses assume a bistatic \ac{OFDM}-based \ac{ISAC} with perfect synchronization and full knowledge of the transmit \ac{OFDM} frame for radar signal processing. Finally, the same \ac{SJ} is assumed for both \ac{DAC} and \ac{ADC}. This is done since, as also demonstrated in \cite{gonem2021} and by the results in Section~\ref{sec:commPerf},  both \ac{DAC} and \ac{ADC} \ac{SJ} yield the same effects. The expression in \eqref{eq:Yl_3} also confirms this, showing that the combined \ac{DAC} and \ac{ADC} \ac{SJ} is the defining factor for the \ac{OFDM}-based \ac{ISAC} system performance.
	
	\subsection{Baseband Sampling}\label{subsec:sensBB}
	
	\textcolor{black}{Assuming \ac{BB} sampling for the considered \ac{OFDM}-based \ac{ISAC} system, \ac{PPLR}, range and Doppler shift \ac{PSLR}, as well as range and Doppler shift \ac{ISLR} were calculated as functions of the normalized \ac{RMS} \ac{SJ} values between $\SI[parse-numbers = false]{10^{-17}}{\second}$ and $\SI[parse-numbers = false]{10^{-10}}{\second}$ (i.e., $0.5\times10^{-8}T_\mathrm{s}$ and $0.5\times10^{-1}T_\mathrm{s}$, respectively) for oversampling factors \mbox{$\eta\in\{1,2,4,8\}$}. In the performed simulations, the considered \ac{RMS} \ac{SJ}  values were assumed to be equal at both \ac{DAC} and \ac{ADC}, and only rectangular windowing was performed during both range and Doppler processing. Since negligible difference between \ac{QPSK} and \mbox{256-\ac{QAM}}, among different $N$ values, and for different \ac{RMS} \ac{SJ} values within the considered range was observed, the obtained results are shown in Table~\ref{tab:PPLR_PSLR_ISLR_BB} as a function of the oversampling factor $\eta$ only. The obtained results show negligible \ac{PPLR} for all cases, and similar overall performance. The most pronounced differences, which are still rather small, are observed in the for range \ac{PSLR} and \ac{ISLR} with $\eta=1$. With this oversampling factor, the worst values are achieves due to the distortion of the edge subcarriers even with low \ac{RMS} \ac{SJ}. It is worth highlighting that, since \ac{CPE} due to \ac{SJ} is not experienced in the \ac{BB} sampling case and the fixed number $M=128$ of \ac{OFDM} symbols is assumed, the same Doppler shift \ac{PSLR} and \ac{ISLR} is observed in all cases.}
	
	\begin{table}[!t]
		\centering
		\captionsetup{width=43pc,justification=centering,labelsep=newline}
		\caption{\textcolor{black}{\textsc{Simulated PPLR, Range and Doppler Shift PSLR, and Range\\ and Doppler Shift ISLR for BB Sampling and RMS SJ between\\ $\SI[parse-numbers = false]{10^{-17}}{\second}$ and $\SI[parse-numbers = false]{10^{-10}}{\second}$ ($0.5\times10^{-8}T_\mathrm{s}$ and $0.5\times10^{-1}T_\mathrm{s}$)}}}
		\label{tab:PPLR_PSLR_ISLR_BB}
			\begin{tabular}{|c|c|c|c|c|c|}
				\hhline{|======|}
				\multirow{2}{*}{\textcolor{black}{$\bm{\eta}$}} & \multirow{2}{*}{\textcolor{black}{\textbf{PPLR}}} & \multirow{2}{*}{\textbf{\begin{tabular}[c]{@{}c@{}}\textcolor{black}{Range}\\ \textcolor{black}{PSLR (dB)}\end{tabular}}} & \multirow{2}{*}{\textbf{\begin{tabular}[c]{@{}c@{}}\textcolor{black}{Doppler}\\ \textcolor{black}{PSLR (dB)}\end{tabular}}} & \multirow{2}{*}{\textbf{\begin{tabular}[c]{@{}c@{}}\textcolor{black}{Range}\\ \textcolor{black}{ISLR (dB)}\end{tabular}}} & \multirow{2}{*}{\textbf{\begin{tabular}[c]{@{}c@{}}\textcolor{black}{Doppler}\\ \textcolor{black}{ISLR (dB)}\end{tabular}}} \\
				& & & & &                   \\\hhline{|======|}
				\textcolor{black}{1} & \textcolor{black}{$\SI{0}{dB}$} & \textcolor{black}{$\SI{-13.44}{dB}$} & \textcolor{black}{$\SI{-13.30}{dB}$} & \textcolor{black}{$\SI{-10.84}{dB}$}  & \textcolor{black}{$\SI{-9.68}{dB}$} \\\hline
				\textcolor{black}{2} & \textcolor{black}{$\SI{0}{dB}$} & \textcolor{black}{$\SI{-13.30}{dB}$} & \textcolor{black}{$\SI{-13.30}{dB}$} & \textcolor{black}{$\SI{-9.68}{dB}$}  & \textcolor{black}{$\SI{-9.68}{dB}$} \\\hline
				\textcolor{black}{4} & \textcolor{black}{$\SI{0}{dB}$} & \textcolor{black}{$\SI{-13.30}{dB}$} & \textcolor{black}{$\SI{-13.30}{dB}$} & \textcolor{black}{$\SI{-9.69}{dB}$}  & \textcolor{black}{$\SI{-9.68}{dB}$} \\\hline
				\textcolor{black}{8} & \textcolor{black}{$\SI{0}{dB}$} &\textcolor{black}{$\SI{-13.27}{dB}$} & \textcolor{black}{$\SI{-13.30}{dB}$} & \textcolor{black}{$\SI{-9.67}{dB}$}  & \textcolor{black}{$\SI{-9.68}{dB}$} \\				
				\hhline{|======|}
			\end{tabular}
	\end{table}

	\begin{figure*}[!t]
		\centering
		\subfloat[ ]{
			
			\psfrag{-17}[c][c]{\scriptsize -$17$}
			\psfrag{-15.25}[c][c]{\scriptsize -$15.25$}
			\psfrag{-13.5}[c][c]{\scriptsize -$13.5$}
			\psfrag{-11.75}[c][c]{\scriptsize -$11.75$}	
			\psfrag{-10}[c][c]{\scriptsize -$10$}
			
			\psfrag{60}[c][c]{\scriptsize $60$}
			\psfrag{90}[c][c]{\scriptsize $90$}
			\psfrag{120}[c][c]{\scriptsize $120$}
			\psfrag{150}[c][c]{\scriptsize $150$}
			
			\psfrag{RMSJ}[c][c]{\footnotesize $\log_{10}\left(\delta_\mathrm{SJ,RMS}\right)$}
			\psfrag{AAAA Image SIR (dB)}[c][c]{\footnotesize Mean image SIR (dB)}
			
			\includegraphics[height=3.6cm]{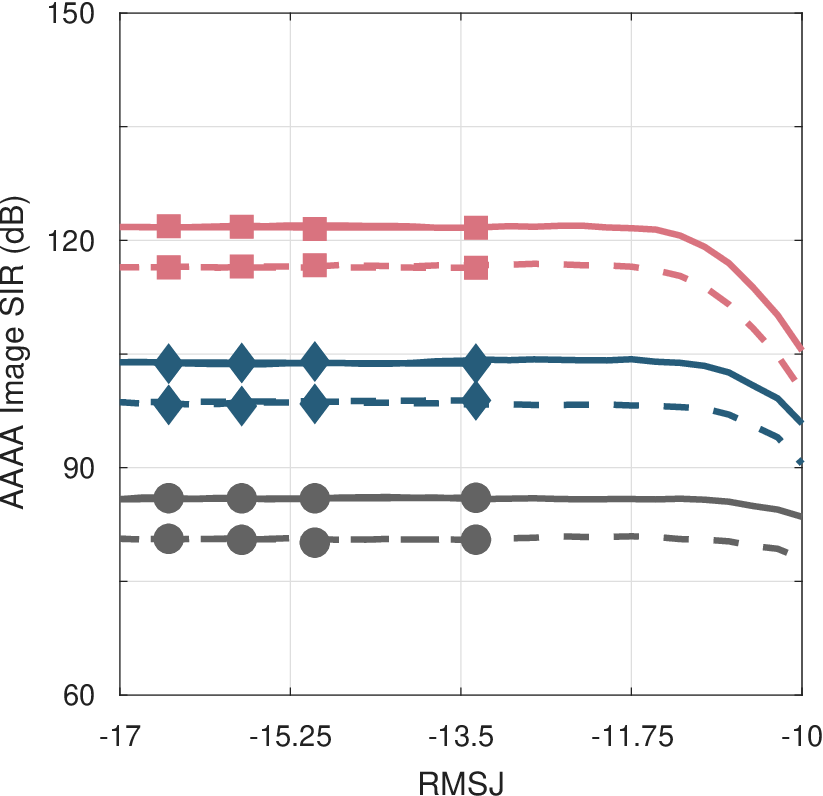}\label{fig:BB_rad_meanSIR_OS1}			
		}\hspace{0.05cm}
		\subfloat[ ]{
			
			\psfrag{-17}[c][c]{\scriptsize -$17$}
			\psfrag{-15.25}[c][c]{\scriptsize -$15.25$}
			\psfrag{-13.5}[c][c]{\scriptsize -$13.5$}
			\psfrag{-11.75}[c][c]{\scriptsize -$11.75$}	
			\psfrag{-10}[c][c]{\scriptsize -$10$}
			
			\psfrag{60}[c][c]{}
			\psfrag{90}[c][c]{}
			\psfrag{120}[c][c]{}
			\psfrag{150}[c][c]{}
			
			\psfrag{RMSJ}[c][c]{\footnotesize $\log_{10}\left(\delta_\mathrm{SJ,RMS}\right)$}
			\psfrag{AAAA Image SIR (dB)}[c][c]{}
			
			\includegraphics[height=3.6cm]{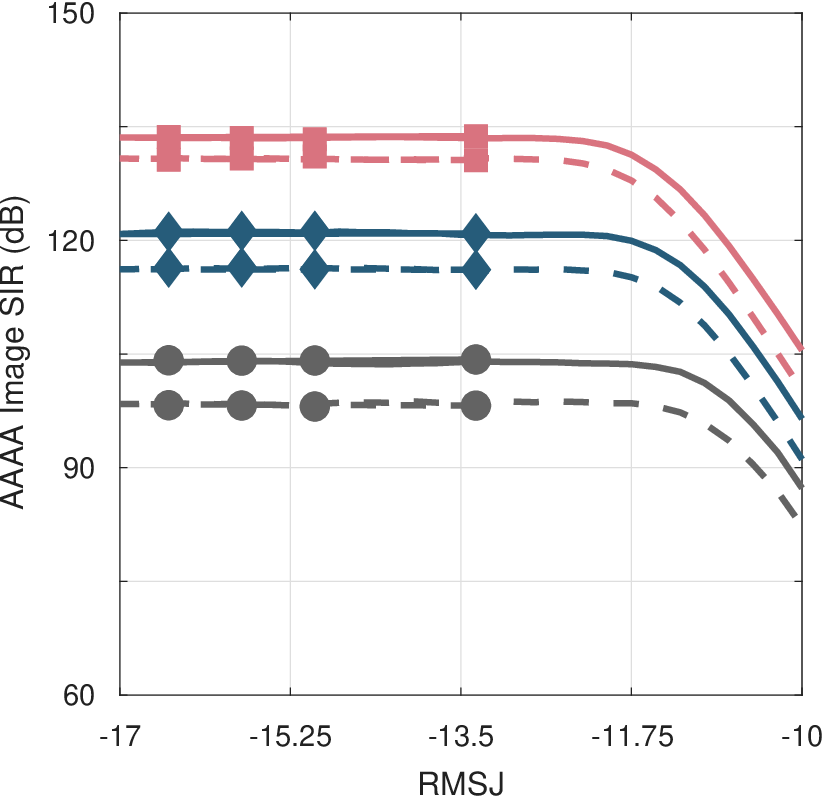}\label{fig:BB_rad_meanSIR_OS2}			
		}\hspace{0.05cm}
		\subfloat[ ]{
			
			\psfrag{-17}[c][c]{\scriptsize -$17$}
			\psfrag{-15.25}[c][c]{\scriptsize -$15.25$}
			\psfrag{-13.5}[c][c]{\scriptsize -$13.5$}
			\psfrag{-11.75}[c][c]{\scriptsize -$11.75$}	
			\psfrag{-10}[c][c]{\scriptsize -$10$}
			
			\psfrag{60}[c][c]{}
			\psfrag{90}[c][c]{}
			\psfrag{120}[c][c]{}
			\psfrag{150}[c][c]{}
			
			\psfrag{RMSJ}[c][c]{\footnotesize $\log_{10}\left(\delta_\mathrm{SJ,RMS}\right)$}
			\psfrag{AAAA Image SIR (dB)}[c][c]{}
			
			\includegraphics[height=3.6cm]{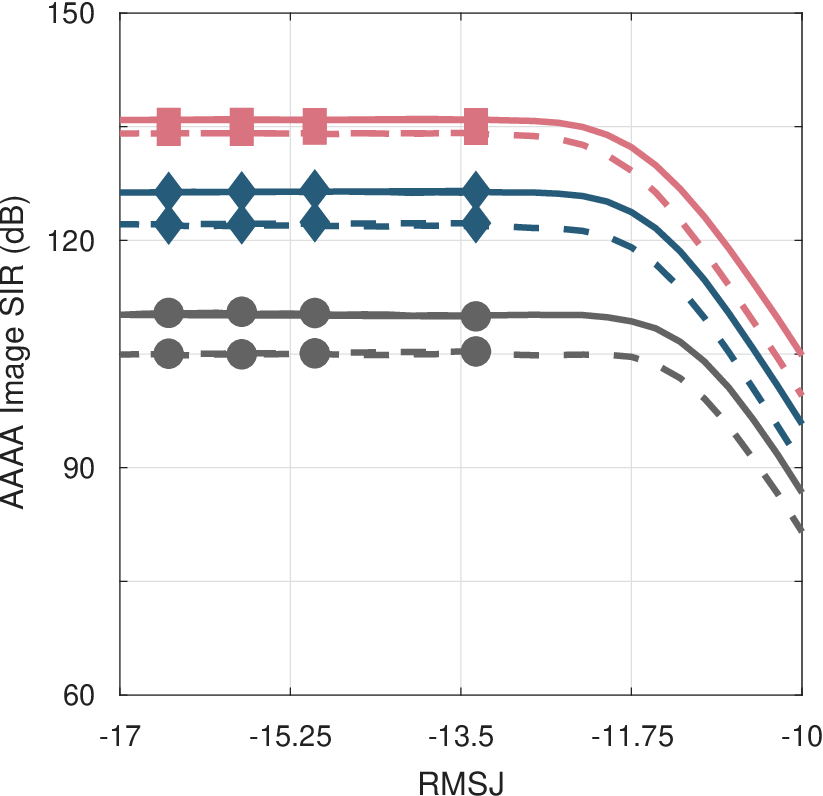}\label{fig:BB_rad_meanSIR_OS4}			
		}\hspace{0.05cm}
		\subfloat[ ]{
			
			\psfrag{-17}[c][c]{\scriptsize -$17$}
			\psfrag{-15.25}[c][c]{\scriptsize -$15.25$}
			\psfrag{-13.5}[c][c]{\scriptsize -$13.5$}
			\psfrag{-11.75}[c][c]{\scriptsize -$11.75$}	
			\psfrag{-10}[c][c]{\scriptsize -$10$}
			
			\psfrag{60}[c][c]{}
			\psfrag{90}[c][c]{}
			\psfrag{120}[c][c]{}
			\psfrag{150}[c][c]{}
			
			\psfrag{RMSJ}[c][c]{\footnotesize $\log_{10}\left(\delta_\mathrm{SJ,RMS}\right)$}
			\psfrag{AAAA Image SIR (dB)}[c][c]{}
			
			\includegraphics[height=3.6cm]{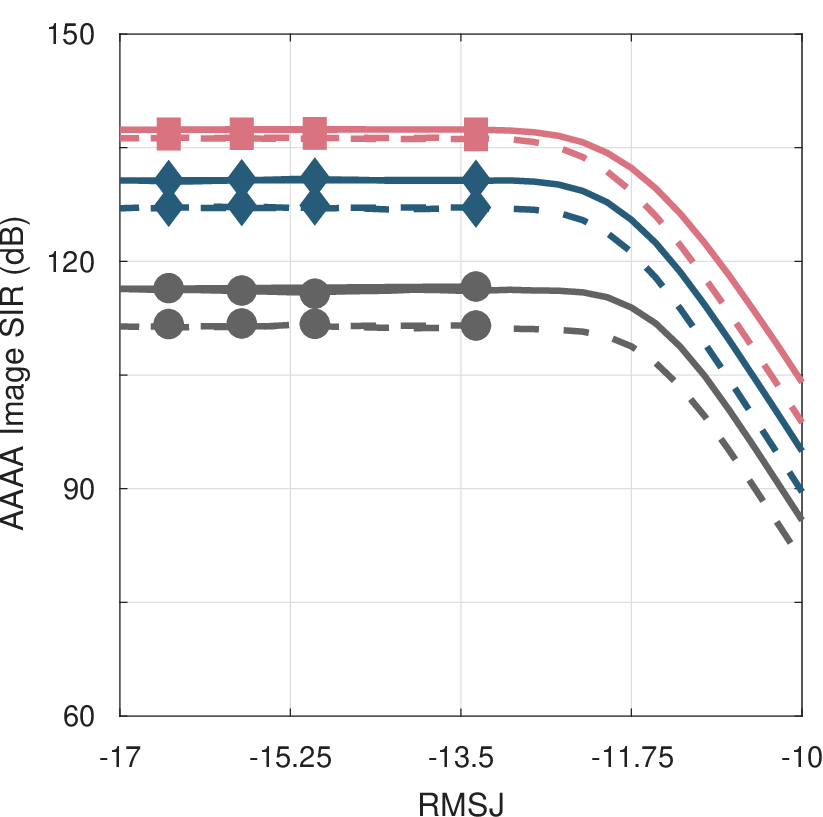}\label{fig:BB_rad_meanSIR_OS8}			
		}\\
		\subfloat[ ]{
			
			\psfrag{-17}[c][c]{\scriptsize -$17$}
			\psfrag{-15.25}[c][c]{\scriptsize -$15.25$}
			\psfrag{-13.5}[c][c]{\scriptsize -$13.5$}
			\psfrag{-11.75}[c][c]{\scriptsize -$11.75$}	
			\psfrag{-10}[c][c]{\scriptsize -$10$}
			
			\psfrag{60}[c][c]{\scriptsize $60$}
			\psfrag{90}[c][c]{\scriptsize $90$}
			\psfrag{120}[c][c]{\scriptsize $120$}
			\psfrag{150}[c][c]{\scriptsize $150$}
			
			\psfrag{RMSJ}[c][c]{\footnotesize $\log_{10}\left(\delta_\mathrm{SJ,RMS}\right)$}
			\psfrag{AAAA Image SIR (dB)}[c][c]{\footnotesize Min. image SIR (dB)}
			
			\includegraphics[height=3.6cm]{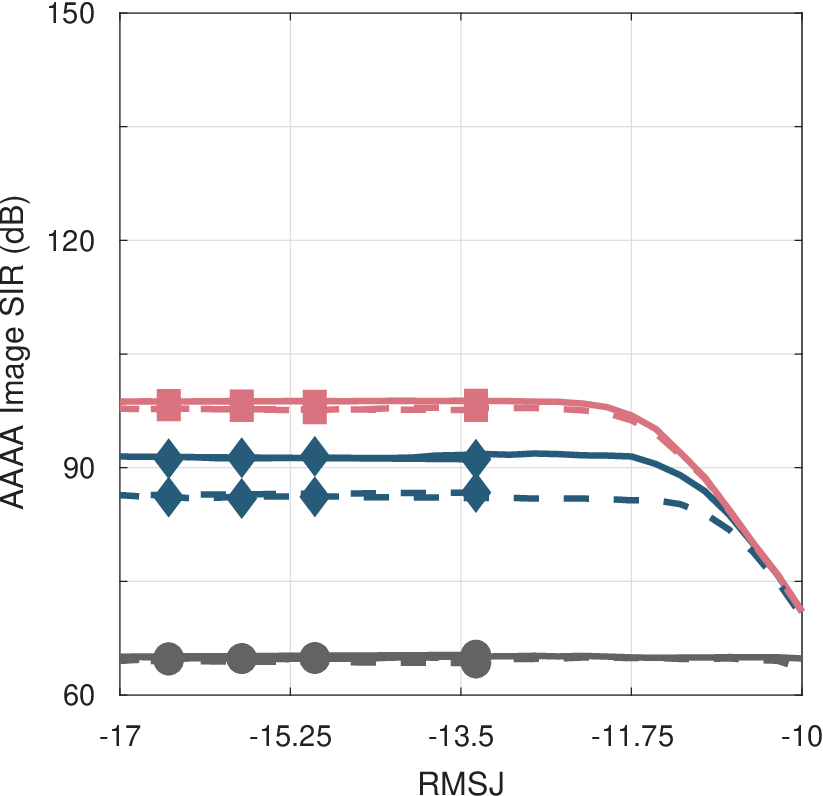}\label{fig:BB_rad_minSIR_OS1}			
		}\hspace{0.05cm}
		\subfloat[ ]{
			
			\psfrag{-17}[c][c]{\scriptsize -$17$}
			\psfrag{-15.25}[c][c]{\scriptsize -$15.25$}
			\psfrag{-13.5}[c][c]{\scriptsize -$13.5$}
			\psfrag{-11.75}[c][c]{\scriptsize -$11.75$}	
			\psfrag{-10}[c][c]{\scriptsize -$10$}
			
			\psfrag{60}[c][c]{}
			\psfrag{90}[c][c]{}
			\psfrag{120}[c][c]{}
			\psfrag{150}[c][c]{}
			
			\psfrag{RMSJ}[c][c]{\footnotesize $\log_{10}\left(\delta_\mathrm{SJ,RMS}\right)$}
			\psfrag{AAAA Image SIR (dB)}[c][c]{}
			
			\includegraphics[height=3.6cm]{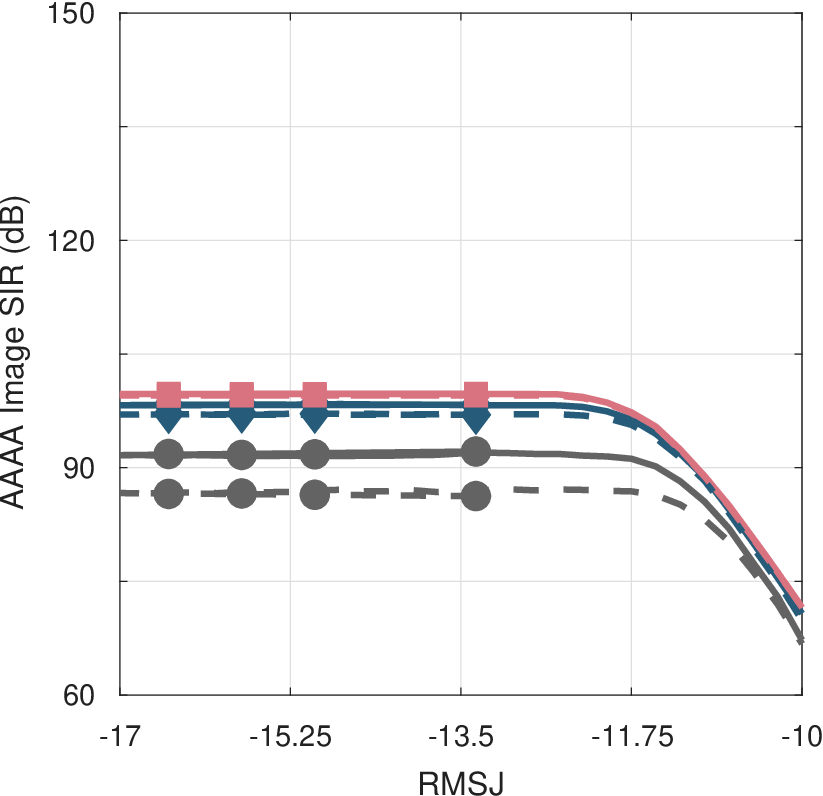}\label{fig:BB_rad_minSIR_OS2}			
		}\hspace{0.05cm}
		\subfloat[ ]{
			
			\psfrag{-17}[c][c]{\scriptsize -$17$}
			\psfrag{-15.25}[c][c]{\scriptsize -$15.25$}
			\psfrag{-13.5}[c][c]{\scriptsize -$13.5$}
			\psfrag{-11.75}[c][c]{\scriptsize -$11.75$}	
			\psfrag{-10}[c][c]{\scriptsize -$10$}
			
			\psfrag{60}[c][c]{}
			\psfrag{90}[c][c]{}
			\psfrag{120}[c][c]{}
			\psfrag{150}[c][c]{}
			
			\psfrag{RMSJ}[c][c]{\footnotesize $\log_{10}\left(\delta_\mathrm{SJ,RMS}\right)$}
			\psfrag{AAAA Image SIR (dB)}[c][c]{}
			
			\includegraphics[height=3.6cm]{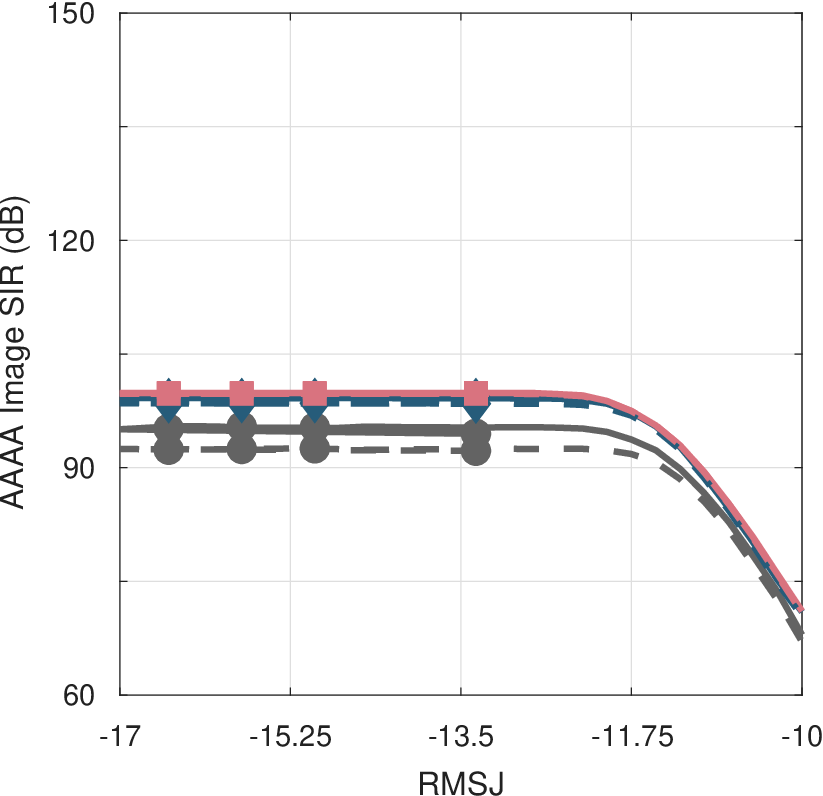}\label{fig:BB_rad_minSIR_OS4}			
		}\hspace{0.05cm}
		\subfloat[ ]{
			
			\psfrag{-17}[c][c]{\scriptsize -$17$}
			\psfrag{-15.25}[c][c]{\scriptsize -$15.25$}
			\psfrag{-13.5}[c][c]{\scriptsize -$13.5$}
			\psfrag{-11.75}[c][c]{\scriptsize -$11.75$}	
			\psfrag{-10}[c][c]{\scriptsize -$10$}
			
			\psfrag{60}[c][c]{}
			\psfrag{90}[c][c]{}
			\psfrag{120}[c][c]{}
			\psfrag{150}[c][c]{}
			
			\psfrag{RMSJ}[c][c]{\footnotesize $\log_{10}\left(\delta_\mathrm{SJ,RMS}\right)$}
			\psfrag{AAAA Image SIR (dB)}[c][c]{}
			
			\includegraphics[height=3.6cm]{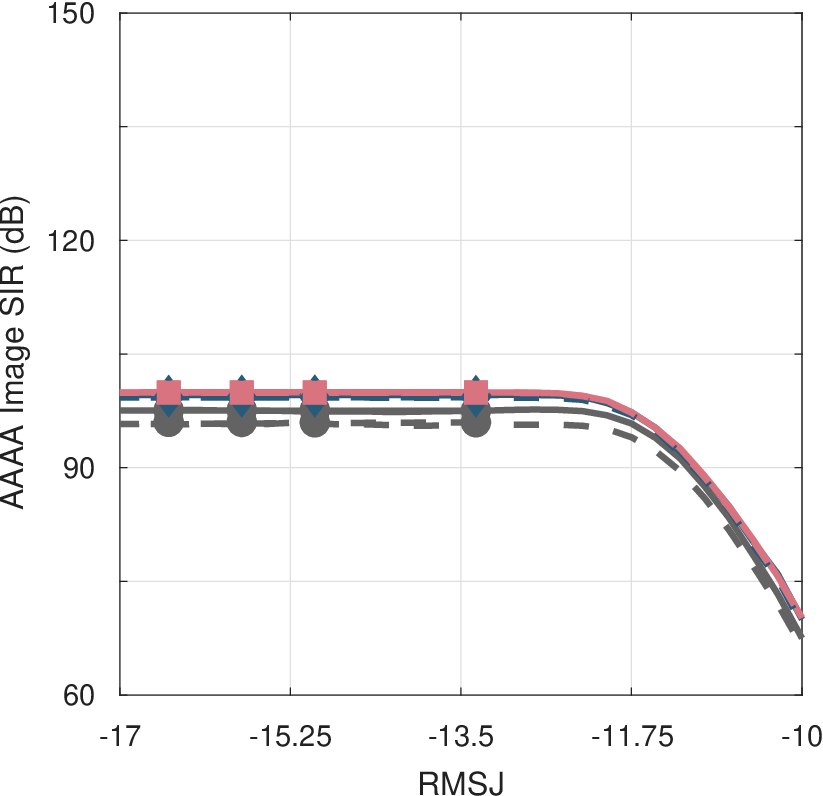}\label{fig:BB_rad_minSIR_OS8}			
		}
		
		\captionsetup{justification=raggedright,labelsep=period,singlelinecheck=false}
		\caption{\ Image SIR as a function of the normalized RMS \ac{SJ} by the critical sampling period $T_\mathrm{s}$ (same at both the DAC and the ADC) jitter obtained with \ac{BB} sampling, $M=128$ OFDM symbols. The simulated mean image SIR is shown for (a) \mbox{$\eta=1$}, (b) \mbox{$\eta=2$}, (c) \mbox{$\eta=4$}, and (d) \mbox{$\eta=8$}, while the simulated minimum image SIR is shown for the same $\eta$ values in (e), (f), (g), and (h), respectively. In all cases, results are shown for \mbox{$N=256$} and QPSK ({\color[rgb]{0.3922,0.3922,0.392}\textbf{\textemdash}} and {\color[rgb]{0.3922,0.3922,0.392}$\CIRCLE$}), \mbox{$N=2048$} and QPSK ({\color[rgb]{0.1490,0.3569,0.4824}\textbf{\textemdash}} and {\color[rgb]{0.1490,0.3569,0.4824}$\blacklozenge$}), and \mbox{$N=16384$} and QPSK ({\color[rgb]{0.8471,0.4510,0.4980}\textbf{\textemdash}} and {\color[rgb]{0.8471,0.4510,0.4980}$\blacksquare$}), as well as \mbox{$N=256$} and 256-QAM ({\color[rgb]{0.3922,0.3922,0.392}\textbf{\textendash~\textendash}} and {\color[rgb]{0.3922,0.3922,0.392}$\CIRCLE$}), \mbox{$N=2048$} and 256-QAM ({\color[rgb]{0.1490,0.3569,0.4824}\textbf{\textendash~\textendash}} and {\color[rgb]{0.1490,0.3569,0.4824}$\blacklozenge$}), and \mbox{$N=16384$} and 256-QAM ({\color[rgb]{0.8471,0.4510,0.4980}\textbf{\textendash~\textendash}} and {\color[rgb]{0.8471,0.4510,0.4980}$\blacksquare$}).}\label{fig:BB_rad_SIR}
		
	\end{figure*}
	
	\begin{figure*}[!t]
		\captionsetup[subfigure]{labelformat=empty}
		\centering
		
		\subfloat[ ]{
			
			\psfrag{55}{\footnotesize (a)}
			\psfrag{22}{\footnotesize (b)}
			\psfrag{33}{\footnotesize (c)}
			
			\psfrag{7.5}[c][c]{\scriptsize $7.5$}
			\psfrag{10}[c][c]{\scriptsize $10$}
			\psfrag{12.5}[c][c]{\scriptsize $12.5$}
			\psfrag{15}[c][c]{\scriptsize $15$}	
			\psfrag{17.5}[c][c]{\scriptsize $17.5$}
			
			\psfrag{-0.1}[c][c]{\scriptsize -$0.1$}
			\psfrag{0}[c][c]{\scriptsize $0$}
			\psfrag{0.1}[c][c]{\scriptsize $0.1$}
			\psfrag{0.2}[c][c]{\scriptsize $0.2$}
			
			\psfrag{0}[c][c]{\scriptsize $0$}
			\psfrag{-15}[c][c]{\scriptsize -$15$}
			\psfrag{-30}[c][c]{\scriptsize -$30$}
			\psfrag{-45}[c][c]{\scriptsize -$45$}
			\psfrag{-60}[c][c]{\scriptsize -$60$}
			
			\psfrag{fD Df}[c][c]{\footnotesize $f_\mathrm{D}/\Delta f$}
			\psfrag{Bistatic range (m)}[c][c]{\footnotesize Bistatic range (m)}
			\psfrag{Norm. mag. (dB)}[c][c]{\footnotesize Norm. mag. (dB)}
			
			\includegraphics[width=12cm]{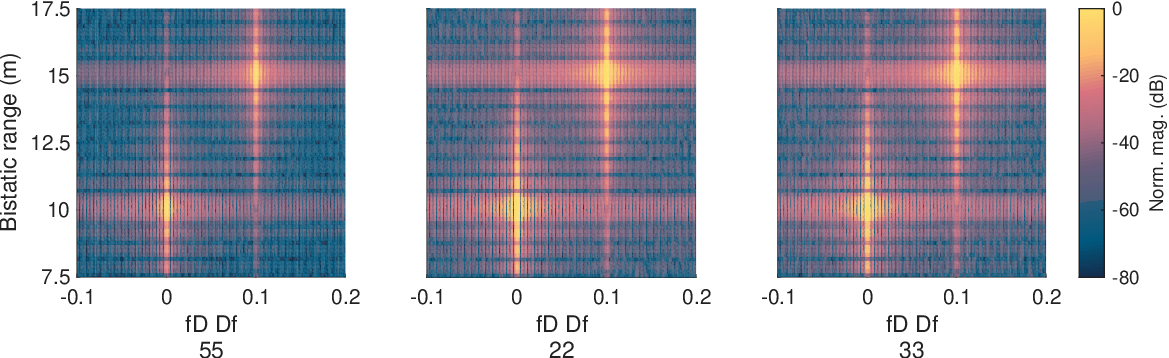}\label{fig:I_rD_BB_rectWin_OS1}		
						
		}\\
		\subfloat[ ]{
			
			\psfrag{55}{\footnotesize (d)}
			\psfrag{22}{\footnotesize (e)}
			\psfrag{33}{\footnotesize (f)}
			
			\psfrag{7.5}[c][c]{\scriptsize $7.5$}
			\psfrag{10}[c][c]{\scriptsize $10$}
			\psfrag{12.5}[c][c]{\scriptsize $12.5$}
			\psfrag{15}[c][c]{\scriptsize $15$}	
			\psfrag{17.5}[c][c]{\scriptsize $17.5$}
			
			\psfrag{-0.1}[c][c]{\scriptsize -$0.1$}
			\psfrag{0}[c][c]{\scriptsize $0$}
			\psfrag{0.1}[c][c]{\scriptsize $0.1$}
			\psfrag{0.2}[c][c]{\scriptsize $0.2$}
			
			\psfrag{0}[c][c]{\scriptsize $0$}
			\psfrag{-15}[c][c]{\scriptsize -$15$}
			\psfrag{-30}[c][c]{\scriptsize -$30$}
			\psfrag{-45}[c][c]{\scriptsize -$45$}
			\psfrag{-60}[c][c]{\scriptsize -$60$}
			
			\psfrag{fD Df}[c][c]{\footnotesize $f_\mathrm{D}/\Delta f$}
			\psfrag{Bistatic range (m)}[c][c]{\footnotesize Bistatic range (m)}
			\psfrag{Norm. mag. (dB)}[c][c]{\footnotesize Norm. mag. (dB)}
			
			\includegraphics[width=12cm]{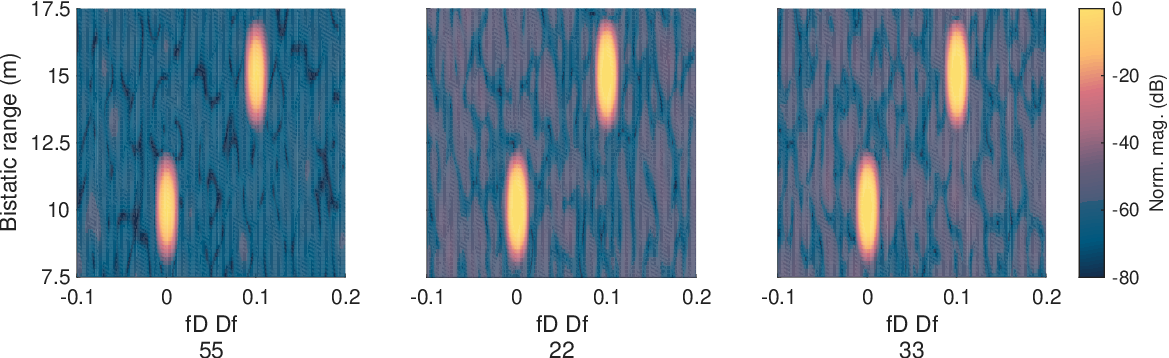}\label{fig:I_rD_BB_chebwin_OS1}		
			
		}
		\captionsetup{justification=raggedright,labelsep=period,singlelinecheck=false}
		\caption{\ Bistatic range-Doppler radar images obtained with BB sampling, $N=2048$, $N_\mathrm{CP}=2048$, $M=128$, QPSK modulation, and $\eta=1$. A target at $\SI{10}{\meter}$ with a Doppler shift of $\SI{0}{\kilo\hertz}$ and another at $\SI{15}{\meter}$ with a Doppler shift of $0.1\Delta f$ were simulated. Results are shown for rectangular windowing for both range and Doppler shift processing and RMS SJ values at both DAC and ADC of \textcolor{black}{(a) $\SI[parse-numbers = false]{10^{-16}}{\second}$, (b) $\SI[parse-numbers = false]{10^{-13}}{\second}$, and (c) $\SI[parse-numbers = false]{10^{-10}}{\second}$ (i.e., $0.5\times10^{-7}T_\mathrm{s}$, $0.5\times10^{-4}T_\mathrm{s}$, and $0.5\times10^{-1}T_\mathrm{s}$, respectively)}, as well as for and Chebyshev windowing with $\SI{100}{dB}$ sidelobe supression for both range and Doppler shift processing and RMS SJ values at both DAC and ADC of \textcolor{black}{(d) $\SI[parse-numbers = false]{10^{-16}}{\second}$, (e) $\SI[parse-numbers = false]{10^{-13}}{\second}$, and (f) $\SI[parse-numbers = false]{10^{-10}}{\second}$.}}\label{fig:I_rD_BB_OS1}
	\end{figure*}
	\begin{figure*}[!t]
		\captionsetup[subfigure]{labelformat=empty}
		\centering
		
		\subfloat[ ]{
			
			\psfrag{55}{\footnotesize (a)}
			\psfrag{22}{\footnotesize (b)}
			\psfrag{33}{\footnotesize (c)}
			
			\psfrag{7.5}[c][c]{\scriptsize $7.5$}
			\psfrag{10}[c][c]{\scriptsize $10$}
			\psfrag{12.5}[c][c]{\scriptsize $12.5$}
			\psfrag{15}[c][c]{\scriptsize $15$}	
			\psfrag{17.5}[c][c]{\scriptsize $17.5$}
			
			\psfrag{-0.1}[c][c]{\scriptsize -$0.1$}
			\psfrag{0}[c][c]{\scriptsize $0$}
			\psfrag{0.1}[c][c]{\scriptsize $0.1$}
			\psfrag{0.2}[c][c]{\scriptsize $0.2$}
			
			\psfrag{0}[c][c]{\scriptsize $0$}
			\psfrag{-15}[c][c]{\scriptsize -$15$}
			\psfrag{-30}[c][c]{\scriptsize -$30$}
			\psfrag{-45}[c][c]{\scriptsize -$45$}
			\psfrag{-60}[c][c]{\scriptsize -$60$}
			
			\psfrag{fD Df}[c][c]{\footnotesize $f_\mathrm{D}/\Delta f$}
			\psfrag{Bistatic range (m)}[c][c]{\footnotesize Bistatic range (m)}
			\psfrag{Norm. mag. (dB)}[c][c]{\footnotesize Norm. mag. (dB)}
			
			\includegraphics[width=12cm]{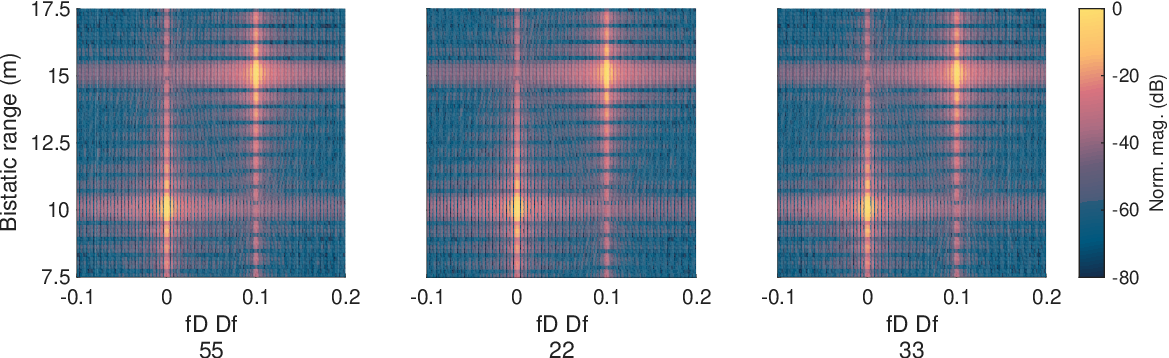}\label{fig:I_rD_BB_rectWin_OS8}		
			
		}\\
		\subfloat[ ]{
			
			\psfrag{55}{\footnotesize (d)}
			\psfrag{22}{\footnotesize (e)}
			\psfrag{33}{\footnotesize (f)}
			
			\psfrag{7.5}[c][c]{\scriptsize $7.5$}
			\psfrag{10}[c][c]{\scriptsize $10$}
			\psfrag{12.5}[c][c]{\scriptsize $12.5$}
			\psfrag{15}[c][c]{\scriptsize $15$}	
			\psfrag{17.5}[c][c]{\scriptsize $17.5$}
			
			\psfrag{-0.1}[c][c]{\scriptsize -$0.1$}
			\psfrag{0}[c][c]{\scriptsize $0$}
			\psfrag{0.1}[c][c]{\scriptsize $0.1$}
			\psfrag{0.2}[c][c]{\scriptsize $0.2$}
			
			\psfrag{0}[c][c]{\scriptsize $0$}
			\psfrag{-15}[c][c]{\scriptsize -$15$}
			\psfrag{-30}[c][c]{\scriptsize -$30$}
			\psfrag{-45}[c][c]{\scriptsize -$45$}
			\psfrag{-60}[c][c]{\scriptsize -$60$}
			
			\psfrag{fD Df}[c][c]{\footnotesize $f_\mathrm{D}/\Delta f$}
			\psfrag{Bistatic range (m)}[c][c]{\footnotesize Bistatic range (m)}
			\psfrag{Norm. mag. (dB)}[c][c]{\footnotesize Norm. mag. (dB)}
			
			\includegraphics[width=12cm]{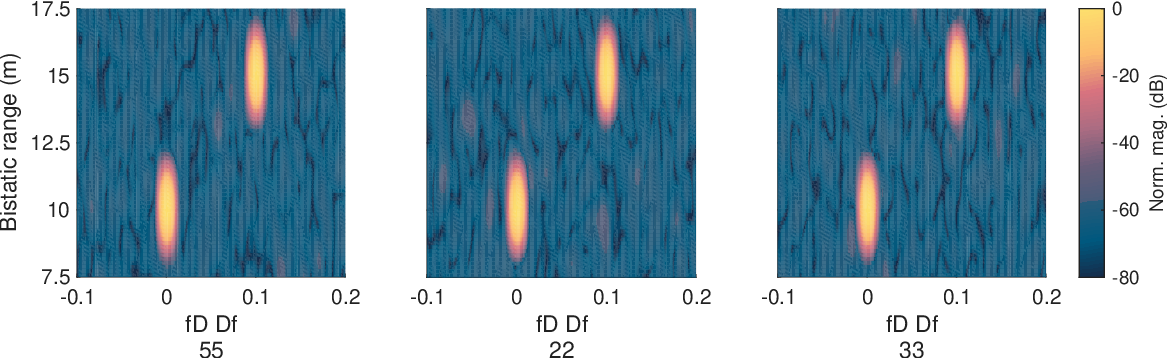}\label{fig:I_rD_BB_chebwin_OS8}		
			
		}
		\captionsetup{justification=raggedright,labelsep=period,singlelinecheck=false}
		\caption{\ Bistatic range-Doppler radar images obtained with BB sampling, $N=2048$, $N_\mathrm{CP}=2048$, $M=128$, QPSK modulation, and $\eta=8$. The same targets as for the radar images shown in Fig.~\ref{fig:I_rD_BB_OS1} were simulated. Results are shown for rectangular windowing for both range and Doppler shift processing and RMS SJ values at both DAC and ADC of \textcolor{black}{(a) $\SI[parse-numbers = false]{10^{-16}}{\second}$, (b) $\SI[parse-numbers = false]{10^{-13}}{\second}$, and (c) $\SI[parse-numbers = false]{10^{-10}}{\second}$ (i.e., $0.5\times10^{-7}T_\mathrm{s}$, $0.5\times10^{-4}T_\mathrm{s}$, and $0.5\times10^{-1}T_\mathrm{s}$, respectively)}, as well as for and Chebyshev windowing with $\SI{100}{dB}$ sidelobe supression for both range and Doppler shift processing and RMS SJ values at both DAC and ADC of \textcolor{black}{(d) $\SI[parse-numbers = false]{10^{-16}}{\second}$, (e) $\SI[parse-numbers = false]{10^{-13}}{\second}$, and (f) $\SI[parse-numbers = false]{10^{-10}}{\second}$.}}\label{fig:I_rD_BB_OS8}		
	\end{figure*}
	
	\textcolor{black}{While \ac{PPLR}, \ac{PSLR}, and \ac{ISLR} do not clearly showcase the influence of \ac{SJ} on the radar sensing performance, the \ac{SIR} distinctly captures this effect as illustrated in Fig.~\ref{fig:BB_rad_SIR}. This figure} shows the achieved mean and minimum image \ac{SIR} as a function of the \ac{RMS} \ac{SJ}. These parameters are defined as the average and minimum power ratio, respectively, that is calculated between the target peak and each point outside the mainlobe. The performed simulations assumed equal \ac{RMS} \ac{SJ} at both \ac{DAC} and \ac{ADC}, as well as the remaining parameters used for the previously discussed \ac{PPLR}, \ac{PSLR}, and \ac{ISLR} results. \textcolor{black}{Overall, a better performance is obtained with \ac{QPSK} compared to \mbox{256-\ac{QAM}}, with a more pronounced difference in the mean image \ac{SIR} case ranging from \SI{3}{dB} to \SI{6}{dB}. In addition, increasing oversampling factors $\eta$ and number of subcarriers $N$ also result in higher \ac{SIR}. While the first parameter is directly proportional to the processing gain experienced against the \ac{SJ}-induced \ac{ICI} after range processing. It can also be seen that the mean image \ac{SIR} values are from around \SI{12}{dB} to \SI{37}{dB} higher than the minimum image \ac{SIR} values, with exact values depending on $\eta$, $N$, and the modulation alphabet. This is due to the \ac{SJ}-induced \ac{ICI}, which may lead to artifacts that cannot be always} suppressed by windowing as later shown in the results in Figs.~\ref{fig:I_rD_BB_OS1} and \ref{fig:I_rD_BB_OS8}. \textcolor{black}{For all considered $\eta$ values, the \ac{SIR} values only start degrading at \ac{RMS} \ac{SJ} equal to or higher than \textcolor{black}{$\SI[parse-numbers = false]{10^{-11}}{\second}$ (i.e., $0.5\times10^{-2}T_\mathrm{s}$)}, which was also the case in the communication analysis in Section~\ref{sec:commPerf}. Unlike in the communication case, a more pronounced image \ac{SIR} improvement is observed when increasing the oversampling factor from \mbox{$\eta=2$} to \mbox{$\eta=4$}, and from \mbox{$\eta=4$} to \mbox{$\eta=8$}. This is particularly visible for mean \ac{SIR} with \mbox{$N\in\{256,2048\}$} (comparing Figs.~\ref{fig:BB_rad_SIR}(b) and \ref{fig:BB_rad_SIR}(c)) in the former case, and with \mbox{$N\in\{256\}$} (comparing Figs.~\ref{fig:BB_rad_SIR}(c) and \ref{fig:BB_rad_SIR}(d)) in the latter. In addition, the fact that maximum observed mean \ac{SIR} is limited at around \mbox{$\SI{136}{dB}$}, as seen for \mbox{$N=16384$} and \ac{QPSK} modulation in Fig.~\ref{fig:BB_rad_SIR}(b), (c), and (d)) is due to the previously discussed limited accuracy in the \ac{SJ} simulation.}

	To illustrate the influence of the degradation of the previously analyzed parameter on obtained radar images, Figs.~\ref{fig:I_rD_BB_OS1} and \ref{fig:I_rD_BB_OS8} show obtained bistatic range-Doppler radar images with $\eta=1$ and $\eta=8$, respectively. Further considered parameters are $N=2048$, $N_\mathrm{CP}=2048$, $M=128$, and QPSK. In addition, two targets were simulated. The first was at \SI{10}{\meter} with a Doppler shift of \SI{0}{\kilo\hertz}, and the second at \SI{15}{\meter} with a Doppler shift of $0.1\Delta f$. In both figures, radar images are shown for the cases where rectangular windowing or Chebyshev windowing with \SI{100}{dB} sidelobe supression are adopted for both range and Doppler shift processing, with RMS SJ values at both DAC and ADC of \textcolor{black}{$\SI[parse-numbers = false]{10^{-16}}{\second}$, $\SI[parse-numbers = false]{10^{-13}}{\second}$, and $\SI[parse-numbers = false]{10^{-10}}{\second}$ (i.e., $0.5\times10^{-7}T_\mathrm{s}$, $0.5\times10^{-4}T_\mathrm{s}$, and $0.5\times10^{-1}T_\mathrm{s}$, respectively)} for both cases. \textcolor{black}{While rectangular windowing results in high range and Doppler shift sidelobes, which partly mask the effect of \ac{SJ} as illustrated by the previously presented discussion on \ac{PPLR}, \ac{PSLR} and \ac{ISLR} results, Chebyshev windowing allows clearly observing the expected \ac{SIR} degradation along with increasing \ac{RMS} \ac{SJ} as previously seen in Fig.~\ref{fig:BB_rad_SIR}.}	

	\subsection{Bandpass Sampling}\label{subsec:sensBP}
	
	For the case where \ac{BP} sampling is performed at a digital \ac{IF} of $f_\mathrm{IF}=\SI{1}{\giga\hertz}$ with oversampling factor $\eta=8$, \textcolor{black}{Table~\ref{tab:PPLR_PSLR_ISLR_BP} shows the simulated \ac{PPLR}, range and Doppler shift \ac{PSLR}, as well as range and Doppler shift \ac{ISLR}. Simulations were performed for both \ac{QPSK} and \mbox{256-\ac{QAM}}, \mbox{$N\in\{256,2048,16384\}$} and $M=128$, within the \ac{RMS} \ac{SJ} range from $\SI[parse-numbers = false]{10^{-17}}{\second}$ to $\SI[parse-numbers = false]{10^{-10}}{\second}$ (i.e., $0.5\times10^{-8}T_\mathrm{s}$ to $0.5\times10^{-1}T_\mathrm{s}$, respectively).  As in the \ac{BB} case in Section~\ref{subsec:sensBP}, negligible differences between all these settings was observed, which is why a single result is shown for each performance parameter. Compared to the \ac{BB} results for $\eta=8$ in  Table~\ref{tab:PPLR_PSLR_ISLR_BB}, only negligible differences of \SI{0.06}{dB} and \SI{0.03}{dB} were observed in range \ac{PSLR} and \ac{ISLR}, respectively.}
	
	\begin{table}[!t]
		\centering
		\captionsetup{width=43pc,justification=centering,labelsep=newline}
		\caption{\textcolor{black}{\textsc{Simulated PPLR, Range and Doppler Shift PSLR, and Range\\ and Doppler Shift ISLR for BP Sampling and RMS SJ between\\ $\SI[parse-numbers = false]{10^{-17}}{\second}$ and $\SI[parse-numbers = false]{10^{-10}}{\second}$ ($0.5\times10^{-8}T_\mathrm{s}$ and $0.5\times10^{-1}T_\mathrm{s}$)}}}
		\label{tab:PPLR_PSLR_ISLR_BP}
			\begin{tabular}{|c|c|c|c|c|c|}
				\hhline{|======|}
				\multirow{2}{*}{\textcolor{black}{$\bm{\eta}$}} & \multirow{2}{*}{\textcolor{black}{\textbf{PPLR}}} & \multirow{2}{*}{\textbf{\begin{tabular}[c]{@{}c@{}}\textcolor{black}{Range}\\ \textcolor{black}{PSLR (dB)}\end{tabular}}} & \multirow{2}{*}{\textbf{\begin{tabular}[c]{@{}c@{}}\textcolor{black}{Doppler}\\ \textcolor{black}{PSLR (dB)}\end{tabular}}} & \multirow{2}{*}{\textbf{\begin{tabular}[c]{@{}c@{}}\textcolor{black}{Range}\\ \textcolor{black}{ISLR (dB)}\end{tabular}}} & \multirow{2}{*}{\textbf{\begin{tabular}[c]{@{}c@{}}\textcolor{black}{Doppler}\\ \textcolor{black}{ISLR (dB)}\end{tabular}}} \\
				& & & & &                   \\\hhline{|======|}
				\textcolor{black}{8} & \textcolor{black}{$\SI{0}{dB}$} &\textcolor{black}{$\SI{-13.33}{dB}$} & \textcolor{black}{$\SI{-13.30}{dB}$} & \textcolor{black}{$\SI{-9.70}{dB}$}  & \textcolor{black}{$\SI{-9.68}{dB}$} \\				
				\hhline{|======|}
			\end{tabular}
	\end{table}

	 \begin{figure}[!t]
	 	\centering
	 	
	 	\subfloat[ ]{
	 		
	 		\psfrag{-17}[c][c]{\scriptsize -$17$}
	 		\psfrag{-15.25}[c][c]{\scriptsize -$15.25$}
	 		\psfrag{-13.5}[c][c]{\scriptsize -$13.5$}
	 		\psfrag{-11.75}[c][c]{\scriptsize -$11.75$}	
	 		\psfrag{-10}[c][c]{\scriptsize -$10$}
	 		
	 		\psfrag{60}[c][c]{\scriptsize $60$}
	 		\psfrag{90}[c][c]{\scriptsize $90$}
	 		\psfrag{120}[c][c]{\scriptsize $120$}
	 		\psfrag{150}[c][c]{\scriptsize $150$}
	 		
	 		\psfrag{RMSJ}[c][c]{\footnotesize $\log_{10}\left(\delta_\mathrm{SJ,RMS}\right)$}
	 		\psfrag{AAAA Image SIR (dB)}[c][c]{\footnotesize Mean image SIR (dB)}
	 		
	 		\includegraphics[height=3.6cm]{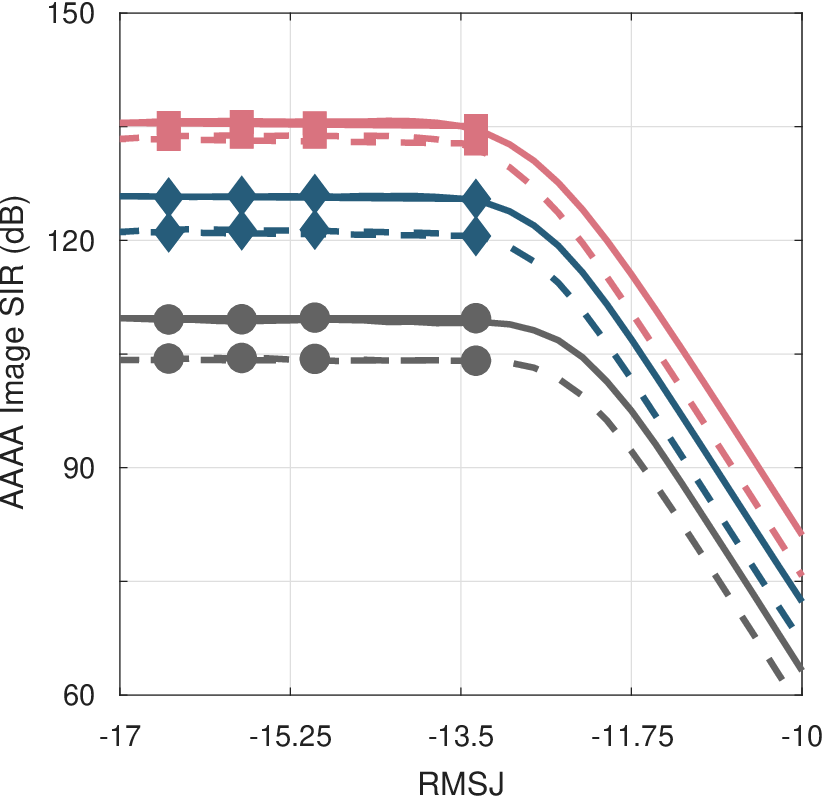}\label{fig:BP_rad_meanSIR}
	 		
	 	}\hspace{0.1cm}
	 	\subfloat[ ]{
	 		
	 		\psfrag{-17}[c][c]{\scriptsize -$17$}
	 		\psfrag{-15.25}[c][c]{\scriptsize -$15.25$}
	 		\psfrag{-13.5}[c][c]{\scriptsize -$13.5$}
	 		\psfrag{-11.75}[c][c]{\scriptsize -$11.75$}	
	 		\psfrag{-10}[c][c]{\scriptsize -$10$}
	 		
	 		\psfrag{60}[c][c]{\scriptsize $60$}
	 		\psfrag{90}[c][c]{\scriptsize $90$}
	 		\psfrag{120}[c][c]{\scriptsize $120$}
	 		\psfrag{150}[c][c]{\scriptsize $150$}
	 		
	 		\psfrag{RMSJ}[c][c]{\footnotesize $\log_{10}\left(\delta_\mathrm{SJ,RMS}\right)$}
	 		\psfrag{AAAA Image SIR (dB)}[c][c]{\footnotesize Min. image SIR (dB)}
	 		
	 		\includegraphics[height=3.6cm]{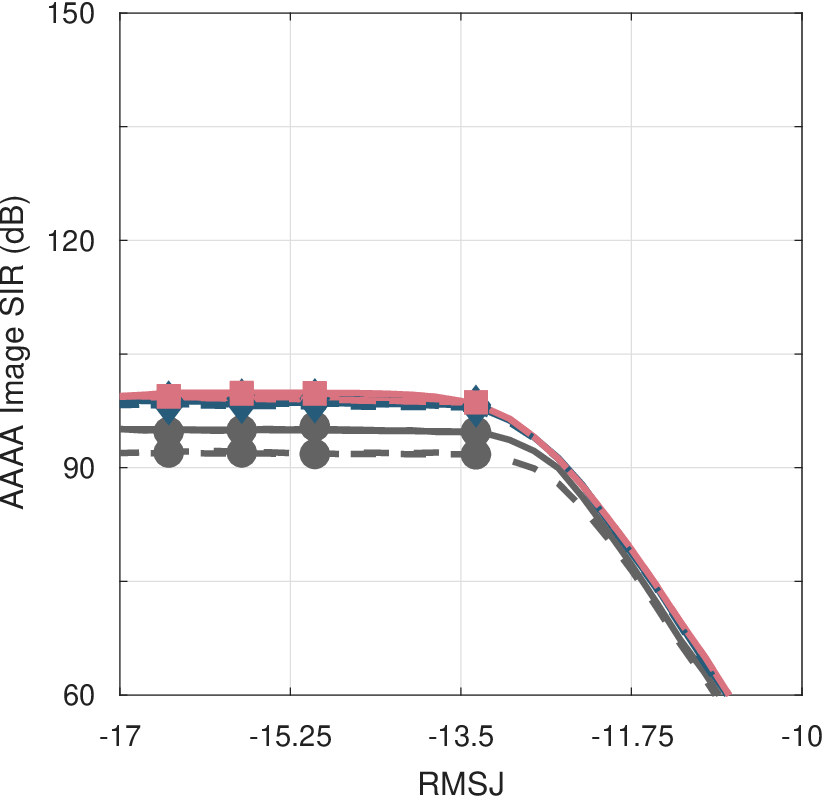}\label{fig:BP_rad_minSIR}		
	 	}
	 	
	 	\captionsetup{justification=raggedright,labelsep=period,singlelinecheck=false}
	 	\caption{\ Mean (a) and minimum (b) image SIR as a function of the normalized RMS \ac{SJ} by the critical sampling period $T_\mathrm{s}$ (same at both the DAC and the ADC) jitter obtained with \ac{BP} sampling at a digital \ac{IF} $f_\mathrm{IF}=\SI{1}{\giga\hertz}$, \mbox{$\eta=8$}, and $M=128$ OFDM symbols. In all cases, results are shown for \mbox{$N=256$} and QPSK ({\color[rgb]{0.3922,0.3922,0.392}\textbf{\textemdash}} and {\color[rgb]{0.3922,0.3922,0.392}$\CIRCLE$}), \mbox{$N=2048$} and QPSK ({\color[rgb]{0.1490,0.3569,0.4824}\textbf{\textemdash}} and {\color[rgb]{0.1490,0.3569,0.4824}$\blacklozenge$}), and \mbox{$N=16384$} and QPSK ({\color[rgb]{0.8471,0.4510,0.4980}\textbf{\textemdash}} and {\color[rgb]{0.8471,0.4510,0.4980}$\blacksquare$}), as well as \mbox{$N=256$} and 256-QAM ({\color[rgb]{0.3922,0.3922,0.392}\textbf{\textendash~\textendash}} and {\color[rgb]{0.3922,0.3922,0.392}$\CIRCLE$}), \mbox{$N=2048$} and 256-QAM ({\color[rgb]{0.1490,0.3569,0.4824}\textbf{\textendash~\textendash}} and {\color[rgb]{0.1490,0.3569,0.4824}$\blacklozenge$}), and \mbox{$N=16384$} and 256-QAM ({\color[rgb]{0.8471,0.4510,0.4980}\textbf{\textendash~\textendash}} and {\color[rgb]{0.8471,0.4510,0.4980}$\blacksquare$}).}\label{fig:BP_rad_SIR}
	 	
	 \end{figure}
	 
	  \textcolor{black}{Next,} the mean and minimum image \ac{SIR} obtained as a function of the \ac{RMS} \ac{SJ} are shown in Figs.~\ref{fig:BP_rad_SIR}(a) and \ref{fig:BP_rad_SIR}(b), respectively. \textcolor{black}{Since negligible improvement is experienced with \ac{CPE} correction} as previously demonstrated by the results in Fig.~\ref{fig:BP_SIR}, \textcolor{black}{only results without \ac{CPE} correction are shown.} As in the \ac{BB} case in Fig.~\ref{fig:BB_rad_SIR}, equal \ac{RMS} \ac{SJ} was assumed for both \ac{DAC} and \ac{ADC}, and the same set of parameters except for a fixed oversampling factor \mbox{$\eta=8$} is adopted. \textcolor{black}{Overall, a similar behavior as in the \ac{BB} case with \mbox{$\eta=8$} is observed, i.e., higher $N$ yields higher image \ac{SIR} and \ac{QPSK} outperforms \mbox{256-\ac{QAM}}. Lower image \ac{SIR} values are obtained in the \ac{BP} sampling case due to the fact that \ac{CPE} is experienced besides \ac{ICI}. Comparing Fig.~\ref{fig:BP_rad_SIR}(a) with the \ac{BB} sampling results from \ref{fig:BB_rad_SIR}(d), lower mean image \ac{SIR} values by \SI{7.11}{dB}, \SI{4.93}{dB}, and \SI{1.79}{dB} are experienced considering \ac{QPSK} modulation and \mbox{$N=256$}, \mbox{$N=2048$}, and \mbox{$N=16384$}, respectively. This aligns with the fact that lower $N$ results in more pronounced \ac{CPE}. For \mbox{256-\ac{QAM}}, these values are increased to \SI{7.24}{dB}, \SI{5.90}{dB}, and \SI{2.36}{dB} as higher modulation order results in more severe effect for the same \ac{CPE}. In terms of minimum image \ac{SIR}, however, a difference between the results for \ac{BP} sampling shown in Fig.~\ref{fig:BP_rad_SIR}(b) are only worse than its \ac{BB} counterparts from Fig.~\ref{fig:BB_rad_SIR}(d) for \mbox{$N=256$}. More specifically, \SI{2.32}{dB}, \SI{3.59}{dB} lower values are obtained for \ac{QPSK} and \mbox{256-\ac{QAM}}, respectively.}
	 
	 \begin{figure*}[!t]
	 	\captionsetup[subfigure]{labelformat=empty}
	 	\centering
	 	
	 	\subfloat[ ]{
	 		
	 		\psfrag{55}{\footnotesize (a)}
	 		\psfrag{22}{\footnotesize (b)}
	 		\psfrag{33}{\footnotesize (c)}
	 		
	 		\psfrag{7.5}[c][c]{\scriptsize $7.5$}
	 		\psfrag{10}[c][c]{\scriptsize $10$}
	 		\psfrag{12.5}[c][c]{\scriptsize $12.5$}
	 		\psfrag{15}[c][c]{\scriptsize $15$}	
	 		\psfrag{17.5}[c][c]{\scriptsize $17.5$}
	 		
	 		\psfrag{-0.1}[c][c]{\scriptsize -$0.1$}
	 		\psfrag{0}[c][c]{\scriptsize $0$}
	 		\psfrag{0.1}[c][c]{\scriptsize $0.1$}
	 		\psfrag{0.2}[c][c]{\scriptsize $0.2$}
	 		
	 		\psfrag{0}[c][c]{\scriptsize $0$}
	 		\psfrag{-15}[c][c]{\scriptsize -$15$}
	 		\psfrag{-30}[c][c]{\scriptsize -$30$}
	 		\psfrag{-45}[c][c]{\scriptsize -$45$}
	 		\psfrag{-60}[c][c]{\scriptsize -$60$}
	 		
	 		\psfrag{fD Df}[c][c]{\footnotesize $f_\mathrm{D}/\Delta f$}
	 		\psfrag{Bistatic range (m)}[c][c]{\footnotesize Bistatic range (m)}
	 		\psfrag{Norm. mag. (dB)}[c][c]{\footnotesize Norm. mag. (dB)}
	 		
	 		\includegraphics[width=12cm]{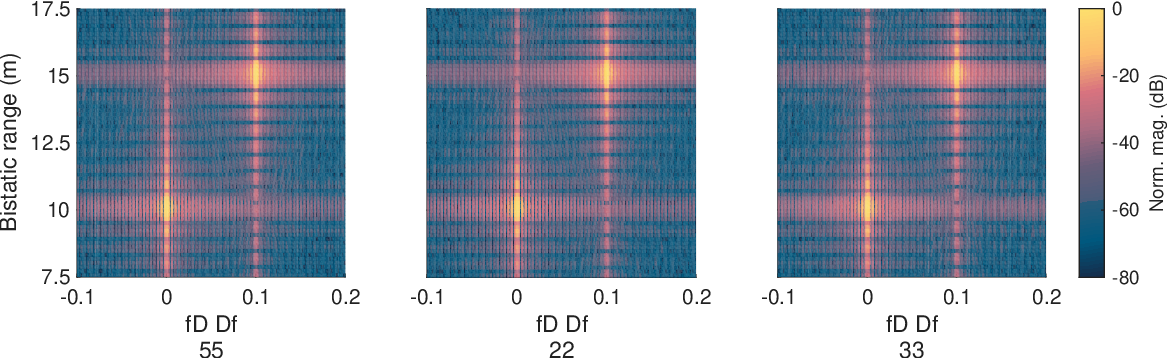}\label{fig:I_rD_BP_rectWin_OS8}		
	 		
	 	}\\
	 	\subfloat[ ]{
	 		
	 		\psfrag{55}{\footnotesize (d)}
	 		\psfrag{22}{\footnotesize (e)}
	 		\psfrag{33}{\footnotesize (f)}
	 		
	 		\psfrag{7.5}[c][c]{\scriptsize $7.5$}
	 		\psfrag{10}[c][c]{\scriptsize $10$}
	 		\psfrag{12.5}[c][c]{\scriptsize $12.5$}
	 		\psfrag{15}[c][c]{\scriptsize $15$}	
	 		\psfrag{17.5}[c][c]{\scriptsize $17.5$}
	 		
	 		\psfrag{-0.1}[c][c]{\scriptsize -$0.1$}
	 		\psfrag{0}[c][c]{\scriptsize $0$}
	 		\psfrag{0.1}[c][c]{\scriptsize $0.1$}
	 		\psfrag{0.2}[c][c]{\scriptsize $0.2$}
	 		
	 		\psfrag{0}[c][c]{\scriptsize $0$}
	 		\psfrag{-15}[c][c]{\scriptsize -$15$}
	 		\psfrag{-30}[c][c]{\scriptsize -$30$}
	 		\psfrag{-45}[c][c]{\scriptsize -$45$}
	 		\psfrag{-60}[c][c]{\scriptsize -$60$}
	 		
	 		\psfrag{fD Df}[c][c]{\footnotesize $f_\mathrm{D}/\Delta f$}
	 		\psfrag{Bistatic range (m)}[c][c]{\footnotesize Bistatic range (m)}
	 		\psfrag{Norm. mag. (dB)}[c][c]{\footnotesize Norm. mag. (dB)}
	 		
	 		\includegraphics[width=12cm]{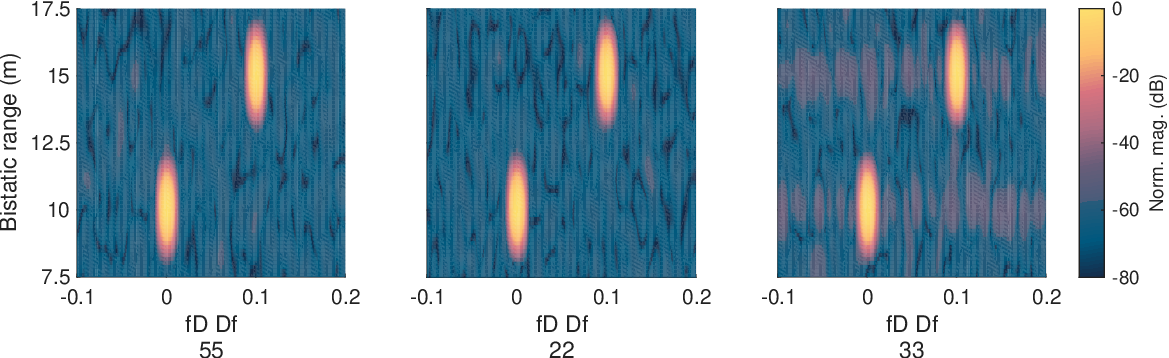}\label{fig:I_rD_BP_chebwin_OS8}		
	 		
	 	}
	 	\captionsetup{justification=raggedright,labelsep=period,singlelinecheck=false}
	 	\caption{\ Bistatic range-Doppler radar images obtained with BP sampling at a digital IF $f_\mathrm{IF}=\SI{1}{\giga\hertz}$, $N=2048$, $N_\mathrm{CP}=2048$, $M=128$, QPSK modulation, $\eta=8$ and Chebyshev windowing with $\SI{100}{dB}$ sidelobe supression for both range and Doppler shift processing. The same targets as for the radar images shown in Figs.~\ref{fig:I_rD_BB_OS1} and \ref{fig:I_rD_BB_OS8} were simulated. Results are shown for rectangular windowing for both range and Doppler shift processing and RMS SJ values at both DAC and ADC of  \textcolor{black}{(a) $\SI[parse-numbers = false]{10^{-16}}{\second}$, (b) $\SI[parse-numbers = false]{10^{-13}}{\second}$, and (c) $\SI[parse-numbers = false]{10^{-10}}{\second}$ (i.e., $0.5\times10^{-7}T_\mathrm{s}$, $0.5\times10^{-4}T_\mathrm{s}$, and $0.5\times10^{-1}T_\mathrm{s}$, respectively)}, as well as for and Chebyshev windowing with $\SI{100}{dB}$ sidelobe supression for both range and Doppler shift processing and RMS SJ values at both DAC and ADC of \textcolor{black}{(d) $\SI[parse-numbers = false]{10^{-16}}{\second}$, (e) $\SI[parse-numbers = false]{10^{-13}}{\second}$, and (f) $\SI[parse-numbers = false]{10^{-10}}{\second}$.}}\label{fig:I_rD_BP_OS8}		
	 \end{figure*}
	 
	 To show how the influence on radar images of the main lobe and sidelobe degradation measured by \ac{PPLR}, \ac{PSLR}, and \ac{ISLR}, as well as the dynamic range reduction measured by the \ac{SIR}, Fig.~\ref{fig:I_rD_BP_OS8} shows simulated bistatic range-Doppler radar images with \ac{BP} sampling at a digital \ac{IF} of $f_\mathrm{IF}=\SI{1}{\giga\hertz}$ and with $\eta=8$. In addition, $N=2048$, $N_\mathrm{CP}=2048$, $M=128$, and QPSK modulation were considered. As for windowing, results are shown for both rectangular window and Chebyshev window with \SI{100}{dB} sidelobe supression applied during both range and Doppler shift processing. The same two targets as for the \ac{BB} case in Figs.~\ref{fig:I_rD_BB_OS1} and \ref{fig:I_rD_BB_OS8} are considered, namely one at \SI{10}{\meter} with a Doppler shift of \SI{0}{\kilo\hertz}, and the another at \SI{15}{\meter} with a Doppler shift of $0.1\Delta f$. Furthermore, \ac{RMS} \ac{SJ} values of \textcolor{black}{$\SI[parse-numbers = false]{10^{-16}}{\second}$, $\SI[parse-numbers = false]{10^{-13}}{\second}$, and $\SI[parse-numbers = false]{10^{-10}}{\second}$ (i.e., $0.5\times10^{-7}T_\mathrm{s}$, $0.5\times10^{-4}T_\mathrm{s}$, and $0.5\times10^{-1}T_\mathrm{s}$, respectively)} were considered, with the same value being adopted for both DAC and ADC at each simulation. The resulting range-Doppler shift radar images from simulations with the aforementioned parameters confirm the findings from the previous results on radar sensing with \ac{BP} sampling in the considered \ac{OFDM}-based \ac{ISAC} system. More specifically, they show that both \ac{ICI} and \ac{CPE} strongly influence the quality of the obtained radar images, which become severely degraded at significantly lower \ac{RMS} \ac{SJ} than in the \ac{BB} case. While the \ac{ICI} can be suppressed to a certain extent due to the radar processing gain, the \ac{CPE} effect is not effectively attenuated, leading to stripes in the Doppler shift direction at the range of the true targets. \textcolor{black}{This is especially visible in Fig.~\ref{fig:I_rD_BP_OS8}(f) where $\SI[parse-numbers = false]{10^{-10}}{\second}$ \ac{RMS} \ac{SJ} and Chebyshev windowing were considered.} Since, in contrast to the range sidelobes, these artifacts cannot be effectively supressed by windowing, they can eventually lead to the detection of ghost targets.
	
	\subsection{Remarks on simulation results}\label{subsec:sensRemarks}
	
	The obtained results have shown that neither the main lobe, measured by \ac{PPLR}, nor the range and Doppler shift sidelobes, measured by \ac{PSLR} and \ac{ISLR}, of a target in a radar image experience significant degradation for \textcolor{black}{the considered \ac{RMS} \ac{SJ} range between $\SI[parse-numbers = false]{10^{-17}}{\second}$ and $\SI[parse-numbers = false]{10^{-10}}{\second}$ (i.e., $0.5\times10^{-8}T_\mathrm{s}$ and $0.5\times10^{-1}T_\mathrm{s}$, respectively) at both \ac{DAC} and \ac{ADC} for any of the considered numbers of subcarriers, i.e., \mbox{$\eta\in\{256,2048,16384\}$}, and oversampling factors, i.e., \mbox{$\eta\in\{1,2,4,8\}$}, both for \ac{BB} and \ac{BP} sampling.}
	
	
	\textcolor{black}{Although no relevant main or sidelobe degradation is measured by the \ac{PPLR}, \ac{PSLR} and \ac{ISLR} parameters,} it was shown that the described \ac{ICI} and \ac{CPE} effects lead to \textcolor{black}{non-negligible} image \ac{SIR} degradation \textcolor{black}{for \ac{RMS} \ac{SJ} above $\SI[parse-numbers = false]{10^{-11}}{\second}$ (i.e., $0.5\times10^{-2}T_\mathrm{s}$).} Notably, it was \textcolor{black}{verified that the \ac{CPE} in the \ac{BP} sampling case may lead to stripes in the Doppler shift direction at the range of the true targets in the radar image. This can lead to ghost targets and, in scenarios with multiple targets,} mask weaker target reflections.
	


	\section{Conclusion}\label{sec:conclusion}
	
	This article analyzed the effects of colored \ac{SJ} resulting from the use of \ac{PLL}-based oscillators to derive the sampling clocks for \acp{DAC} and \acp{ADC} on the communication and sensing performance of \ac{OFDM}-based \ac{ISAC} for both \ac{BB} and \ac{BP} sampling strategies. After a mathematical formulation of the influence of \ac{SJ} at the \ac{DAC} and \ac{ADC} at the transmitter and receiver sides, respectively, on an oversampled \ac{OFDM} signal in Section~\ref{sec:sysModel}, communication and radar sensing performance analyses were presented.
	
	The claim from \cite{gonem2021} that \ac{DAC} and \ac{ADC} \ac{SJ} have similar effects was also confirmed with the derived expression in \eqref{eq:Yl_3}, and the obtained results which showed that the combined \ac{RMS} \ac{SJ}, and not the individual \ac{DAC} and \ac{ADC} contributions, drive the performance of \ac{OFDM}-based \ac{ISAC} systems. Furthermore, simulation results indicated that the degradation due to the considered colored \ac{SJ} becomes noticeable when \ac{RMS} \ac{SJ} reaches approximately \textcolor{black}{$\SI[parse-numbers = false]{10^{-11}}{\second}$, which corresponds to $0.5\times10^{-2}$ times the critical sampling period $T_\text{s}$, at both \ac{DAC} and \ac{ADC} for both \ac{BB} and \ac{BP} sampling. At this} or higher \ac{SJ} levels, non-negligible communication performance degradation is observed due to \ac{ICI} in the \ac{BB} sampling case and both \ac{ICI} and \ac{CPE} in the \ac{BP} sampling case. Comparing this value \textcolor{black}{with the \SI{49.44}{\femto\second} \ac{RMS} \ac{SJ} calculated based on the model for the Texas Instruments LMX2594 \ac{RF} synthesizer \cite{ti2019} show in Fig.~\ref{fig:samplingJitter_PNpsd_RMS}, it can be concluded that sufficient communication and radar sensing performance robustness against \ac{SJ} can be achieved with state-of-the-art hardware.}
	
	\textcolor{black}{While not covered by the analysis in this article, \ac{SJ} may also bias the estimation and cause additional time and frequency synchronization offsets. This is, however, beyond the scope of this article and can be compensated, e.g., through synchronization offset estimation fine-tuning, as discussed and demonstrated in \cite{brunner2024} within a bistatic \ac{OFDM}-based \ac{ISAC} context.}

	\section*{Acknowledgment}
	
	L. Giroto would like to thank Benedikt Geiger from the Karlsruhe Institute of Technology (KIT), Germany, for the valuable discussions, as well as Christian Maximilian Karle, also from KIT, for his support with the server PC used to conduct the simulations that produced the results presented in this article.
	%
	
	\bibliographystyle{IEEEtran}
	\bibliography{./References/OverviewPapers,./References/RadCom_Enablement,./References/B5G_6G,./References/Interference,./References/Automotive,./References/RadarNetworks,./References/ChirpSequence,./References/PMCW,./References/OFDM,./References/OCDM,./References/OFDM_Variations,./References/CS_OCDM_Variations,./References/CP_DSSS,./References/BandwidthEnlargement_DigitalRadars,./References/CompressedSensing_DigitalRadars,./References/RadarTargetSimulator,./References/Parameters,./References/HardwareImplementation,./References/FirstRadCom,./References/Interference_CS,./References/ResourceAllocation,./References/SFO,./References/Bistatic,./References/RIS,./References/Nokia,./References/5G_NR,./References/CoMP,./References/PhaseNoise,./References/Jitter}

@ARTICLE{lima2021,
	author={C. {de Lima et al.}},
	journal={IEEE Access}, 
	title={Convergent Communication, Sensing and Localization in {6G} Systems: An Overview of Technologies, Opportunities and Challenges}, 
	year={2021},
	month={Jan.},
	volume={9},
	number={},
	pages={26902-26925},
	doi={10.1109/ACCESS.2021.3053486}
}

@ARTICLE{viswanathan2020,
	author={Viswanathan, Harish and Mogensen, Preben E.},
	journal={IEEE Access}, 
	title={Communications in the {6G} Era}, 
	year={2020},
	month={Mar.},
	volume={8},
	number={},
	pages={57063-57074}
}

@ARTICLE{chafii2023,	
	author={Chafii, Marwa and Bariah, Lina and Muhaidat, Sami and Debbah, Merouane},	
	journal={IEEE Commun. Surv. Tut.},	
	title={Twelve Scientific Challenges for {6G}: Rethinking the Foundations of Communications Theory},	
	year={2023},
	month={Second Quarter},
	volume={25},
	number={2},
	pages={868-904},
	doi={10.1109/COMST.2023.3243918}
}

@ARTICLE{nagulu2024,	
	author={Aravind {Nagulu et al.}},
	journal={Proc. IEEE}, 
	title={Doubling Down on Wireless Capacity: A Review of Integrated Circuits, Systems, and Networks for Full Duplex}, 	
	year={2024},
	month={May},
	volume={112},
	number={5},
	pages={405-432}
}

@ARTICLE{smida2024,	
	author={Smida, Besma and Wichman, Risto and Kolodziej, Kenneth E. and Suraweera, Himal A. and Riihonen, Taneli and Sabharwal, Ashutosh},
	journal={Proc. IEEE},
	title={In-Band Full-Duplex: The Physical Layer}, 
	year={2024},
	month={May},
	volume={112},
	number={5},
	pages={433-462}
}

@article{rajatheva2020,
	title={White Paper on Broadband Connectivity in {6G}}, 
	author={Nandana {Rajatheva et al.}},
	year={2020},
	month={Apr.},
	journal={arXiv preprint arXiv:2004.14247 [eess.SP]},
	url = {https://arxiv.org/abs/2004.14247}
}

@article{parssinen2021,
	title={White paper on {RF} enabling {6G}: opportunities and challenges from technology to spectrum},
	author={Aarno {P{\"a}rssinen et al.}},
	year={2021},
	journal={University of Oulu},
	url={https://urn.fi/URN:ISBN:9789526228419}
}

@ARTICLE{pegoraro2024,	
	author={Jacopo {Pegoraro et al.}},	
	journal={IEEE Trans. Wireless Commun. (Early Access)},	
	title={{JUMP}: Joint communication and sensing with Unsynchronized transceivers Made Practical},	
	year={2024},
	month={Feb.},
	volume={},
	number={},
	pages={1-16},
	doi={10.1109/TWC.2024.3365853}
}

@article{brunner2024,
	title={Bistatic {OFDM}-based {ISAC} with Over-the-Air Synchronization: System Concept and Performance Analysis}, 
	author={D. {Brunner et al.}},
	year={2024},
	month={May},
	journal={arXiv preprint arXiv:2405.04962 [eess.SP]},
	url = {https://arxiv.org/abs/2405.04962}
}

@article{giroto2024,
	title={Pilot-Based {SFO} Estimation for {OFDM}-based Bistatic Integrated Sensing and Communication}, 
	author={L. {Giroto de Oliveira et al.}},
	year={2024},
	month={Jul.},
	journal={arXiv preprint arXiv:2407.07567 [eess.SP]},
	url = {https://arxiv.org/abs/2407.07567}
}

@ARTICLE{aguilar2024,	
	author={Aguilar, Julian and Werbunat, David and Janoudi, Vinzenz and Bonfert, Christina and Waldschmidt, Christian},	
	journal={IEEE J. Microw.}, 	
	title={Uncoupled Digital Radars Creating a Coherent Sensor Network}, 	
	year={2024},
	month={Jun.},	
	volume={5},	
	number={3},	
	pages={459-472},
	doi={10.1109/JMW.2024.3405633}
}

@ARTICLE{thomae2019,	
	author={{Thom\"a et al.}, Reiner S.},	
	journal={IEEE Commun. Mag.}, 	
	title={Cooperative Passive Coherent Location: A Promising {5G} Service to Support Road Safety}, 
	year={2019},
	month={Sept.},
	volume={57},	
	number={9},	
	pages={86-92},
	doi={10.1109/MCOM.001.1800242}
}

@ARTICLE{alian2015,
	author={Haj Mirza Alian, Ehsan and Mitran, Patrick},
	journal={IEEE Trans. Commun.}, 
	title={Jitter-Robust Spectral Shaping in {OFDM}}, 
	year={2015},
	month={},	
	volume={63},	
	number={4},	
	pages={1282-1290},
	doi={10.1109/TCOMM.2015.2409259}
}

@INPROCEEDINGS{manoj2003,	
	author={Manoj, K.N. and Thiagarajan, G.},	
	booktitle={2003 IEEE Int. Conf. Commun.},	
	title={The effect of sampling jitter in {OFDM} systems}, 	
	year={2003},
	month={May},
	volume={3},
	number={},	
	pages={2061-2065},
	doi={10.1109/ICC.2003.1203983}
}

@article{syrjala2010,
	author = {Syrj\"al\"a, Ville and Valkama, Mikko},
	year = {2010},
	month = {Apr.},
	pages = {193-202},
	title = {Analysis and mitigation of phase noise and sampling jitter in {OFDM} radio receivers},
	volume = {2},
	journal = {Int. J. Microw. Wireless Technol.},
	doi = {10.1017/S1759078710000309}
}

@ARTICLE{gonem2021,	
	author={Gonem, Omaro Fawzi Abdelhamid and Giddings, Roger Philip and Tang, Jianming},	
	journal={IEEE Photon. J.},	
	title={Timing Jitter Analysis and Mitigation in Hybrid {OFDM}-{DFMA} {PONs}},	
	year={2021},
	month={Dec.},	
	volume={13},	
	number={6},	
	pages={1-13},
	doi={10.1109/JPHOT.2021.3121168}
}

@article{loehning2007,
	author = {L\"ohning, Michael and Fettweis, Gerhard},
	year = {2007},
	month = {Jan.},
	pages = {11-18},
	title = {The effects of aperture jitter and clock jitter in wideband {ADCs}},
	volume = {29},
	issue={1},
	journal = {Comput. Standards Interfaces},
	doi = {10.1016/j.csi.2005.12.005}
}

@INPROCEEDINGS{putra2009,	
	author={Putra, Bakti Darma and Fettweis, Gerhard},	
	booktitle={2009 IEEE 10th Workshop Signal Process. Advances Wireless Commun.},	
	title={Clock jitter estimation and suppression in {OFDM} systems employing bandpass {$\Sigma\Delta$} {ADC}},	
	year={2009},
	month={Jun.},
	volume={},	
	number={},	
	pages={623-627},
	doi={10.1109/SPAWC.2009.5161860}
}

@ARTICLE{onunkwo2006,
	author={Onunkwo, U. and Ye Li and Swami, A.},
	journal={IEEE J. Sel. Areas Commun.}, 
	title={Effect of timing jitter on {OFDM}-based {UWB} systems}, 
	year={2006},
	month={Apr.},
	volume={24},
	number={4},
	pages={787-793},
	doi={10.1109/JSAC.2005.863830}
}

@ARTICLE{yang2010,	
	author={Yang, Lei and Armstrong, Jean},	
	journal={IEEE Commun. Lett.},	
	title={Oversampling to reduce the effect of timing jitter on high speed {OFDM} systems}, 
	year={2010},
	month={Mar.},
	volume={14},	
	number={3},	
	pages={196-198},
	doi={10.1109/LCOMM.2010.03.091963}
}

@electronic{ti2016,
	author = {{Texas Instruments}},
	title = {\textit{LMK04208 Low-Noise Clock Jitter Cleaner with Dual Loop PLLs} (2016). {Accessed}: Sept. 4, 2024. {[Online]}. {Available}:},
	howpublished  = {\url{https://www.ti.com/document-viewer/lmk04208/datasheet}},
}

@electronic{ti2019,
	author = {{Texas Instruments}},
	title = {\textit{LMX2594 15-GHz Wideband PLLATINUM\textsuperscript{TM} RF Synthesizer With Phase Synchronization and JESD204B Support} (2019). {Accessed}: Sept. 4, 2024. {[Online]}. {Available}:},
	howpublished  = {\url{https://www.ti.com/document-viewer/lmx2594/datasheet}},
}

@electronic{rfsoc2021,
	author = {{AMD}},
	title = {\textit{Zynq UltraScale+ RFSoC RF Data Converter Evaluation Tool (ZCU111)} (2021). {Accessed}: Sept. 4, 2024. {[Online]}. {Available}:},
	howpublished  = {\url{https://docs.amd.com/v/u/en-US/ug1287-zcu111-rfsoc-eval-tool}},
}

@INPROCEEDINGS{wild2023,
	author={Wild, Thorsten and Grudnitsky, Artjom and Mandelli, Silvio and Henninger, Marcus and Guan, Junqing and Schaich, Frank},
	booktitle={2023 20th Eur. Radar Conf.}, 
	title={{6G} Integrated Sensing and Communication: From Vision to Realization}, 
	year={2023},
	month={Sept.},
	volume={},	
	number={},	
	pages={355-358},
	doi={10.23919/EuRAD58043.2023.10289474}
}

@ARTICLE{hwang2009,
	author={T. {Hwang} and C. {Yang} and G. {Wu} and S. {Li} and G. {Ye Li}},
	journal={IEEE Trans. on Veh. Technol.}, 
	title={{OFDM} and Its Wireless Applications: A Survey}, 
	year={2009},
	month={Aug.},
	volume={58},
	number={4},
	pages={1673-1694},
	doi={10.1109/TVT.2008.2004555}
}

@ARTICLE{speth1999,	
	author={Speth, M. and Fechtel, S.A. and Fock, G. and Meyr, H.},	
	journal={IEEE Trans. Commun.}, 	
	title={Optimum receiver design for wireless broad-band systems using {OFDM} — {Part I}}, 	
	year={1999},
	month={Nov.},
	volume={47},	
	number={11},	
	pages={1668-1677},
	doi={10.1109/26.803501}
}

@INPROCEEDINGS{nuss2018,
	author={B. {Nuss} and J. {Mayer} and T. {Zwick}},
	booktitle={2018 IEEE MTT-S Int. Conf. Microw. Intell. Mobility}, 
	title={Limitations of {MIMO} and Multi-User Access for {OFDM} Radar in Automotive Applications}, 
	year={2018},
	month={Apr.},
	volume={},
	number={},
	pages={1-4},
	doi={10.1109/ICMIM.2018.8443533}
}

@ARTICLE{barneto2019,
	author={C. B. {Barneto et al.}},
	journal={IEEE Trans. Microw. Theory Tech.}, 
	title={Full-Duplex {OFDM} Radar With {LTE} and {5G} {NR} Waveforms: Challenges, Solutions, and Measurements}, 
	year={2019},
	month={Oct.},
	volume={67},
	number={10},
	pages={4042-4054},
	doi={10.1109/TMTT.2019.2930510}
}

@ARTICLE{hakobyan2018,
	author={G. {Hakobyan} and B. {Yang}},
	journal={IEEE Trans. Veh. Technol.}, 
	title={A Novel Intercarrier-Interference Free Signal Processing Scheme for {OFDM} Radar}, 
	year={2018},
	month={Jun.},
	volume={67},
	number={6},
	pages={5158-5167},
	doi={10.1109/TVT.2017.2723868}
}

@INPROCEEDINGS{mandelli2023survey,
	author={Mandelli, Silvio and Henninger, Marcus and Bauhofer, Maximilian and Wild, Thorsten},	
	booktitle={2023 2nd Int. Conf. 6G Netw.},	
	title={Survey on Integrated Sensing and Communication Performance Modeling and Use Cases Feasibility}, 	
	year={2023},
	month={Oct.},
	volume={},	
	number={},	
	pages={1-8},	
	doi={10.1109/6GNet58894.2023.10317691}
}

@INPROCEEDINGS{kadelka2023,
	author={Kadelka, Arndt and Zimmermann, Gerd and Plach\'y, Jan and Holschke, Oliver},
	booktitle={2023 IEEE 3rd Int. Symp. Joint Commun. Sens.}, 	
	title={A {CSP}'s View on Opportunities and Challenges of Integrated Communications and Sensing}, 
	year={2023},
	month={Mar.},
	volume={},
	number={},
	pages={1-6}
}

@ARTICLE{liu2022,
	author={Fan {Liu et al.}},
	journal={IEEE J. Sel. Areas Commun.}, 
	title={Integrated Sensing and Communications: Towards Dual-functional Wireless Networks for {6G} and Beyond}, 
	year={2022},
	month={Jun.},
	volume={40},
	number={6},
	pages={1728-1767}
}

@ARTICLE{giroto2021_tmtt,
	author={L. {Giroto de Oliveira} and B. {Nuss} and M. B. {Alabd} and A. {Diewald} and M. {Pauli} and T. {Zwick}},
	journal={IEEE Trans. Microw. Theory Tech.}, 
	title={Joint Radar-Communication Systems: Modulation Schemes and System Design}, 
	year={2022},
	month={Mar.},
	volume={70},	
	number={3},	
	pages={1521-1551},
	doi={},
}

@ARTICLE{wild2021,
	author={T. {Wild} and V. {Braun} and H. {Viswanathan}},
	journal={IEEE Access}, 
	title={Joint Design of Communication and Sensing for Beyond {5G} and {6G} Systems}, 
	year={2021},
	volume={9},
	month={Feb.},
	number={},
	pages={30845-30857},
	doi={10.1109/ACCESS.2021.3059488}
}

@INPROCEEDINGS{thomae2021,
	author={Reiner Thom\"a and Thomas Dallmann and Snezhana Jovanoska and Peter Knott and Anke Schmeink},
	booktitle={2021 15th European Conf. Antennas Propag.}, 
	title={Joint Communication and Radar Sensing: An Overview},
	year={2021},
	month={Mar.},
	volume={},
	number={},
	pages={1-5},
}

@ARTICLE{shatov2024,	
	author={V. {Shatov et al.}},	
	journal={IEEE Access}, 	
	title={Joint Radar and Communications: Architectures, Use Cases, Aspects of Radio Access, Signal Processing, and Hardware}, 
	year={2024},
	volume={},
	number={},
	pages={1-1},	
	doi={10.1109/ACCESS.2024.3383771}
}

@INPROCEEDINGS{wymeersch2024,	
	author={Henk {Wymeersch et al.}},	
	booktitle={2024 IEEE 4th Int. Symp. Joint Commun. Sens.},	
	title={Joint Communication and Sensing for {6G} - A Cross-Layer Perspective},	
	year={2024},
	month={Mar.},
	volume={},
	number={},	
	pages={01-06},	
	doi={10.1109/JCS61227.2024.10646326}
}

@ARTICLE{lellouch2016,
	author={Lellouch, Gabriel and Mishra, Amit Kumar and Inggs, Michael},
	journal={IEEE Trans. Aerosp. Electron. Syst.}, 
	title={Design of {OFDM} radar pulses using genetic algorithm based techniques}, 
	year={2016},
	month={Aug.},
	volume={52},
	number={4},
	pages={1953-1966},
	doi={10.1109/TAES.2016.140671}
}

@article{liao2024,
	title={Pulse Shaping for Random {ISAC} Signals: The Ambiguity Function Between Symbols Matters}, 
	author={Zihan {Liao et al.}},
	year={2024},
	month={Jul.},
	journal={arXiv preprint arXiv:2407.15530 [eess.SP]},
	url = {https://arxiv.org/abs/2407.15530}
}

@ARTICLE{khanzadi2014,	
	author={Khanzadi, M. Reza and Kuylenstierna, Dan and Panahi, Ashkan and Eriksson, Thomas and Zirath, Herbert},	
	journal={IEEE Trans. Circuits Syst. I, Reg. Papers}, 	
	title={Calculation of the Performance of Communication Systems From Measured Oscillator Phase Noise}, 	
	year={2014},
	month={May},
	volume={61},
	number={5},
	pages={1553-1565},	
	doi={10.1109/TCSI.2013.2285698}
}

@ARTICLE{armada2001,	
	author={Garcia Armada, A.},
	journal={IEEE Trans. Broadcast}, 
	title={Understanding the effects of phase noise in orthogonal frequency division multiplexing ({OFDM})}, 
	year={2001},
	month={Jun.},
	volume={47},
	number={2},
	pages={153-159},
	doi={10.1109/11.948268}
}

@INPROCEEDINGS{schweizer2018,	
	author={Schweizer, Benedikt and Schindler, Daniel and Knill, Christina and Hasch, Jurgen and Waldschmidt, Christian},	
	booktitle={2018 IEEE/MTT-S Int. Microw. Symp.}, 
	title={On Hardware Implementations of Stepped-Carrier {OFDM} Radars}, 
	year={2018},
	month={Jun.},
	volume={},
	number={},
	pages={891-894},
	doi={10.1109/MWSYM.2018.8439179}
}

@INPROCEEDINGS{syrjala2009,
	author={Syrjala, Ville and Valkama, Mikko and Tchamov, Nikolay N. and Rinne, Jukka},
	booktitle={2009 Wireless Telecommun. Symp.},
	title={Phase noise modelling and mitigation techniques in {OFDM} communications systems},
	year={2009},
	month={Apr.},
	volume={},
	number={},
	pages={1-7},
	doi={10.1109/WTS.2009.5068965}
}

@ARTICLE{walt2023,	
	author={van der Walt, Pieter W. and Steyn, Werner},	
	journal={IEEE Aerosp. Electron. Syst. Mag.},	
	title={Characterizing Phase Noise in Self-Referencing Radar}, 	
	year={2023},
	month={Dec.},
	volume={38},
	number={12},
	pages={4-13},
	doi={10.1109/MAES.2023.3323919}
}

@article{giroto2024PN,
	title={On the Sensing Performance of {OFDM}-based {ISAC} under Influence of Oscillator Phase Noise}, 
	author={L. {Giroto de Oliveira et al.}},
	year={2024},
	month={Oct.},
	journal={arXiv preprint arXiv:2410.13336 [eess.SP]},
	url = {https://arxiv.org/abs/2410.13336}
}

@ARTICLE{li2024,
	author={Yueheng {Li et al.}},	
	journal={IEEE Trans. Wireless Commun. (Early Access)}, 
	title={User Detection in {RIS}-based {mmWave} {JCAS}: Concept and Demonstration},	
	year={2024},
	month={Feb.},
	volume={},
	number={},
	pages={1-16},
	doi={10.1109/TWC.2024.3364048}
}

@BOOK{bookSFO,
	title = {{RF} Analog Impairments Modeling for Communication Systems Simulation: Application to {OFDM}-based Transceivers},
	author = {Lydi Smaini},
	year = {2012},
	publisher = {John Wiley \& Sons, Ltd},
	address = {New York, NY, USA}
}

@ARTICLE{tsai2005,
	author={Pei-Yun {Tsai} and Hsin-Yu {Kang} and Tzi-Dar {Chiueh}},
	journal={IEEE Trans. Veh. Technol.}, 
	title={Joint weighted least-squares estimation of carrier-frequency offset and timing offset for {OFDM} systems over multipath fading channels}, 
	year={2005},
	month={Jan.},
	volume={54},
	number={1},
	pages={211-223},
	doi={10.1109/TVT.2004.838891}
}

@Article{wu2012,
	author={Wu, Y. and Zhao, Y. and Li, D.},
	journal={Wireless Pers. Commun.}, 
	title={Sampling Frequency Offset Estimation for Pilot-Aided {OFDM} Systems in Mobile Environment}, 
	year={2012},
	month={Jan.},
	volume={62},	
	pages={215-226}
}

@ARTICLE{erup1993,
	author={Erup, L. and Gardner, F.M. and Harris, R.A.},
	journal={IEEE Trans. Commun.}, 
	title={Interpolation in digital modems — Part {II}: Implementation and performance}, 
	year={1993},
	month={Jun.},
	volume={41},
	number={6},
	pages={998-1008},
	doi={10.1109/26.231921}
}

@INPROCEEDINGS{farrow1988,
	author={Farrow, C.W.},
	booktitle={1988 IEEE Int. Symp. Circuits Syst.}, 
	title={A continuously variable digital delay element}, 
	year={1988},
	month={Jun.},
	volume={3},
	number={},
	pages={2641-2645},
	doi={10.1109/ISCAS.1988.15483}
}

@INPROCEEDINGS{nguyen2002,	
	author={Van Duc Nguyen and Kuchenbecker, H.-P.},	
	booktitle={13th IEEE Int. Symp. Pers., Indoor Mobile Radio Commun.}, 	
	title={Intercarrier and intersymbol interference analysis of {OFDM} systems on time-invariant channels}, 
	year={2002},
	month={Sept.},
	volume={4},	
	number={},	
	pages={1482-1487},	
	doi={10.1109/PIMRC.2002.1045425}
}

@INPROCEEDINGS{wang2023,	
	author={Wang, Lin and Wei, Zhiqing and Su, Liyan and Feng, Zhiyong and Wu, Huici and Xue, Dongsheng},	
	booktitle={2023 21st Int. Symp. Model. Omptim. Mobile, Ad Hoc, Wireless Netw.}, 
	title={Coherent Compensation Based {ISAC} Signal Processing for Long-Range Sensing: (Invited Paper)}, 	
	year={2023},
	month={Aug.},	
	volume={},	
	number={},	
	pages={689-695},	
	doi={10.23919/WiOpt58741.2023.10349853}
}

\end{document}